\documentclass{emulateapj}

\usepackage{morefloats}

\shorttitle{Variation of BAL Troughs}
\shortauthors{Filiz Ak et~al.}

\usepackage{natbib}
\begin{document}

\title{Broad Absorption Line Variability \\ on Multi-Year Timescales in a 
Large Quasar Sample}

\author{N. Filiz Ak\altaffilmark{1,2,3},
W.~N. Brandt\altaffilmark{1,2}, 
P. B. Hall\altaffilmark{4},  
D.~P. Schneider\altaffilmark{1,2}, 
S. ~F. Anderson \altaffilmark{5}, 
F. Hamann\altaffilmark{6}, 
B.~F. Lundgren\altaffilmark{7,8}, 
Adam~D. Myers\altaffilmark{9},
I. P\^aris \altaffilmark{10},
P. Petitjean\altaffilmark{11}, 
Nicholas~P. Ross \altaffilmark{12},   
Yue Shen\altaffilmark{13,14,15},
Don York\altaffilmark{16,17}}
\altaffiltext{1}{Department of Astronomy \&   Astrophysics, Pennsylvania 
State University,  University Park, PA, 16802, USA}
\altaffiltext{2}{Institute for Gravitation and the Cosmos, Pennsylvania 
State University, University Park, PA 16802, USA}
\altaffiltext{3}{Faculty of Sciences, Department of Astronomy and Space 
Sciences, Erciyes University, 38039 Kayseri, Turkey}
\altaffiltext{4}{Department of Physics and Astronomy, York University, 
4700 Keele St., Toronto, Ontario, M3J 1P3, Canada}
\altaffiltext{5}{Astronomy Department, University of Washington, Seattle, WA 98195, USA}
\altaffiltext{6}{Department of Astronomy, University of Florida, Gainesville, FL 32611-2055, USA}
\altaffiltext{7}{Department of Astronomy, University of Wisconsin, Madison, WI
53706, USA}
\altaffiltext{8}{NSF Astronomy and Astrophysics Postdoctoral Fellow}
\altaffiltext{9}{Department of Physics and Astronomy, University of Wyoming, Laramie, WY 82071, USA}
\altaffiltext{10}{Departamento de Astronom\'ia, Universidad de Chile, Casilla 36-D, Santiago, Chile}
\altaffiltext{11}{Universite Paris 6, Institut d'Astrophysique de Paris, 75014, 
Paris, France}
\altaffiltext{12}{Lawrence Berkeley National Laboratory, 1 Cyclotron Road,
Berkeley, CA 92420, USA}
\altaffiltext{13}{Harvard-Smithsonian Center for Astrophysics, 60 Garden St., 
MS-51, Cambridge, MA 02138, USA}
\altaffiltext{14}{Carnegie Observatories, 813 Santa Barbara Street, Pasadena, CA 91101, USA}
\altaffiltext{15}{Hubble Fellow}
\altaffiltext{16}{The University of Chicago, Department of Astronomy and Astrophysics, Chicago, IL 60637, USA}
\altaffiltext{17}{The University of Chicago, Enrico Fermi Institute, Chicago, IL 60637, USA}

\email{nfilizak@astro.psu.edu}

\begin{abstract}

We present a detailed investigation of the variability of 428 C\,{\sc iv} and 
235 Si\,{\sc iv} Broad Absorption Line (BAL) troughs identified in multi-epoch 
observations of 291 quasars by the Sloan Digital Sky Survey-I/II/III. These
observations primarily sample rest-frame timescales of \hbox{1--3.7~yr} over 
which significant rearrangement of the BAL wind is expected. 
We derive a number of observational results on, e.g.,  
the frequency of BAL variability, 
the velocity range over which BAL variability occurs, 
the primary observed form of BAL-trough variability, 
the dependence of BAL variability upon timescale, 
the frequency of  BAL strengthening vs. weakening, 
correlations between BAL variability and BAL-trough profiles, 
relations between C\,{\sc iv} and Si\,{\sc iv} BAL variability, 
coordinated multi-trough variability, and 
BAL variations as a function of quasar properties.
We assess implications of these observational results for quasar
winds. Our results support models where most BAL absorption
is formed within an order-of-magnitude of the wind-launching radius, 
although a significant minority of BAL troughs may arise on larger scales. We estimate 
an  average lifetime for a BAL trough along our line-of-sight of a few 
thousand years. 
BAL disappearance and emergence events appear to be extremes of general BAL 
variability, rather than being qualitatively distinct phenomena. 
We derive the parameters of a random-walk model for BAL EW 
variability, finding that this model can acceptably describe some 
key aspects of EW variability. The coordinated trough variability of BAL 
quasars with multiple troughs suggests that changes in ``shielding gas''
may play a significant role in driving general BAL variability.
\end{abstract}

\section{Introduction}\label{vintro}

The high-velocity winds from quasars are important for several related 
reasons. First, these winds can significantly affect observed quasar 
properties via, e.g., ultraviolet (UV) line absorption, high-ionization line 
emission, optical/UV reddening, and \hbox{X-ray} absorption 
\citep[e.g.,][]{wcs81,turnshek88,leighly04,collin06,gal02,gal06,gibson09,richards11}. 
Second, wind absorption lines are observed frequently, indicating that 
winds have a high covering factor and are a substantial part of quasar 
nuclear regions \citep[e.g.,][]{gb08,gibson09,allen11}. 
Third, winds might improve the efficiency of  accretion onto the central 
supermassive black hole (SMBH) by removing angular momentum from 
the accretion disk \citep[e.g.,][]{emmering92,konigl94}.
Finally, winds can evacuate gas from the host galaxy, perhaps shaping 
SMBH growth and galaxy evolution 
\citep[e.g.,][]{dimatteo05,chartas09,rupke11,sturm11,borquet13}.

The strongest absorption lines created by quasar winds are Broad 
Absorption Line (BAL) troughs with velocity widths greater than 
2000~km~s$^{-1}$ and typical outflow velocities of 
\hbox{1000--30000~km~s$^{-1}$} \citep[e.g.,][]{wey91}. 
Many BALs are believed to be formed in an equatorial wind that is 
launched from the accretion disk at \hbox{10--100} light days from 
the SMBH \citep[e.g.,][]{murray95,proga00}. If the BALs are formed
in the vicinity of the launching region, then the timescale for 
wind material to cross the region of interest is about \hbox{1--10~yr}, 
and this is a reasonable characteristic timescale over which flow 
structures might be expected to change. This is also the characteristic 
timescale for significant angular rotation of the accretion disk at 
the wind-launching radius. Assessments of the transverse
velocities of BAL material indicate these are often comparable 
to the aforementioned outflow velocities \citep[e.g.,][]{cap11,hall11}, 
and characteristic variability timescales of years 
are again deduced for material moving transversely through our 
line-of-sight. Thus, studies of multi-year BAL variability 
can provide useful insights into the nature of quasar winds.

The existence of BAL variability has been known for over two decades 
\citep[e.g.,][]{smith88,turnshekea88,barlow92}.
Early investigations of this phenomenon were generally single-object studies
with \hbox{2--4} observational epochs, although \citet{barlow93} performed 
an early spectroscopic monitoring survey of 23 BAL quasars. In recent 
years, systematic sample-based studies of BAL variability, investigating 5--30 
objects,  have become increasingly common 
\citep[e.g.,][]{lundgren07,gibson08,gibson10,cap11,cap12,ak12,haggard12,miller12,vivek12}; 
see Table~\ref{varet1} for a summary of the basic properties of these samples. 
Sample-based studies have the advantage of allowing broadly applicable and 
statistically reliable conclusions about BAL variability to be drawn. BAL variability 
has been found to be a complex and diverse phenomenon. Generally, changes 
in the residual flux in portions of BAL troughs are observed, while detections of 
BAL acceleration/deceleration are much rarer \citep[e.g.,][]{vilkov01,gibson08,gibson10,cap12}. 

We have been using observations taken as part of the ongoing Baryon
Oscillation Spectroscopic Survey \citep[BOSS;][]{dawson13} of the Sloan 
Digital Sky Survey-III \citep[SDSS-III;][]{eisen11} to perform the largest survey 
of multi-year BAL variability to date (see Section~\ref{obs2} and \citealp{ak12} 
for further description). 
Our final sample will include $\approx$~2100 BAL quasars with high-quality
spectra providing multi-year variability coverage in the rest frame; 
this size is about two orders of magnitude larger than other samples being 
used to investigate multi-year BAL variability (see Table~\ref{varet1}). In 
\citet{ak12} we presented some first results from the survey focused on 
C~{\sc iv} BAL disappearance events. In this paper, we provide more general 
findings regarding the variability of C~{\sc iv} and Si~{\sc iv} BAL troughs, 
primarily on multi-year timescales but also extending to much shorter timescales. 
These findings are based upon 291 BAL quasars selected from our full sample
to have particularly high-quality spectroscopic coverage of these troughs; 
428 distinct C~{\sc iv} and 235 distinct Si~{\sc iv} troughs are utilized. 
Our overall approach is first to characterize systematically how BAL
troughs vary on multi-year timescales and then to use this characterization to 
derive physical implications for BAL outflows. For example, our results provide 
insights into the radial distance where most BAL troughs are formed, the lifetimes 
of BAL troughs along our line of sight, the connection between BAL 
disappearance/emergence events and general BAL variability, and the 
driving mechanisms of BAL variability.

In Section~\ref{os2} we describe the observations and sample selection underlying
this work, and in Section~\ref{dpa2} we describe data preparation and analysis 
approaches. Our observational results on BAL variability are presented in 
Section~\ref{vres}. In Section~\ref{disc} we provide a discussion of implications for quasar 
winds, and in Section~\ref{vsummary} we  present a summary and describe promising 
future avenues of relevant research. 

Throughout this work we use a cosmology with 
\hbox{$H_0=70$~km~s$^{-1}$~Mpc$^{-1}$}, 
\hbox{$\Omega_M=0.3$}, and 
\hbox{$\Omega_{\Lambda}=0.7$}. 
All time intervals and EWs are in the rest frame of the quasar unless stated 
otherwise. Negative signs for velocities indicate that a BAL trough is blueshifted
with respect to the systemic velocity. We define EWs of absorption features to 
be positive. Positive EW variations indicate strengthening, and 
negative values indicate weakening.

\section{Observations and Sample Selection}\label{os2}
\subsection{Observations}\label{obs2}

We have utilized  data from the  Sloan Digital Sky Survey-I/II \citep[hereafter 
``SDSS";][]{york00} and BOSS that use a mosaic CCD camera \citep{gunn98} 
plus multi-object spectrograph on a dedicated 2.5~m telescope \citep{gunn06} 
at Apache Point Observatory. Between 2000--2008, the SDSS I/II completed 
spectroscopy over 9380 deg$^{2}$ and obtained over 1.6 million spectra in total, 
including more than 105000 quasars \citep[e.g.,][]{richards02,aba09,schneider10}. BOSS is 
observing a sample of $\approx$~210000 quasars, the majority of which are  
at $z > 2.2$ \citep{ross12,paris12},  with the main scientific motivation being to measure 
the baryon acoustic oscillation (BAO) feature in the Lyman-$\alpha$ forest 
\citep[e.g.,][]{busca12,slosar13}.  Using an improved spectrograph, BOSS spectra have 
coverage between \hbox{3600--10000}~\AA\ at a resolution of \hbox{1300--3000} 
\citep[see][]{dawson13,smee13}.

In addition to its primary quasar program, BOSS is executing several ancillary 
projects \citep[see][]{dawson13} including one focused on investigating the 
dynamics of quasar winds over multi-year timescales. This project re-observes 
selected bright BAL quasars that have previous SDSS spectral observations to enable 
a high-quality and relatively unbiased study of BAL variability over multi-year 
timescales in the rest frame of the quasar. The project targets were selected 
using information from the catalog of BAL quasars for SDSS DR5 \citep{gibson09} 
and the SDSS DR5 quasar catalog \citep{schneider07}. The details of the BAL 
target selection are described in Section~2.1 of \citet{ak12}. Briefly, the 2005 selected 
targets are optically bright ($i < 19.3$) BAL quasars with redshifts 
\hbox{$0.48\leq z \leq4.65$}. The observed SDSS spectra of these targets have 
a signal-to-noise ratio per 0.4~\AA\, pixel at 1650--1750~\AA\, of 
\hbox{SN$_{1700}>$~6}; \hbox{SN$_{1700}$} is defined for the continuum 
redward of C\,{\sc iv} and should not be affected by BAL absorption. The targets 
were chosen to have a modified balnicity index BI$_0 > 100~\rm{km\,s^{-1}}$. 
BI$_0$ is defined by \cite{gibson08} using the following equation:
\begin{equation}
\mbox{BI$_0$} \equiv \int_{0}^{-25000} \left( {1-\frac{f(v)} {0.9}} \right) C dv.
\label{eqv1}
\end{equation}
where $f(v)$ is the normalized flux density  as a function of velocity, $v$, and 
$C$ is a constant which is equal to 1.0 only when a trough is wider than 
$2000~\rm{km\,s^{-1}}$, it is  otherwise 0.0.

In this study, we use SDSS spectra observed between MJD~51602 and 
54557 (2000 February 28 to 2008 January 04) and BOSS spectra 
observed between MJD~55176 and 56109 (2009 December 11 to 
2012 July 01); i.e., we utilize spectra taken after the completion of hardware 
commissioning for both SDSS and BOSS. Between these dates, 1087 of the 
2005 targets from the ancillary project were observed by BOSS. 

\subsection{Sample Selection}\label{ss2}

In this section we present the  selection criteria used to create a ``main sample'' 
for studying C\,{\sc iv} and Si\,{\sc iv} BAL variability on multi-year 
timescales in the rest frame. We select our main-sample BAL quasars from the 
targets observed via the ancillary project based on the following criteria:  

\begin{enumerate}

\item

To enable more robust continuum fits, we select only the quasars that have 
spectral coverage of the relatively line free (RLF, see Section~\ref{spa2}) windows 
blueward of the Si\,{\sc iv} line as well as redward of the C\,{\sc iv} line; these 
windows play a key role in constraining the fitted continuum. Thus, we utilize 
the quasars that have $z>2$ from the sample of 1087 observed objects. 
(619 quasars)

\item
The targeted sample of quasars was required to have SN$_{1700}>$~6 for 
the SDSS spectrum, although higher values of SN$_{1700}$ are advantageous 
for the study of moderate or weak BAL variations. We have thus chosen to 
utilize quasars that have SN$_{1700} > 10$ for both the SDSS and BOSS spectra; 
visual inspection shows that this choice provides a good balance between high 
spectral quality and large sample size.
(356 quasars out of 619)

\item
As in \citet{ak12}, to avoid confusion between emission-line and BAL variability, 
we  consider only  the BAL troughs that are significantly detached from the 
C\,{\sc iv} and Si\,{\sc iv} emission lines by setting  velocity limits for BAL 
troughs. We consider the BAL regions of each transition that lie between 
$-$3000 and $-$30000~$\rm{km\,s^{-1}}$. To select the quasars with 
moderate-to-strong BAL troughs, we require  that the  balnicity index of both 
the SDSS and BOSS spectra of the main-sample quasars have 
\hbox{$\rm{BI'} > 100~\rm{km\,s^{-1}}$}; a consistent threshold for both 
SDSS and BOSS is required to avoid biases in our later analyses.
\hbox{$\rm{BI'}$} is defined by \cite{ak12} using the following equation:
\begin{equation}
{\rm BI'} \equiv \int_{-3000}^{-30000} \left( {1-\frac{f(v)} {0.9}} \right) C dv.
\label{eeqv1}
\end{equation}
Similar to the BI$_0$ definition, in this equation $f(v)$ is the normalized flux density  
as a function of velocity, $v$, and $C$ is a constant which is equal to 1.0 only 
when a trough is wider than $2000~\rm{km\,s^{-1}}$, it is  otherwise 0.0.
(297 quasars out of 356)

\item 
We have rejected six quasars ($2\%$) from our main sample because of 
difficulties in defining the continua and/or emission lines in their spectra. 
Spectra of these quasars  possess strong absorption lines of many transitions 
causing large systematic uncertainties in BAL measurements. 
(291 quasars out of 297)
\end{enumerate}

Based on these criteria,  we selected 291 BAL quasars as our main sample. 
All  291 quasars have at least one observation from SDSS and one 
from BOSS, and 22\% have additional SDSS and/or BOSS observations. 
In total, our sample contains  699 spectra  of main-sample quasars that cover 
rest-frame timescales from 5.9 hr to 3.7~yr. 

We have cross-matched our 291 main-sample quasars with the catalog of 
quasar properties from SDSS DR7 \citep{shen11} and obtained their absolute 
$i$-band magnitudes, $M_{i}$, estimated bolometric 
luminosities, $L_{\rm{Bol}}$, Eddington-luminosity ratios, 
$L_{\rm{Bol}}/L_{\rm{Edd}}$, and  virial black-hole mass estimates, 
$M_{\rm BH}$. In addition we use the radio-loudness parameter, $R$, defined as 
$R = f_{\rm 6cm}/f_{\rm 2500{\rm \AA}}$, where  $f_{\rm 6cm}$ is the radio flux 
density at rest-frame 6 cm and $f_{2500{\rm \AA}}$ is the optical flux density at 
rest-frame 2500 $\rm \AA$. \cite{shen11} calculated  the $R$ parameter using 
radio emission detected in Very Large Array (VLA) Faint Images of the Radio 
Sky at Twenty-Centimeters \citep[FIRST;][]{becker95} observations. We obtain 
redshift values from \cite{hw10}, and these are used throughout.

\begin{figure*}[!t!]
\epsscale{0.6}
\plotone{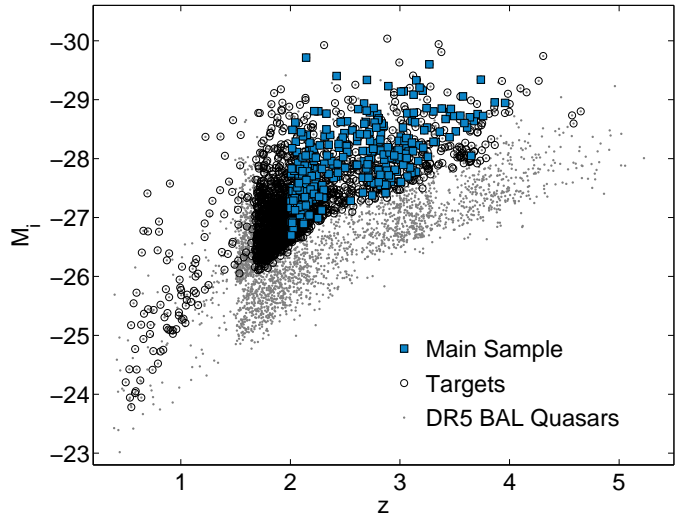} 
\caption{Absolute $i$-band magnitude, $M_{i}$, vs. redshift for the main
sample of this paper (blue squares), the 2005 targets of the BOSS
ancillary project on BAL quasars (open circles), and all SDSS DR5
BAL quasars (dots).}
\label{vare1}
\end{figure*}

\begin{figure*}[!t!]
\epsscale{0.6}
\plotone{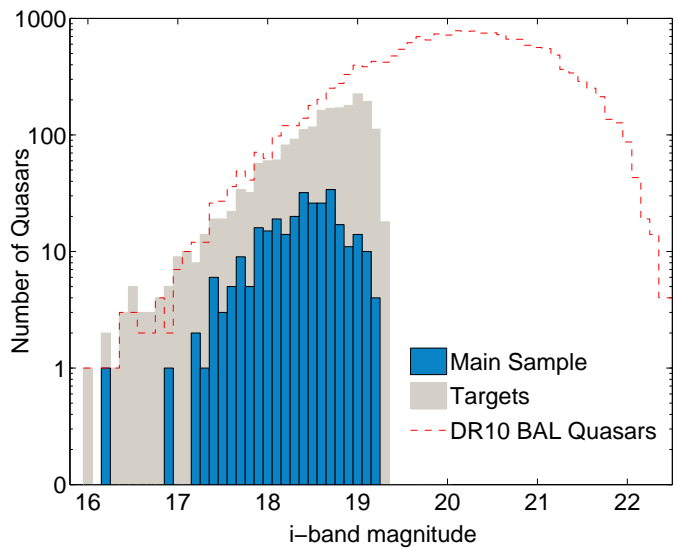} 
\caption{Comparison of the $i$-band apparent magnitude distributions of the
main sample of this paper (blue), the 2005 targets of the BOSS
ancillary project on BAL quasars (gray), and the BAL quasars identified
in the SDSS DR10 quasar catalog (dashed red line). Note that we are
targeting the brightest BAL quasars in the SDSS sky area in order to
obtain spectra of the highest possible quality.}
\label{vare2}
\end{figure*}

We illustrate some of the basic properties of our sample in Figures~\ref{vare1}
and \ref{vare2}. Figure~\ref{vare1} shows $M_{i}$ vs. redshift for 
the main sample of this paper, the 2005 targets of the BOSS ancillary project 
on BAL quasars, and all SDSS DR5 BAL quasars. The $M_{i}$ distribution of 
the main-sample quasars spans about the same range as that of the general 
ancillary program targets from \hbox{$z=2$--4}, covering a factor of $\approx 5$ 
in luminosity at any given redshift. Figure~\ref{vare2} compares the $i$-band 
apparent magnitude distributions of our main sample, the 2005 BOSS ancillary
project targets, and the BAL quasars identified in the SDSS DR10 quasar 
catalog \citep{paris13}. It is clear that the main sample and ancillary project 
effectively cover the brightest BAL quasars that generally provide the highest 
quality SDSS and BOSS spectra. The mean $i$-band magnitude is 18.4 
for the main-sample quasars and 18.7 for the ancillary project targets. 
Using the catalog of BAL quasars for SDSS DR5 \citep{gibson09}, we found 
that 22 of the 291 main-sample quasars possess Al\,{\sc iii} BALs and thus are 
identified as low-ionization BAL quasars. Additional unidentified low-ionization 
BAL quasars may be present; e.g., among objects that lack spectral coverage
of the important Mg\,{\sc ii} low-ionization transition. 

\section{Data Preparation and Analysis}\label{dpa2}
\subsection{Basic Spectral Preparation}\label{spa2}

For the purpose of investigating BAL variability, we compared the multi-epoch 
spectral observations of our main-sample quasars. We have normalized each 
spectrum by a model for the continuum following the procedure in  Section~ 3.1 of  
\citet{ak12}. Briefly, we first correct the spectra for Galactic extinction using 
the $A_{V}$ values from \cite{schlegel98} and then transform from the observed 
frame to the rest frame using the redshift values from \cite{hw10}. We remove 
the pixels from the spectra that contain significant night-sky line residuals that 
are flagged by the SDSS and BOSS data-reduction pipelines. 

To reconstruct the underlying continuum, as in \citet{gibson08, gibson09} and \citet{ak12}, 
we define RLF windows to be the  following spectral regions: 
\hbox{1250--1350}~\AA\,, 1700--1800~\AA\,, 1950--2200~\AA\,, 
2650--2910~\AA\,, 3950--4050~\AA. The RLF windows were selected to be  
relatively free from  emission and absorption lines considering the composite 
quasar spectra of \citet{vanden01}. However, some weak emission and/or absorption 
lines can be present in  these spectral regions. Thus, to exclude the data points that deviate 
from the fit by more than $3\sigma$, we apply an iterative sigma-clipping algorithm 
using a non-linear least squares fit. We fit the RLF windows of each spectrum 
with an intrinsically reddened power-law continuum model where we use 
Small-Magellanic-Cloud type reddening. We calculate the continuum uncertainties  
using $\Delta \chi^2$ confidence-region estimation for 68.3\% confidence bounds. 
Throughout this work, we propagate  the continuum uncertainties into the  
uncertainties on rest-frame EW measurements. As in previous studies 
\citep[e.g.,][]{lundgren07,ak12}, we do not model the emission lines since 
investigation of emission-line characteristics is beyond the scope of this study 
(also see Section~\ref{ss2}).

\subsection{Identification and Measurements of BAL Troughs}\label{bal2}

As is  common practice \citep[e.g.,][]{trump06,gibson08,gibson09,allen11,ak12}, 
we smoothed each spectrum using a Savitzky-Golay algorithm to perform local 
linear regression for three consecutive data points \citep[see Section~ 3.2 of][]{ak12}. 
We utilize normalized and smoothed spectra only for identification of BAL troughs; 
unsmoothed spectra are used for further calculations. Only BAL troughs in the 
velocity range $-$3000 to $-30000~\rm{km\,s^{-1}}$ are considered (see 
Section~\ref{ss2}); the small fraction ($\approx2.5$\%) of extremely high-velocity 
C\,{\sc iv} BAL troughs exceeding the $-30000~\rm{km\,s^{-1}}$ limit were 
removed by considering Si\,{\sc iv} BAL troughs at corresponding velocities.

\begin{figure}[t!]
\epsscale{1.15}
\plotone{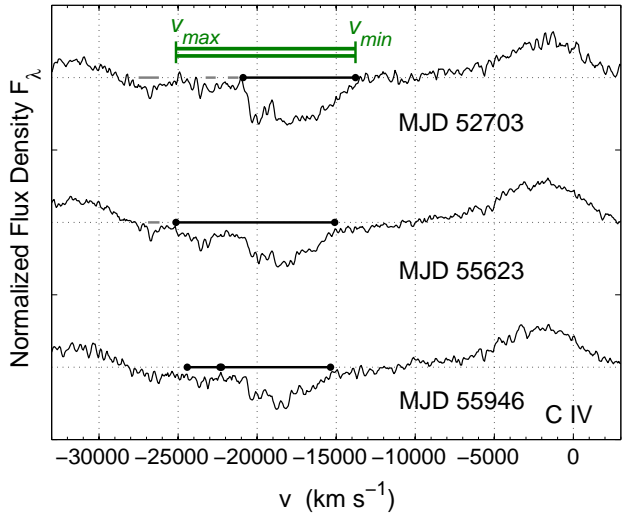} %10413
\caption{Example of our adopted BAL-trough definition illustrated using the 
three available spectra of the quasar SDSS~J090944.05+363406.7. The 
three normalized spectra for this quasar are arbitrarily offset in flux 
for clarity of presentation. Horizontal dotted lines show the continuum
levels for each spectrum, and the tick marks on the $y$-axis show the 
zero level for each spectrum. Horizontal black bars show absorption lines 
with $\Delta v \geq 2000~\rm{km\,s^{-1}}$, corresponding to the 
traditional BAL definition, and horizontal gray bars show absorption lines 
with \hbox{$\Delta v = 500$--2000~km~s$^{-1}$}. The variable absorption 
complex is complicated; in some spectra only part of it is classified 
as a BAL under the canonical definition. We use all 
three available spectra to define the minimum and maximum velocities of 
the trough. The horizontal double green bar shows the resulting BAL trough 
lying between $v_{\rm min}$ and $v_{\rm max}$ (see Section~\ref{bal2}).}
\label{varf1}
\end{figure}

The canonical definition of BAL troughs (see Equation~\ref{eeqv1}) was developed 
for the purpose of finding 
BAL troughs in a single-epoch spectrum. However, our primary purpose here is 
investigating BAL variability in multi-epoch spectra. Therefore, we adopt a modified 
BAL-trough definition more appropriate for our purpose that is strongly motivated 
by the canonical definition and reduces to it for single-epoch data. Our adopted 
BAL-trough definition utilizes the information from all available spectral 
observations of a quasar.  As is well known, absorption troughs are sometimes 
isolated and sometimes appear in complexes in which single troughs may split or 
adjacent troughs may merge over time. To address these complications, we treat 
each BAL complex as a single BAL trough. We identify BAL complexes using the 
following algorithm (see Figure~\ref{varf1}):
\begin{enumerate}

\item
We first identify BAL and mini-BAL troughs (hereafter just ``troughs'') under the 
canonical definition in each single-epoch spectrum of a quasar.\footnote{Mini-BALs 
are defined using Equation~\ref{eeqv1} but for a trough width of 
500--2000$~\rm{km\,s^{-1}}$ \citep[cf.][]{hs04}.} We set the 
maximum velocity of a trough to be $v_{\rm max,t}$ and the minimum velocity to 
be $v_{\rm min,t}$.

\item
We select $v_{\rm min,t}$ of the highest velocity trough in the first-epoch 
spectrum and  compare the corresponding velocities in all the available spectra. 
If the $v_{\rm min,t}$ velocity intersects any trough region in the other 
available epochs, we re-assign $v_{\rm min,t}$ to be the lowest velocity of this 
intersecting trough and repeat the comparison to the other available spectra.  
If the $v_{\rm min,t}$ velocity does not intersect any trough region in the other 
available spectra, we set $v_{\rm min,t}$ to be the minimum red-edge velocity 
of the BAL complex, $v_{\rm min}$.

\item
To define the maximum velocity of the complex, we take the $v_{\rm max,t}$
of the  lowest-velocity trough associated with the complex and compare with 
the other available epochs. If the $v_{\rm max,t}$ velocity in the other spectra 
intersects a trough, we set $v_{\rm max,t}$ to be the highest velocity of 
this intersecting trough. If the $v_{\rm max,t}$  velocity does not intersect  
any trough region in the other available spectra, we set $v_{\rm max,t}$ to be 
the maximum blue-edge velocity of the BAL complex, $v_{\rm max}$.

\end{enumerate}
\noindent
In this algorithm  each trough can be associated with only one trough complex. 
Each trough complex includes at least one trough which is wider than 
$2000~\rm{km\,s^{-1}}$ lying between $-$3000 and $-30000~\rm{km\,s^{-1}}$.

After implementing the above algorithm, we define each BAL complex lying 
between $v_{\rm max}$ and $v_{\rm min}$ as a distinct BAL trough. Here, 
$v_{\rm max}$ is the maximum velocity taken to be the velocity at the blue edge 
for any trough associated with the complex across all available spectra of each 
quasar. Similarly, $v_{\rm min}$ is defined using the red-edge velocities. After 
an automated identification of BAL troughs using  our algorithm above, we 
visually inspect all the available spectra of each main-sample quasar. The 
inspection shows that our approach for BAL-trough identification is appropriately
 implemented for both C\,{\sc iv} and Si\,{\sc iv} BAL troughs. We found that, in 
the majority of cases, the complex would be identified as a single BAL trough 
under the canonical definition in at least one of our epochs. Moreover, our 
adopted BAL-trough definition produces the same results as  the canonical 
definition for non-merging and non-splitting BAL troughs that lie between 
constant $v_{\rm max}$ and $v_{\rm min}$ in all the available spectral 
observations. Using our adopted BAL-trough definition we identified a total of 
428 distinct C\,{\sc iv} and 235 distinct Si\,{\sc iv} BAL troughs in the 699 
main-sample spectra. 

Figure~\ref{varf1} illustrates our adopted BAL-trough definition and the canonical 
one using all the available spectra of the quasar SDSS~J090944.05+363406.7. 
If we apply  the canonical BAL-trough definition in this example,  the two adjacent
BAL troughs seen in the last epoch  would be identified as two distinct BALs. 
However, the same structure appears as one distinct BAL trough in the 
second-epoch spectrum and in the first-epoch spectrum appears as two 
mini-BALs along with one BAL trough.  As  is clearly seen in this example, our 
adopted BAL-trough definition produces more physically meaningful results than 
the canonical definition for the purpose of studying variability in multi-epoch 
spectra. Moreover, we select the $v_{\rm max}$ and $v_{\rm min}$ velocities 
using information from all the available spectra instead of only a single-epoch 
observation. If we were to take the $v_{\rm max}$ and $v_{\rm min}$ velocities 
as the blue-edge and red-edge velocities of the complex where all absorption has 
merged to one BAL trough (i.e., the second-epoch spectrum in this example), we 
would lose the pertinent information from the part of the BAL troughs which 
extend  beyond these velocities in the other-epoch spectra (i.e., the first-epoch 
spectrum in this example).

\begin{figure*}[]
\epsscale{0.75}
\plotone{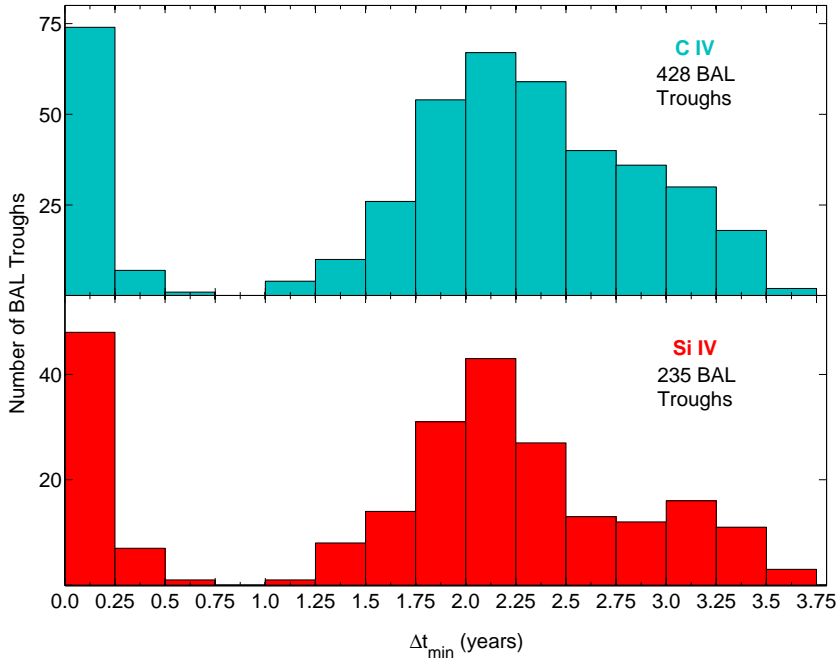}
\caption{Distributions of the minimum sampled rest-frame timescale, 
$\Delta t_{\rm min}$, for 428 distinct C\,{\sc iv} (upper panel) and 235 
distinct Si\,{\sc iv} (lower panel) BAL troughs present in the spectra of 
the main-sample quasars. We have significant trough statistics 
on timescales as long as \hbox{3--3.5~yr}.}
\label{varf2}
\end{figure*}

In this study, we investigate BAL-trough variability on a large range of rest-frame 
timescales. The spectral observations from SDSS-I/II and BOSS provide 
coverage of long timescales, typically longer than 1~yr.  To sample shorter 
timescales, we use the additional observations from SDSS and/or BOSS that 
are available for more than 20\% of our main-sample quasars. To avoid the 
repeat examination of BAL troughs and the associated multi-counting biases, 
we utilize only the two-epoch spectra for each quasar that give the \textit{minimum} 
sampled rest-frame timescale, $\Delta t_{\rm min}$.\footnote{For example, 
if a quasar has observations that sample rest-frame timescales of 
0.01, 0.02, 0.03, 2.00, 2.01, and 2.03~yr, we select only the two-epoch spectra that  
sample the 0.01~yr timescale. Note that the 2.00, 2.01, and 2.03~yr timescales 
are nearly the same (agreeing to within 1.5\%) and provide little independent information. 
Thus, 
using all three of these timescales would result in this object being inordinately weighted 
in statistical characterizations of BAL variability on \hbox{$\approx2$}~yr timescales 
(i.e., causing multi-counting bias). \label{fot19}} 
 Although we use only such two-epoch spectra for our calculations below, 
we still consider all available observations in our BAL-trough identification algorithm
 as we have found this provides the most physically meaningful results. 
By selecting $\Delta t_{\rm min}$, we sample rest-frame timescales from a few hours to a 
few years. Figure~\ref{varf2} shows the distribution of minimum sampled 
rest-frame timescales, $\Delta t_{\rm min}$, for distinct BAL troughs identified 
in our main-sample spectra. The $\Delta t_{\rm min}$ values range between 5.9~hr 
and 3.7~yr with a median of 2.1~yr. Given that C\,{\sc iv} BAL troughs are not 
always accompanied by Si\,{\sc iv} BAL troughs, we ran a two-sample 
Kolmogorov-Smirnov (KS) test to compare the sampled timescale distributions 
of  C\,{\sc iv} and Si\,{\sc iv} BAL troughs and found no significant difference.

In some sections of this study, we focus on  BAL-trough variability characteristics 
solely on multi-year timescales. Therefore, we define another timescale of 
$\Delta t_{\rm min,1}$ to sample  minimum rest-frame timescales of more than 
1 yr. By this definition, we select the same non-repeating sample of distinct BAL 
quasars as with the $\Delta t_{\rm min}$ selection.
$\Delta t_{\rm min,1}$ ranges between 1--3.7~yr with a median of 2.3~yr. 
Compared to previous studies \citep[e.g.,][]{lundgren07,gibson08,gibson09,
cap11, cap12}, and especially those focusing on multi-year timescales, we have 
a significantly larger (by about an order-of-magnitude) BAL-trough sample.

We  measure the rest-frame EW of each BAL trough in each epoch and calculate
the uncertainties on EWs using Equations~1 and 2 of \citet{kaspi02}, where 
uncertainties are derived by propagating the continuum-estimation errors (see 
Section~\ref{spa2}) and the observational errors of each contributing pixel.  In addition, 
we measure  the weighted  centroid velocity, $v_{\rm cent}$, for each BAL trough; 
i.e., the mean of the velocities where each data point is weighted with its distance 
from the normalized continuum level. We also calculate an average BAL-trough 
depth, $d_{\rm{BAL}}$, which is the mean distance from the normalized 
continuum level for each data point of a BAL trough.

\subsection{Identification and Measurements of Variable BAL Troughs}\label{vr2}

As one approach to identify variable BAL troughs in our main sample, we select 
BAL troughs showing significant EW variations. To calculate EW variations, 
$\Delta\rm{EW}$, and uncertainties on this quantity, $\sigma_{\Delta\rm{EW}}$, 
we use the following equations:
\begin{eqnarray}
\Delta\rm{EW} = \rm{EW}_2 - \rm{EW}_1,  ~~~~~~
\sigma_{\Delta\rm{EW}} = \sqrt{\sigma_{\rm{EW}_2}^2 + \sigma_{
\rm{EW}_1}^2}
\label{eqv2}
\end{eqnarray}
where $\rm{EW}_1$  and $\rm{EW}_2$ are the EWs measured from two-epoch
spectra that are observed at times $t_1$ and $t_2$.  In our sample, the mean 
$\sigma_{\Delta\rm{EW}}$ is 0.5~\AA\, for C\,{\sc iv} and 0.4~\AA\, for Si\,{\sc iv} 
BAL troughs. Similarly, we calculate fractional EW variations,
$\frac {\Delta\rm{EW}} {\langle \rm{EW} \rangle}$, and corresponding 
uncertainties, $\sigma_{\frac {\Delta\rm{EW}}  {\langle \rm{EW} \rangle}}$, with 
the following equations:
\begin{eqnarray}
\frac {\Delta\rm{EW}}  {\langle \rm{EW} \rangle} = {\frac {(\rm{EW}_2 - 
\rm{EW}_1)} {(\rm{EW}_2 + \rm{EW}_1) \times 0.5}},
\nonumber \\ \nonumber \\
\sigma_{\frac {\Delta\rm{EW}}  {\langle \rm{EW} \rangle}} = \frac 
{4 \times (\rm{EW}_2 \sigma_{\rm{EW}_1} + \rm{EW}_1 
\sigma_{\rm{EW}_2} )} {(\rm{EW}_2 + \rm{EW}_1)^2}
\label{eqv3}
\end{eqnarray}

We identify 248 variable  C\,{\sc iv} BAL troughs and 119 variable Si\,{\sc iv} BAL 
troughs  showing EW variations at a significance level of more than $3\sigma$ 
on timescales of more than 1~yr ($\Delta t_{\rm min,1}$). Similarly, we identify 
223 variable C\,{\sc iv} 
BAL troughs and 99 variable Si\,{\sc iv} BAL troughs by comparing the two-epoch 
spectra that sample the $\Delta t_{\rm min}$ timescales in the rest frame. 

Considering that variations tend to occur in portions of BAL troughs \citep[e.g.,][]
{gibson08,cap11}, as an alternative approach, we define a variable BAL trough 
to have at least one variable region. For the purpose of determining regions in 
each BAL trough where a variation has occurred, we compare two-epoch spectra 
of each quasar. Since a proper comparison requires consideration of the 
signal-to-noise ratio of each spectrum, we define a measurement of the deviation 
between two observations for each pixel in units of $\sigma$ using the  following 
equation:
\begin{equation}
N_{\sigma}(\lambda) =  {\frac { f_2 - f_1 } 
{\sqrt {\sigma_2^2  + \sigma_1^2 }}}
\label{eqv10}
\end{equation}
where $f_1$  and $f_2$ are the normalized flux densities and  $\sigma_1$ and 
$\sigma_2$ are the normalized flux-density standard deviations at wavelength 
$\lambda$. Both $\sigma_1$ and $\sigma_2$ include observational errors and 
uncertainties on the estimated continuum model. Similarly to \citet{gibson08}, we 
identify variable regions of BAL troughs to be where an absorption feature 
is detected with $N_\sigma \geq 1$ or $N_\sigma \leq -1$ for at least five 
consecutive data points. This requirement allows detection of  variable regions 
wider than $\approx275~\rm{km\,s^{-1}}$. Selection using a smaller number of 
consecutive data points as the requirement may cause non-physical observational 
errors to be indistinguishable from the variable regions of BAL troughs. On the 
other hand, requiring a larger number of consecutive data points will cause 
non-detection of narrow variable regions. By requiring the number of data 
points to be $\geq5$, we require the significance of variations to be $>$99.9\%.

We have identified 903 variable regions for C\,{\sc iv}  BAL troughs and 294 
variable regions for Si\,{\sc iv} BAL troughs for  variations on timescales of more 
than 1~yr ($\Delta t_{\rm min,1}$). We also identified 757 variable regions for 
C\,{\sc iv}  BAL troughs and 232 variable regions for Si\,{\sc iv} BAL troughs by 
comparing the two-epoch spectra that sample the $\Delta t_{\rm min}$ timescales 
in the rest frame. The number of BAL troughs having at least one variable region 
is 294 for C\,{\sc iv} and 119 for Si\,{\sc iv} on timescales of more than 1~yr. 

Comparing the two approaches to variable BAL-trough identification, we found 
that 26 C\,{\sc iv} BAL troughs showing EW variations at more than 3$\sigma$ 
significance do not have a variable region satisfying our requirements, although 
several narrow variable regions in these BAL troughs collectively produce EW 
variations at more than 3$\sigma$ significance.  
We also found that 72 C\,{\sc iv} BAL troughs having one variable 
region do not show EW variations at more than 3$\sigma$ significance. In these 
72 cases, a narrow variable region in a strong BAL trough cannot produce an 
EW variation at more than 3$\sigma$ significance due to statistical dilution by 
the rest of the trough.  Similarly, we found that 29 Si\,{\sc iv} BAL troughs 
showing EW variations at more than 3$\sigma$ significance do not have a 
variable region satisfying our requirements, and 29 Si\,{\sc iv} BAL troughs 
having one variable region do not show EW variation at more than 3$\sigma$ 
significance.

We will refer to both of these approaches to variable BAL trough identification 
in the following sections. We present our measurements for C\,{\sc iv} and 
Si\,{\sc iv} BAL troughs in Tables~\ref{balc} and \ref{balsi} and for C\,{\sc iv} 
and Si\,{\sc iv} variable regions in Tables~\ref{varrc} and \ref{varrsi}, respectively.

\section{Results on BAL Variability}\label{vres}

In this section, we present the results of our BAL variability investigations 
utilizing the multi-epoch observations of  428 distinct C\,{\sc iv} and 235 distinct 
Si\,{\sc iv} BAL troughs in the 699 main-sample spectra of 291 quasars. We 
examine the fraction of variable BAL troughs and BAL quasars (Section~\ref{vfrac}), the 
velocity widths of the variable regions of BAL troughs (Section~\ref{width}), EW variations 
as a function of timescales (Section~\ref{EWt}), the distribution of EW variations (Section~\ref{vdist}), 
EW variations as a function of BAL profile properties (Section~\ref{ewew}), relative EW 
variations between  C\,{\sc iv} and Si\,{\sc iv} BAL troughs (Section~\ref{csibal}), 
correlated EW variations in BAL quasars with multiple troughs (Section~\ref{multiple}), 
and EW variations as a function of quasar properties (Section~\ref{qparam}). 

\subsection{Fraction of BAL Troughs  and BAL Quasars Showing Variability}
\label{vfrac}

\begin{figure*}[]
\epsscale{0.9}
\plotone{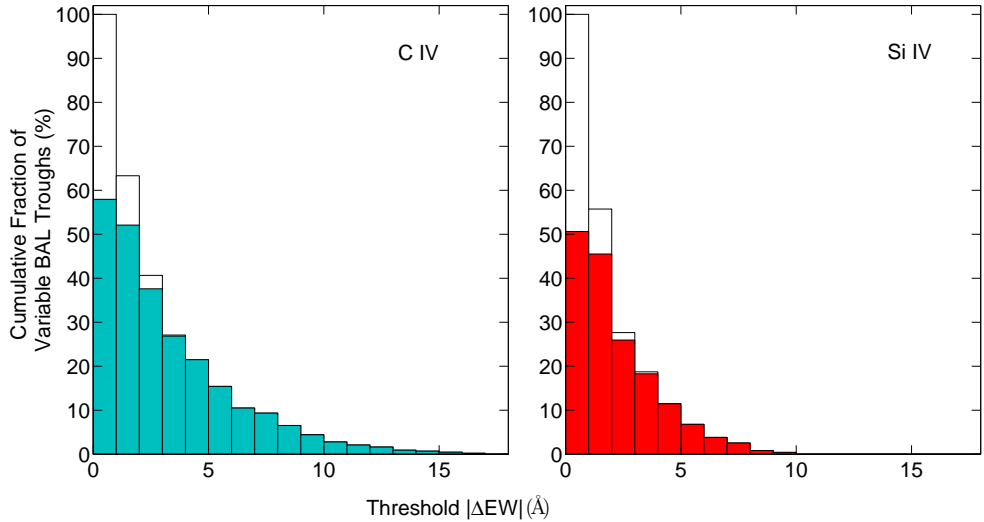}
\caption{Cumulative fraction of BAL troughs with a given threshold 
$|\Delta$EW$|$ for C\,{\sc iv} and Si\,{\sc iv} troughs. The open histograms 
show the cumulative fraction of BAL troughs with EW variations of more than a 
given threshold. The shaded parts of the histograms indicate the fraction of BAL 
troughs showing EW variations of more than 3$\sigma$ significance on 
timescales of 1--3.7~yr.  }
\label{varef1}
\end{figure*}

We calculate the fraction of  BAL troughs  showing variability and the fraction 
of quasars showing BAL-trough variability in our main sample considering the 
two different variability  identification approaches explained in Section~\ref{vr2}. First, 
requiring a variable BAL trough to show an EW variation at more than $3\sigma$ 
significance, we find that the fraction of variable BAL troughs is 
$57.9^{+3.9}_{-3.7}$\% for C\,{\sc iv}  and $50.6^{+5.1}_{-4.6}$\% for  Si\,{\sc iv} 
on timescales of 1--3.7~yr. Figure~\ref{varef1} presents the cumulative fraction of variable BAL 
troughs; the $y$-axis shows the cumulative fraction of BAL troughs with EW 
variations of more than a given threshold $|\Delta$EW$|$. Although the cumulative 
fraction of variable BAL troughs decreases for large $|\Delta$EW$|$, it remains 
significant even for threshold $|\Delta$EW$|$ values as large as 5~\AA.

We also compare the fraction of variable C\,{\sc iv} and Si\,{\sc iv} BAL troughs 
from the same absorbing material. Given that the Si\,{\sc iv} region between 
$-13000$ and $-30000$~km~s$^{-1}$ can have contamination by emission 
and/or absorption lines such as C\,{\sc ii} (1335~$\rm{\AA}$), O\,{\sc i} 
(1306~$\rm{\AA}$), and Ly$\alpha$+N\,{\sc v}, we select a sample of 136 
C\,{\sc iv} BAL troughs lying between $-3000$ and $-13000$~km~s$^{-1}$ 
that are accompanied by  Si\,{\sc iv} BAL troughs at corresponding velocities. 
In this sample, we found that the fraction of BAL troughs showing EW variations 
at more than $3\sigma$ significance is $50.0^{+6.8}_{-6.0}$\% for C\,{\sc iv} and 
$53.7^{+6.3}_{-5.4}$\% for Si\,{\sc iv}  on timescales of 1--3.7~yr. These fractions 
indicate that  C\,{\sc iv} and Si\,{\sc iv} BAL troughs at corresponding 
velocities are about equally likely to vary.

In addition to considering the fraction of variable BAL troughs, we also calculate 
the fraction of quasars showing BAL-trough variability. Requiring a variable BAL 
trough to show an EW variation at more than $3\sigma$ significance, we found 
the fraction of quasars showing C\,{\sc iv} BAL variations to be 
$62.2^{+4.9}_{-4.6}$\% (181/291). Considering a total of 181 quasars showing 
Si\,{\sc iv} BAL absorption, we found the fraction of quasars showing Si\,{\sc iv} 
BAL variations to be $59.1^{+6.3}_{-5.7}$\% (107/181). 
Using multi-epoch observations of 24 quasars, \citet{cap11} found that the fraction 
of quasars showing C\,{\sc iv} absorption variations is 39\% on timescales of 
0.35--0.75~yr and 65\% on timescales of 3.8--7.7~yr. \citet{cap12} reported that 11 
out of 19 (58\%)  quasars exhibited Si\,{\sc iv} absorption variations on 
timescales of 3.8--7.7~yr. Considering that our data sample timescales of 
1--3.7~yr with an average of 2.3~yr, both of our results broadly show consistency 
with the results of the \citet{cap11} study.

Alternatively, we consider variable BAL troughs to be those with at least one 
variable region detected in the trough  (see Section~\ref{vr2}). We find that the fraction 
of BAL troughs showing variability is  $68.6^{+4.3}_{-4.0}$\% for C\,{\sc iv} and  
$50.6^{+5.1}_{-4.6}$\% for Si\,{\sc iv}  on timescales of 1--3.7~yr. This approach 
is more sensitive to local variations in BAL troughs; a narrow variable region in a 
wide BAL trough may not produce an EW variation at more than 3$\sigma$ 
significance.

\begin{figure*}[t!]
\epsscale{0.8}
\plotone{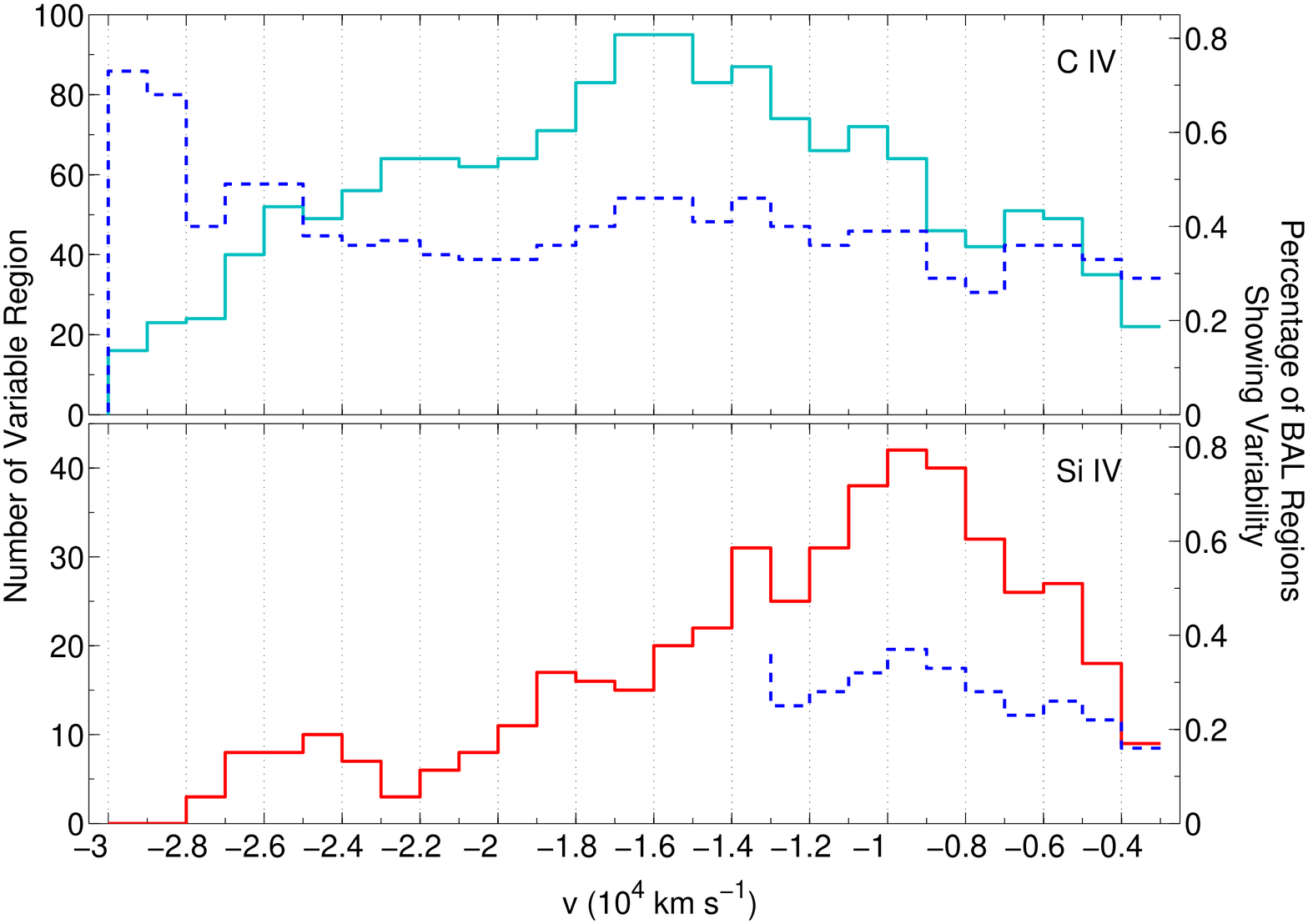}
\caption{Number of times a variable region is found at a particular velocity in 
C\,{\sc iv} (upper panel) and Si\,{\sc iv} (lower panel) BAL troughs for variations 
on timescales of more than 1~yr. In the upper panel, the dashed dark-blue line 
along with the right $y$-axis shows the percentage of BAL regions showing 
variability in C\,{\sc iv}. C\,{\sc iv} variable regions are found across a wide range 
of velocities. The Si\,{\sc iv} region can have contamination by emission lines 
such as C\,{\sc ii} (1335~$\rm{\AA}$), O\,{\sc i} (1306~$\rm{\AA}$), and 
Ly$\alpha$+N\,{\sc v}, causing the apparent decrease in the number of variable 
Si\,{\sc iv} regions between $-13000$ and $-30000$~km~s$^{-1}$; 
the histogram for Si\,{\sc iv} should not be interpreted physically in this velocity 
range. Therefore, we do not show the percentage of variable regions for 
Si\,{\sc iv} between $-13000$ and $-30000$~km~s$^{-1}$.}
\label{varf3}
\end{figure*}

We further investigate the number of variable regions as a function of velocity, 
$v$, that is measured relative to the quasar redshift. Figure~\ref{varf3} presents 
the number of variable regions found at a particular velocity for C\,{\sc iv}  and 
Si\,{\sc iv}  BAL troughs for  variations on timescales of more than 1~yr ($\Delta t_{\rm min,1}$).  
C\,{\sc iv} variable regions are found across a wide range of velocities, and the 
number of variable regions appears to peak in the range between 
\hbox{$-9000$ and $-$21000~km~s$^{-1}$} in concert with the number 
of BAL troughs.
Figure~\ref{varf3} also displays the percentage of BAL regions showing variability 
for C\,{\sc iv} and  Si\,{\sc iv} that is calculated from the ratio of the number of 
variable regions to the number of BAL troughs found at a particular velocity. We 
found that the percentage of C\,{\sc iv} BAL regions showing variability is roughly  
constant  at around 30--40\% at  velocities of $-3000$~km~s$^{-1}$ to 
$-25000$~km~s$^{-1}$ and rises in the few highest-velocity bins; we have 
verified that this trend is statistically significant. 
\citet{cap11} also investigated the fraction of variations as a function of outflow 
velocity on short (0.35--0.75~yr) and long (3.8--7.7~yr) timescales. Consistent 
with our results, they found that C\,{\sc iv} BAL troughs tend to be more variable 
at higher velocities.

The Si\,{\sc iv} region can have contamination by emission and/or absorption 
lines, and the superposition of these emission lines and Si\,{\sc iv} BALs may 
prevent BAL troughs from continuously lying at least 10\% under the continuum 
level and thus may lead to non-detection. A visual inspection showed that in 
some cases BAL troughs appear to be broken into two or more narrow sections 
by such emission lines, causing the apparent absorption features to fail to satisfy 
the criteria to be identified as a BAL trough.  Consequently, these effects also 
cause a decrease in the number of variable Si\,{\sc iv} BAL troughs. Due to these 
effects, the number of variable Si\,{\sc iv} BAL troughs and therefore variable 
Si\,{\sc iv} regions decreases between $-13000$ and $-30000$~km~s$^{-1}$; 
the histogram for Si\,{\sc iv} should not be interpreted physically in this velocity 
range. Thus, we show  the  percentage  distribution of  Si\,{\sc iv} BAL troughs 
only for a region between $-3000$ and $-13000$~km~s$^{-1}$.

\subsection{Velocity Widths of Variable Regions of BALs}\label{width}

\begin{figure*}[t!]
\epsscale{0.75}
\plotone{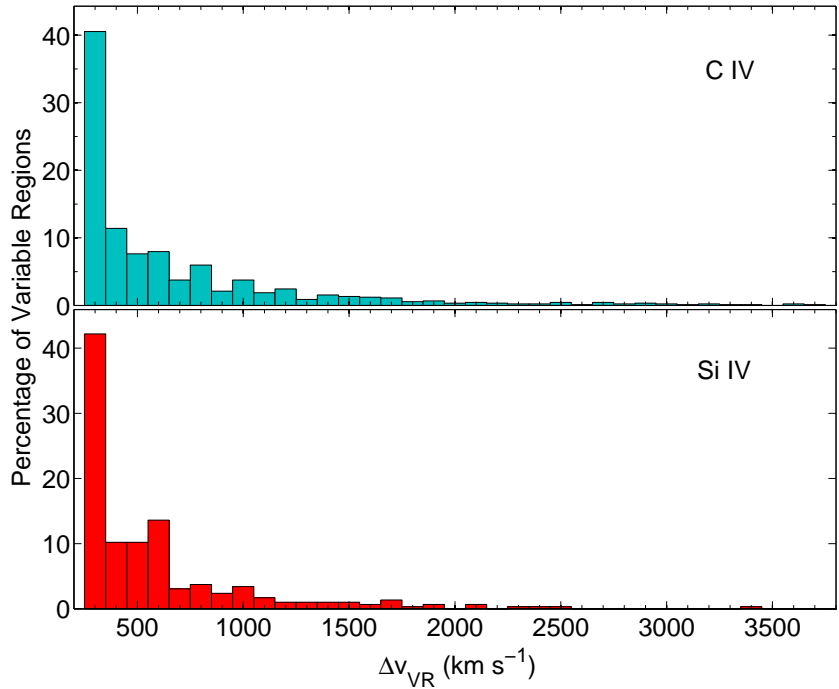}
\caption{Percentage of variable regions with a given velocity width in 
C\,{\sc iv} (upper panel) and Si\,{\sc iv} (lower panel) BAL troughs
for variations on timescales of more than 1~yr. We found seven 
variable regions for C\,{\sc iv} and one variable region for Si\,{\sc iv}
with velocity widths of 4500--7500$~\rm{km\,s^{-1}}$ that we do not 
show in this figure. For both C\,{\sc iv} and Si\,{\sc iv}, the number of 
variable regions rises rapidly toward small velocity widths down to our 
velocity width measurement limit.}
\label{varf4}
\end{figure*}

\citet{gibson08} showed that variations tend to occur only in portions of BAL 
troughs. To investigate the distribution of velocity widths of variable BAL regions, 
we calculate $\Delta v_{\rm{VR}}$ values for 903 variable regions in C\,{\sc iv} 
BAL troughs and 294 variable regions in Si\,{\sc iv} BAL troughs for  variations 
on timescales of more than 1~yr, where $\Delta v_{\rm{VR}}$ is the velocity 
width of a variable region. Figure~\ref{varf4} shows the $\Delta v_{\rm{VR}}$ 
distributions of variable regions detected in C\,{\sc iv} and Si\,{\sc iv}  BAL troughs. 
Consistent with \citet{gibson08}, we find that  the number of variable regions, in 
both C\,{\sc iv} and Si\,{\sc iv}, rises toward small velocity widths down to our 
velocity width measurement limit of $\approx 275~\rm{km\,s^{-1}}$ (see 
Section~\ref{vr2}). We found that the mean of the $\Delta v_{\rm{VR}}$ measurements 
for C\,{\sc iv} and Si\,{\sc iv} BAL troughs are $713.6~\rm{km\,s^{-1}}$ and 
$592.8~\rm{km\,s^{-1}}$, respectively. 

\begin{figure*}[t!]
\epsscale{0.8}
\plotone{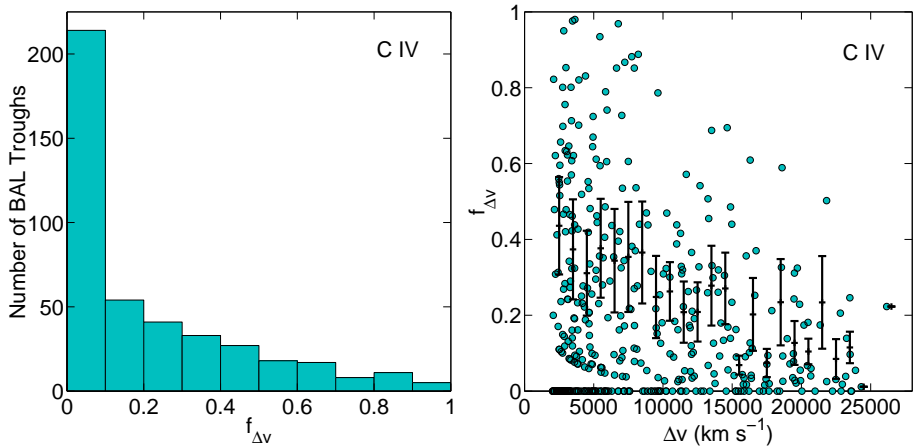}
\caption{Distribution of $f_{\Delta v}$ (left panel), and $f_{\Delta v}$ distribution 
as a function of BAL trough width $\Delta v$ (right panel). Vertical black bars in 
the right panel show the standard deviation around the mean for non-zero 
$f_{\Delta v}$ values in given 100$~\rm{km\,s^{-1}}$ wide $\Delta v$ bins.
Both of the panels are for C\,{\sc iv} BAL troughs on timescales of 1--3.7~yr. }
\label{varef2}
\end{figure*}

To examine the fraction of a BAL trough that is variable, we define $f_{\Delta v}$  
as the sum of the velocity widths of all varying regions in a BAL trough divided by 
the  BAL trough velocity width, $\Delta v$. Figure~\ref{varef2} shows the 
$f_{\Delta v}$ distribution  for C\,{\sc iv} BAL troughs on timescales of 1--3.7~yr 
and  the $f_{\Delta v}$ distribution as a function of $\Delta v$ for C\,{\sc iv}. 
We find that the mean of $f_{\Delta v}$ is 0.20 for C\,{\sc iv} and 0.13 for 
Si\,{\sc iv}, indicating that Si\,{\sc iv} variable regions on average tend to be 
narrower than C\,{\sc iv} variable regions. Figure~\ref{varef2} also suggests 
that narrow C\,{\sc iv} BAL troughs tend to have a larger fraction of 
variable regions compared to wide C\,{\sc iv} BAL troughs. 

\begin{figure}[h!]
\epsscale{1}
\plotone{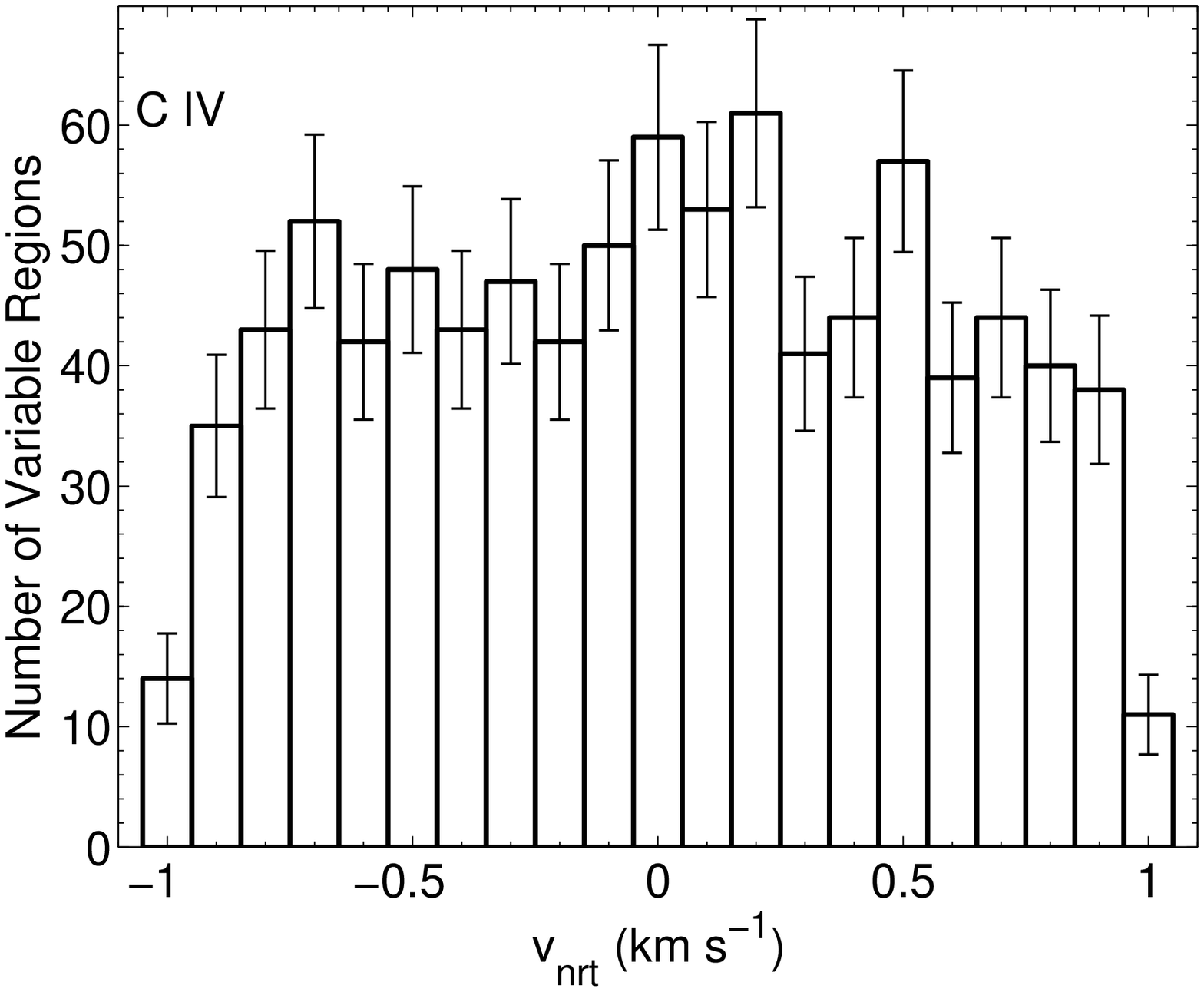}
\caption{The number  of variable regions found at a given $v_{\rm nrt}$. 
The distribution is relatively constant across the entire trough width. }
\label{varef3}
\end{figure}

We also investigate the position of each variable region in a given BAL trough to 
assess if the incidence of variability depends upon the relative velocity within the 
trough. We calculate the normalized relative velocity in the trough,  $v_{\rm nrt}$. 
Here $v_{\rm nrt}$ is $\frac{v_{\rm cent}-v_{\rm mid}}{|v_{\rm cent}-v_{\rm max}|}$ 
for the blue part of the trough and 
$\frac{v_{\rm cent} - v_{\rm mid}}{ |v_{\rm cent} - v_{\rm min}|}$ for  the red part 
of the trough, where $v_{\rm mid}$ is the mid velocity of a variable region. 
Figure~\ref{varef3} shows the number of variable regions found at a given 
$v_{\rm nrt}$. A non-parametric triples test  \citep{triples} shows that the 
distribution of $v_{\rm nrt}$ shows no significant evidence of asymmetry 
($P = 0.55$), and the distribution is relatively constant across almost the entire 
width of a trough. Consistent with our results, \citet{cap11} found no evidence for 
a higher incidence of variability with positive or negative velocity offset, but we  
establish this result with substantially better statistics. 

\subsection{EW Variations as a Function of Timescale}\label{EWt}

\begin{figure*}[t!]
\epsscale{0.8}
\plotone{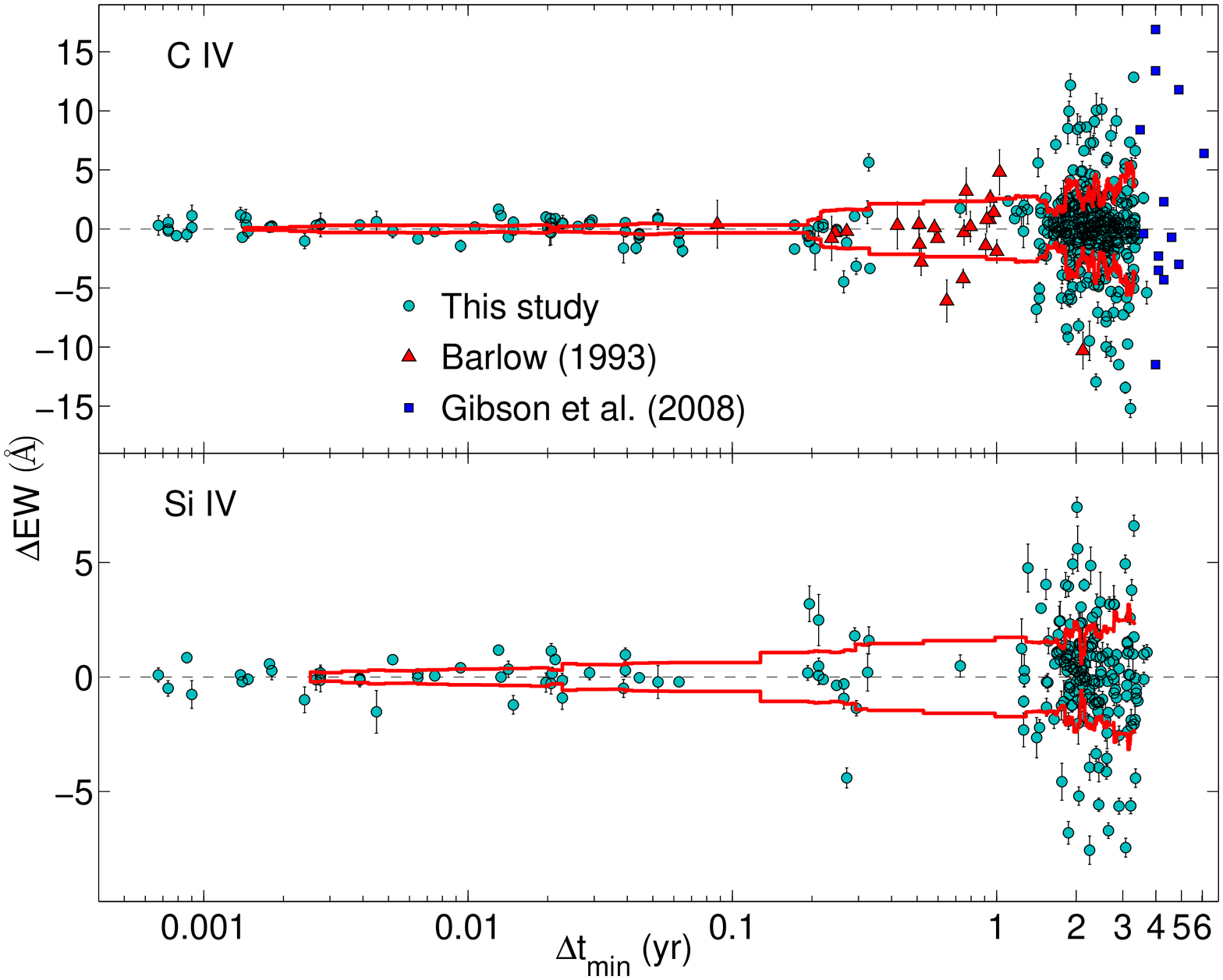}
\caption{EW variation, $\Delta$EW, vs. the minimum sampled rest-frame 
timescale, $\Delta t_{\rm min}$, for C\,{\sc iv} (upper panel) and Si\,{\sc iv} 
(lower panel) BAL troughs. The data are from this study (light blue 
circles), \citeauthor{barlow93} (1993; red triangles), and \citeauthor{gibson08} 
(2008; dark blue squares). 
The red solid curves indicate the standard deviation derived from the data 
in this study, calculated using a sliding window containing 20 time-ordered 
data points. The standard deviation of $\Delta$EW increases with increasing 
rest-frame timescale.}
\label{varf5}
\end{figure*}

Previous BAL-variability studies \citep[e.g.,][]{gibson08,gibson10,cap11} have 
found that C\,{\sc iv} BAL-trough variability is larger for longer timescales. This is 
expected since, e.g., quasar variability  in general is larger on longer timescales 
\citep[e.g.,][]{vanden04}. In order to investigate EW variations as a function of 
timescale with a larger sample over a wide range of rest-frame timescales, we 
utilize the $\Delta$EW and  $\Delta$EW/$\langle$EW$\rangle$  values of each 
distinct BAL trough from the two-epoch spectra for timescales of $\Delta t_{\rm min}$ 
(see Equations~\ref{eqv2} and \ref{eqv3}). In Figure~\ref{varf5}, we show EW variations 
for C\,{\sc iv} and Si\,{\sc iv} BAL troughs as functions of  $\Delta t_{\rm min}$. For 
comparison,  we also include the data from \citet{barlow93} that correspond to a 
timescale range mainly between 0.2--1~yr and from \citet{gibson08} that extend 
the timescales up to 6.1~yr. To display the spread of $\Delta$EW, we calculate 
the standard deviation of our data using a sliding window containing 20 
time-ordered data points; we statistically remove the mean EW error in each 
window from the standard deviation (via standard error propagation). The 
curves of standard deviation indicate an increase 
of EW variations with increasing rest-frame timescale both for C\,{\sc iv} and 
Si\,{\sc iv}  BAL troughs. This trend is consistent with that of previous 
BAL-variability studies. The majority of the \citet{barlow93} data lie between 
the standard-deviation curves  that are calculated from our data. Although the 
data from \citet{gibson08} sample longer timescales than our data, the trend of 
the standard-deviation curves shows general agreement regarding 
the increase of EW variations with increasing timescale. 

\begin{figure*}[t!]
\epsscale{0.8}
\plotone{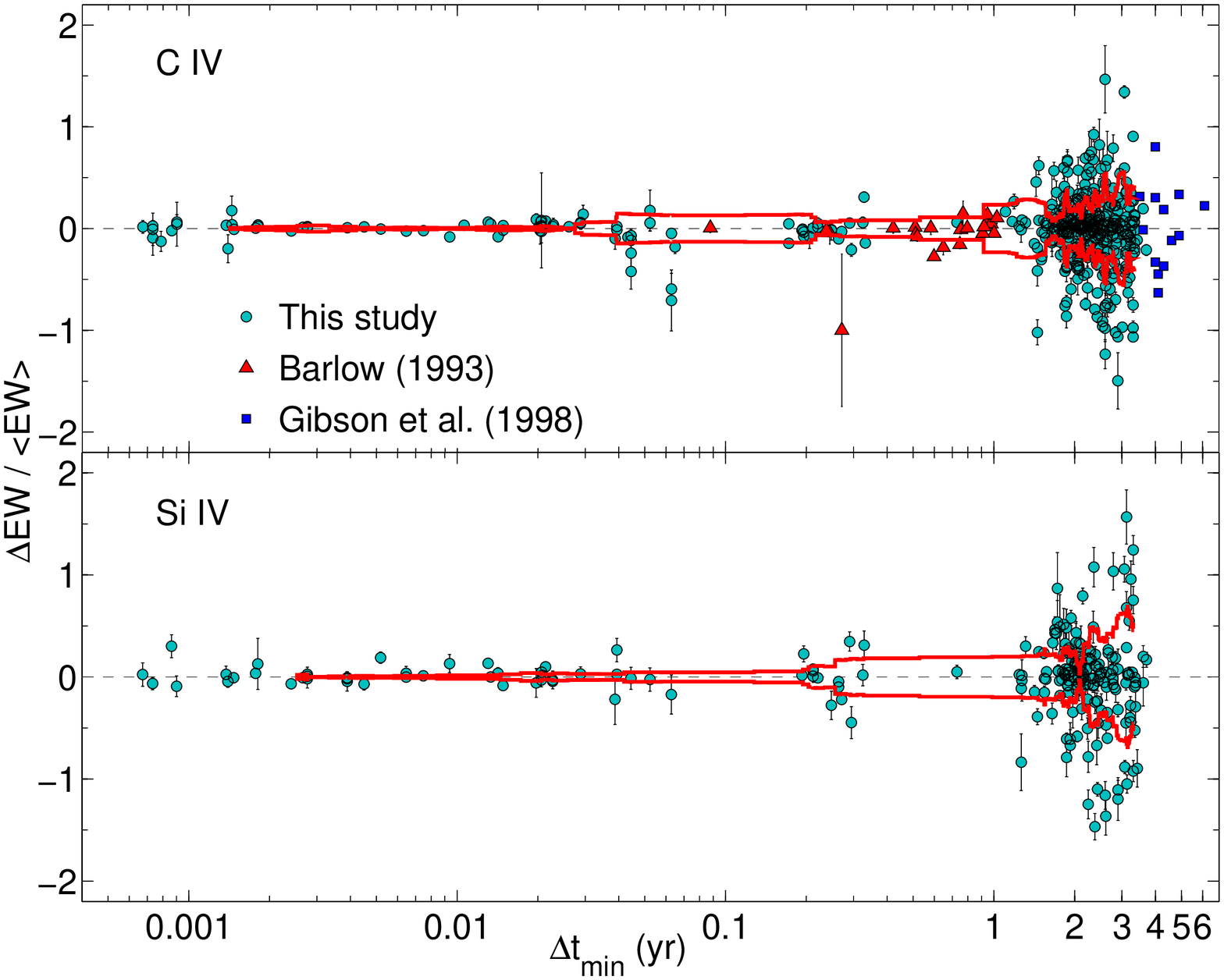}
\caption{Same as Figure~5 but for fractional EW variation, 
$\Delta \rm {EW}/\langle \rm {EW} \rangle$. \\ \\}
\label{varf6}
\end{figure*}

Figure~\ref{varf6} shows fractional EW variations for C\,{\sc iv} and Si\,{\sc iv} 
BAL troughs against $\Delta t_{\rm min}$. The curves of standard deviation are 
illustrated both for C\,{\sc iv} and Si\,{\sc iv} BAL troughs. The spread of the 
curves indicates an increase of fractional EW variations with increasing 
rest-frame timescales.  As for Figure~\ref{varf5}, we include the data from 
\citet{barlow93} and \citet{gibson08} for comparison purposes.

Figures~\ref{varf5} and \ref{varf6} demonstrate that our data points are 
consistent with  zero variation  for timescales $\Delta t_{\rm min} < 0.01$ 
yr (i.e., 3.6 days) in the rest frame given the measurement errors.  This result 
is expected given the fact that troughs are unlikely to show significant 
variations on timescales of hours or days. Therefore, any observed variations 
on these timescales provides an empirical estimate of the total systematic and 
measurement errors of our data and data-processing methods. For a more 
quantitative assessment, we calculated the median deviation in units of 
$\sigma$ for $\Delta t_{\rm min} < 0.01$ yr and found that it  is $\approx1$ 
for both C\,{\sc iv} and Si\,{\sc iv}. 

To quantify the relationship between EW variations and timescale, we calculate the 
mean of $|\Delta$EW$|$ and $\Delta t_{\rm min}$ for bins containing 15 
time-ordered data points. From a robust linear-regression model using the 
bisquare weight function \citep{press}, we found a fit of
\begin{eqnarray}
\log |\Delta{\rm EW_{C\,IV}}| = (0.258 \pm 0.031) \times \log \Delta t_{\rm{min}} 
\nonumber  \\ + (0.289 \pm 0.027) 
\label{eqv4}
\end{eqnarray}
where the units of $|\Delta$EW$|$ and  $\Delta t_{\rm min}$ are  \AA\, and 
yr, respectively. Similarly, for fractional EW variations we found a fit of
\begin{eqnarray}
\log \left| \frac{\Delta{\rm EW}}{\langle\rm{EW}\rangle} \right|_{\rm{C\,IV}} = 
(0.283 \pm 0.042) \times \log \Delta t_{\rm{min}}  \nonumber \\
+ (-0.737 \pm 0.037).
\label{eqv5}
\end{eqnarray}

\begin{figure}[t!]
\epsscale{1}
\plotone{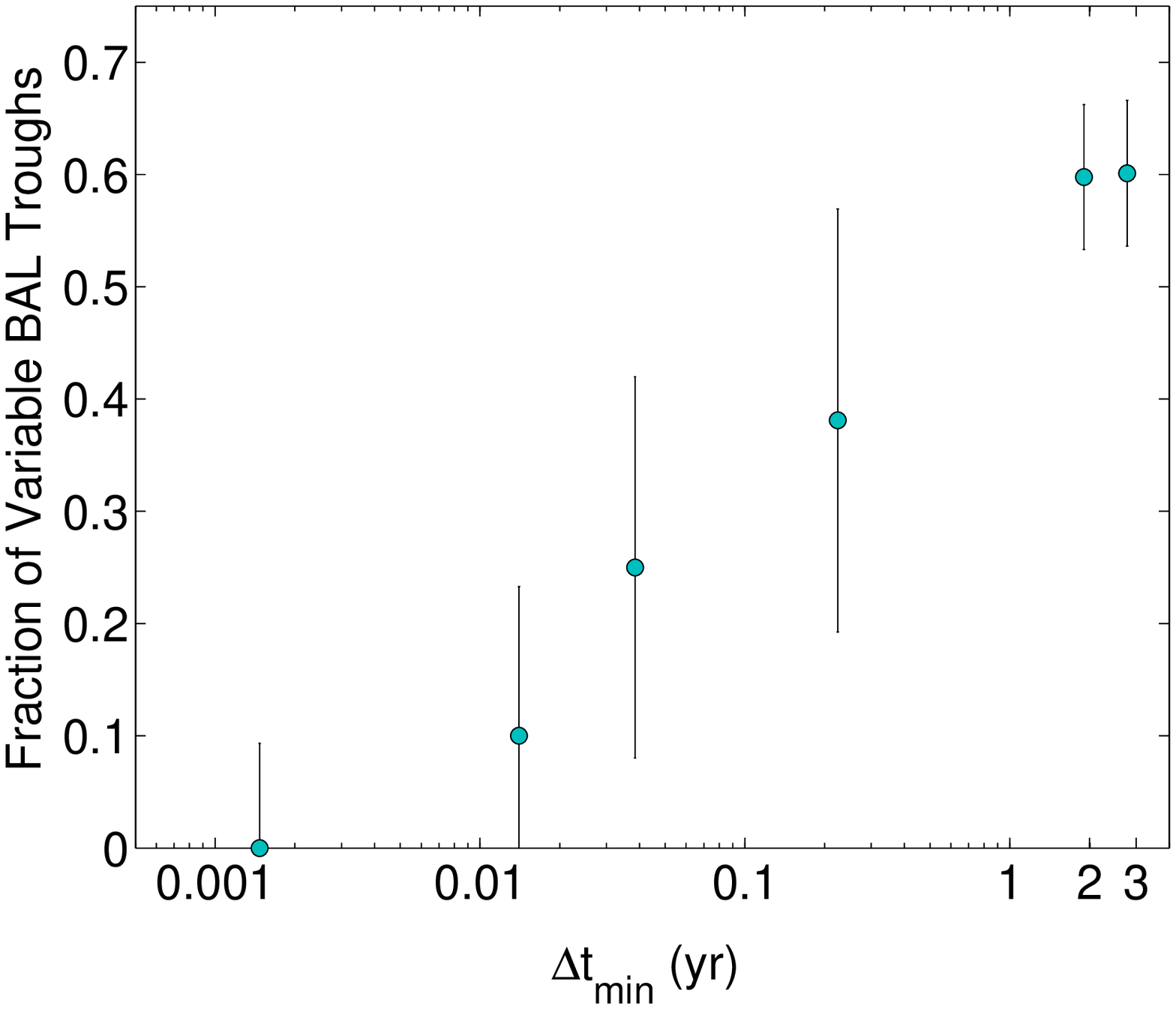}
\caption{The fraction of C\,{\sc iv}  BAL troughs showing variability at more than 
3$\sigma$ significance as a function of timescale. We calculate the counting 
errors following \citet{gehrel}.}
\label{varex8}
\end{figure}

In Figure~\ref{varex8}, we display the fraction of C\,{\sc iv}  BAL troughs showing 
variability at more than 3$\sigma$ significance as a function of timescale. We 
calculate the mean timescale and the fraction of variable BAL troughs for 20 
time-ordered data points for variations on timescales of less than 1~yr, and for 
two equal-size bins for variations of more than 1~yr. Consistent with \citet{gibson10}
and \citet{cap13}, our results indicate the incidence of variability increases with time.  

\begin{figure*}[t!]
\epsscale{0.8}
\plotone{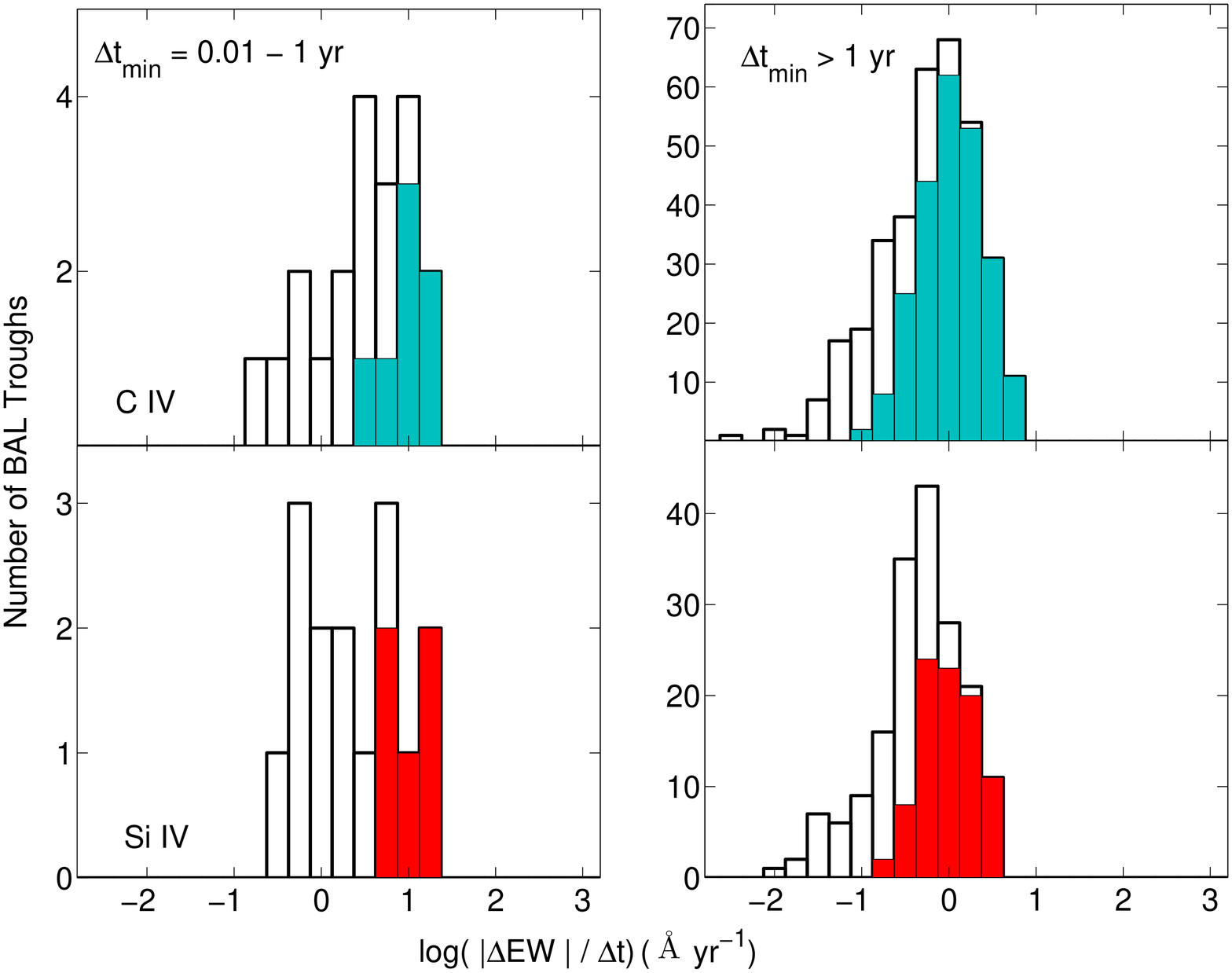}
\caption{Distributions of rate-of-change of $|\rm{EW}|$, $|\Delta$EW$|/\Delta t$, 
for C\,{\sc iv} (upper panels) and Si\,{\sc iv} (lower panels) BAL troughs both 
on short (left panels) and long  (right panels) timescales. The shaded parts 
of the histograms show the BAL troughs with $\geq3\sigma$ significance 
variations of EW.}
\label{varf7}
\end{figure*}

We also investigate the rate-of-change of EW variations on short  and long 
timescales by calculating $|\Delta$EW$|/\Delta t$ for C\,{\sc iv}  and Si\,{\sc iv} 
BAL troughs on timescales of \hbox{$\Delta t_{\rm min}= 0.01-1$~yr} and   
$\Delta t_{\rm min}~>~1$~yr. Figure~\ref{varf7} displays the distribution of 
$|\Delta$EW$|/\Delta t$ for C\,{\sc iv} and Si\,{\sc iv}  BAL troughs on both  
short  and long timescales. It also presents the distribution of 
$|\Delta$EW$|/\Delta t$ for BAL troughs with $\geq3\sigma$  significance 
variations of EW in each panel. The distributions for variations with 
$\geq3\sigma$ significance show that the average rate-of-change of EW 
variations is larger on short timescales. \citet{gibson10} found a similar result 
from the comparison of variations of BAL troughs for eight individual sources on 
short and long timescales (see their Figure~9). 
Such behavior would be expected, for example, if BAL EWs execute a simple 
random walk with a step timescale of $\lesssim 1$~yr (see Section~\ref{ranw}).
On timescales longer than the step timescale then the observed 
\hbox{$|\Delta$EW$|/\Delta t \propto 1/\sqrt{n}$}, where $n$ is the number of 
steps during the observed period.

\subsection{Distribution of EW Variations} \label{vdist}

\begin{figure*}[t!]
\epsscale{0.8}
\plotone{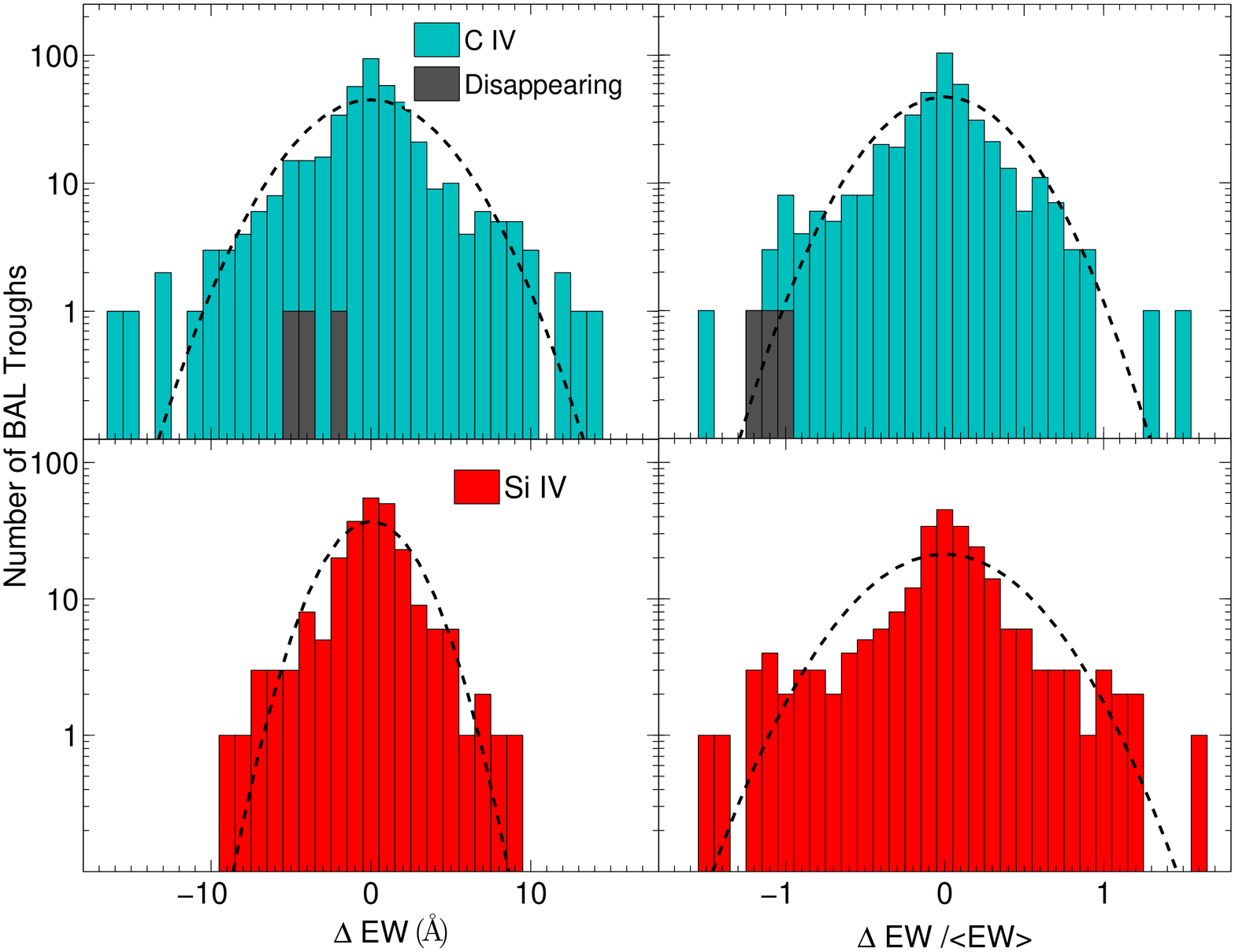} %10592
\caption{Distributions of BAL-trough EW variations, $\Delta$EW, and fractional 
EW variations, $\Delta$EW/$\langle$EW$\rangle$,  for C\,{\sc iv} (upper panels) 
and Si\,{\sc iv} (lower panels) for variations on timescales of more than 1~yr. Black 
dashed lines show the best Gaussian fits to the data. Our main sample includes 
three of the disappearing C\,{\sc iv} BAL troughs described in \citet{ak12}; these 
are plotted with a dark gray histogram in the upper panels (see Section~\ref{disem2} 
for further discussion). Our main sample also includes one additional  new case 
of  C\,{\sc iv} BAL  disappearance that satisfies the disappearance 
criteria used in \citet{ak12}; this case is plotted in the upper panel at 
$\Delta$EW/$\langle$EW$\rangle=-1.5$ and is found in the quasar 
SDSS~J095901.24+550408.2. The cases of BAL disappearance appear 
to be an extreme of the overall distribution of BAL variability, rather than a 
distinct population of variability events. None of the disappearing C\,{\sc iv} 
BAL troughs has a Si\,{\sc iv} BAL trough at corresponding velocities at our 
observed epochs.}
\label{varf8}
\end{figure*}

In this section, we investigate several characteristics of BAL EW variability distributions. 
We first examine  the symmetry of distributions of BAL EW variations to assess and 
constrain any differences between the formation and decay timescales of  BAL 
troughs.  
Figure~\ref{varf8} presents  the distributions of 428 C\,{\sc iv} and 235 Si\,{\sc iv} 
BAL-trough EW variations, $\Delta$EW, and fractional EW variations, 
$\Delta$EW/$\langle$EW$\rangle$, for variations on timescales of more than 
1 yr. To maintain an equilibrium of BAL troughs in quasar spectra, we expect 
the weakening and strengthening of BAL troughs in a large sample to be  
balanced. To assess whether BAL-trough variations are balanced, we examine 
the mean values of the $\Delta$EW and $\Delta$EW/$\langle$EW$\rangle$ 
distributions. We find that the mean of the $\Delta$EW distribution for C\,{\sc iv}  
BAL troughs is $-0.082\pm0.184$~\AA\, and for Si\,{\sc iv}  BAL troughs is 
$0.101\pm0.163$~\AA. The error on the mean is calculated following 
$\sigma / \sqrt{N}$  where $N$ is the number of BAL troughs.  Similarly, we 
find the mean of the $\Delta$EW/$\langle$EW$\rangle$ distribution for 
C\,{\sc iv}  BAL troughs is $-0.032\pm0.018$ and for Si\,{\sc iv}  BAL troughs is 
$-0.003\pm0.029$. These results indicate that the mean values of the 
$\Delta$EW and $\Delta$EW/$\langle$EW$\rangle$ distributions for C\,{\sc iv} 
and Si\,{\sc iv}  BAL troughs are broadly consistent with zero.

We next investigated if the $\Delta$EW and $\Delta$EW/$\langle$EW$\rangle$ 
distributions are symmetric. An asymmetric distribution could be seen, for instance, 
if the strengthening and weakening of BAL troughs occur at different rates. For 
example, if a typical BAL trough forms rapidly and decays slowly, the distribution of BAL 
variations would be skewed to  negative  $\Delta$EW (i.e., the number of 
weakening BAL troughs would be larger at any given time).  \citet{gibson10} 
showed that the distribution of $\Delta$EW for 23 C\,{\sc iv}  BAL troughs is 
reasonably symmetric.

As can be seen in Figure~\ref{varf8}, the $\Delta$EW and 
$\Delta$EW/$\langle$EW$\rangle$ distributions do not show clear evidence of 
asymmetry. For a more quantitative examination, we use a non-parametric triples 
test. The results of this test on the $\Delta$EW distributions for 
C\,{\sc iv} and Si\,{\sc iv} BAL troughs show no significant evidence of asymmetry
($P = 0.45$ and $P= 0.40$, respectively). Similarly, we found no significant 
evidence of asymmetry for the $\Delta$EW/$\langle$EW$\rangle$ distributions for  
C\,{\sc iv} and Si\,{\sc iv} BAL troughs ($P = 0.25$ and $P= 0.04$, respectively).

As another approach to assess asymmetry, we compare the distributions of 
strengthening and weakening BAL troughs by  running  a two-sample KS test 
for the distributions of BAL troughs with increasing and decreasing EWs. The 
test results show that the positive and negative parts of the $\Delta$EW 
distributions for C\,{\sc iv} and Si\,{\sc iv} BAL troughs do not show significant 
inconsistency ($P = 0.58$ and $P= 0.83$, respectively). Similarly, we find that 
the positive and negative parts of the $\Delta$EW/$\langle$EW$\rangle$  
distributions for C\,{\sc iv} and Si\,{\sc iv} BAL troughs do not show significant 
inconsistency ($P= 0.12$ and $P = 0.60$, respectively).

Given that the test results show no significant evidence of asymmetry,  we 
calculate the skewness of the $\Delta$EW and $\Delta$EW/$\langle$EW$\rangle$ 
distributions to establish a basic limit upon the amount of asymmetry allowed 
by our data. We found that the skewness of the $\Delta$EW distribution is 
$-0.17 \pm 0.12$ for C\,{\sc iv} BAL troughs and $-0.17 \pm 0.16$ for Si\,{\sc iv} 
BAL troughs. For the calculation of the error on skewness, we follow the standard 
approximate $\sqrt{6/N}$ formula \citep{press}. The skewness of the 
$\Delta$EW/$\langle$EW$\rangle$ distribution is $-0.33\pm 0.12$ for 
C\,{\sc iv} BAL troughs and $-0.32 \pm 0.16$ for Si\,{\sc iv} BAL troughs.

We next investigated whether $\Delta$EW and $\Delta$EW/$\langle$EW$\rangle$ 
are Gaussian distributed; following the central-limit theorem, many random 
variability processes will generate Gaussian distributions.  Although their sample 
of BAL troughs was not large enough to draw firm conclusions about the 
characteristics of BAL variation distributions, \citet{gibson10} suggested that 
$\Delta$EW may not be normally distributed over time spans of months-to-years. 
We analyze the shape of  the $\Delta$EW and $\Delta$EW/$\langle$EW$\rangle$ 
distributions for C\,{\sc iv} and Si\,{\sc iv} BAL troughs considering variations longer 
than 1 yr using  the Lilliefors normality test \citep{lillie}. The test shows that both 
the $\Delta$EW and $\Delta$EW/$\langle$EW$\rangle$ distributions for 
C\,{\sc iv} and Si\,{\sc iv} BAL troughs are non-Gaussian (at a significance level 
of $>99.9\%$). 

Figure~\ref{varf8} shows the best Gaussian model distributions for comparison 
purposes in each panel. Since the mean values of the $\Delta$EW distributions 
are consistent with zero, we calculate the best Gaussian models using $\mu=0$ 
and  the measured standard deviation of each distribution. For both C\,{\sc iv} and Si\,{\sc iv} 
BAL troughs, the distributions of $\Delta$EW have a stronger central peak, and 
both of the distributions are weaker than the Gaussian distributions for 
\hbox{$|\Delta$EW$| \approx$ 2--7~\AA}. Similarly, the 
$\Delta$EW/$\langle$EW$\rangle$ distributions for C\,{\sc iv} and Si\,{\sc iv} 
BAL troughs are stronger for  \hbox{$|\Delta$EW/$\langle$EW$\rangle| <$ 0.2} 
and weaker for a range of 
\hbox{$|\Delta$EW/$\langle$EW$\rangle| \approx$ 0.2--0.6} compared to 
the Gaussian distributions. 

A non-Gaussian distribution of EW variations on timescales of more than 1~yr 
may appear as a result of the superposition of several Gaussian distributions, 
each of which characterizes the EW variations in a small time interval (see 
Figure~\ref{varf5}). In order to test this hyphothesis, we ran a 
Lilliefors normality test for the $\Delta$EW and $\Delta$EW/$\langle$EW$\rangle$ 
distributions in a small time interval. We selected the time interval of 2.0--2.5~yr, 
since a large number of BAL troughs are sampled in this range (see 
Figures~\ref{varf5} and \ref{varf6}); 156 C\,{\sc iv} and 90 Si\,{\sc iv} BAL troughs 
are sampled. The  Lilliefors test 
showed that the $\Delta$EW and $\Delta$EW/$\langle$EW$\rangle$ distributions 
for C\,{\sc iv} and Si\,{\sc iv} BAL-trough variations on 
\hbox{$\Delta t_{\rm{min,1}}$ = 2.0--2.5~yr} are also non-Gaussian 
distributions (significance level of $>$95\%). 

We have also investigated the $\Delta$EW and $\Delta$EW/$\langle$EW$\rangle$ 
distributions for C\,{\sc iv} and Si\,{\sc iv} BAL troughs on timescales of 
$\Delta t_{\rm min}<1$~yr. Performing the triples and Lilliefors tests, we found 
that the $\Delta$EW and $\Delta$EW/$\langle$EW$\rangle$ distributions are 
also symmetric and non-Gaussian for both  C\,{\sc iv} and Si\,{\sc iv} BAL troughs 
(significance level of $>$99.9\%).

\subsection{EW Variations as a Function of BAL-Profile Properties}\label{ewew}

\begin{figure*}[t!]
\epsscale{0.8}
\plotone{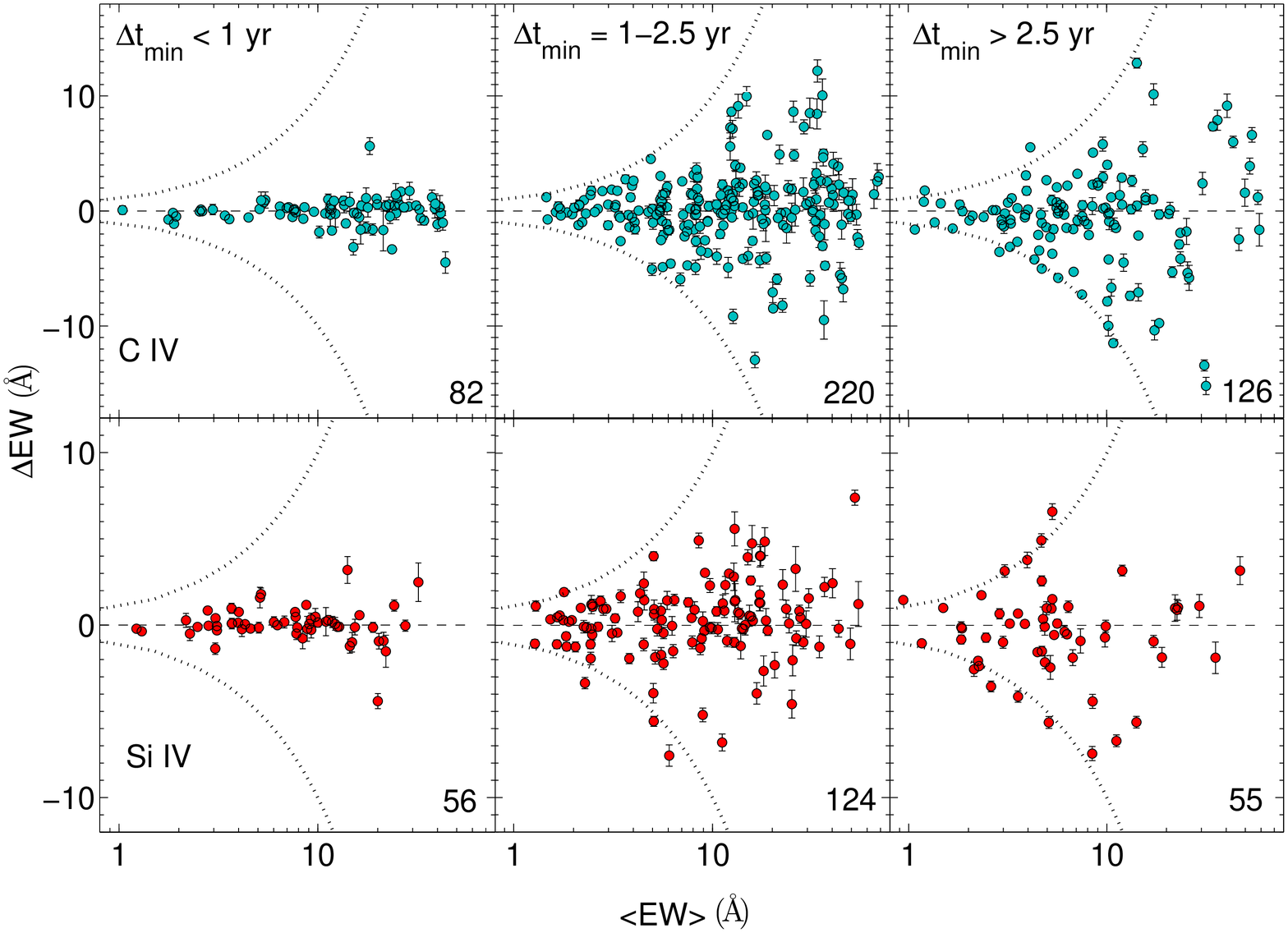}
\caption{EW variation, $\Delta$EW, vs. average EW over the two relevant 
epochs, $\langle$EW$\rangle$, for C\,{\sc iv} (blue) and Si\,{\sc iv} (red) BAL 
troughs for three different timescales as labeled. The $x$-axis is logarithmic. 
The black dotted curves denote where $|\Delta$EW$|$ is equal to 
$\langle$EW$\rangle$. The number of data points are given in the lower right 
of each panel.}
\label{varf12}
\end{figure*}

In this section, we investigate EW variations of BAL troughs as a function of 
BAL-profile properties, such as average EW, velocity width, depth, and centroid 
velocity. Although \citet{gibson08} found no significant correlation between the 
magnitude of EW variations and the average EW of C\,{\sc iv} BAL troughs, we 
search for correlations for both C\,{\sc iv} and Si\,{\sc iv} BAL troughs with a 
much larger sample for variations in three different timescale ranges. 
Figure~\ref{varf12} shows $\Delta$EW as a function of $\langle$EW$\rangle$ 
for short ($<1$~yr), moderate (1--2.5~yr), 
and long ($>2.5$~yr) timescales. Here the average EW, 
$\langle$EW$\rangle$, indicates the average of the measured EWs of each 
BAL trough in two-epoch spectra. The number of data points used is given in 
the lower right of each panel. We illustrate with dotted 
curves  where the EW variation is equal to the average EW. We search for 
correlations using the Spearman rank-correlation test between  
$|\Delta$EW$|$ and $\langle$EW$\rangle$.\footnote{
Although we assess correlations between $|\Delta$EW$|$ and 
BAL-trough profile properties, we prefer to use $\Delta$EW in our 
figures instead. This allows us to examine  possible dependencies 
of strengthening vs. weakening (for example, see the middle panel of Figure~\ref{varf14}) 
and associated discussion.}
The test results show that the correlations are 
highly significant ($>$99\%) for C\,{\sc iv} BAL-trough EW variations on short, 
moderate and long timescales. The $|\Delta$EW$|$ for Si\,{\sc iv} BAL troughs 
also likely correlates with $\langle$EW$\rangle$ on moderate timescales with 
a significance level of $>$95\%. We do not detect significant correlations for 
variations of Si\,{\sc iv} BAL troughs on short and long timescales, although 
these cases have the poorest trough statistics. These results suggest that 
weak BAL troughs tend to have small EW variations, an expected finding given 
that BAL troughs cannot weaken by more than their initial EWs measured in 
the first-epoch spectra. Remarkably, it is apparent in Figure~\ref{varf12} that 
weak BAL troughs also do not rise above the equality line.  In addition, 
Figure~\ref{varf12}  indicates that the EW variations of weak BAL troughs 
can be close to their average EWs.   

\begin{figure*}[t!]
\epsscale{0.8}
\plotone{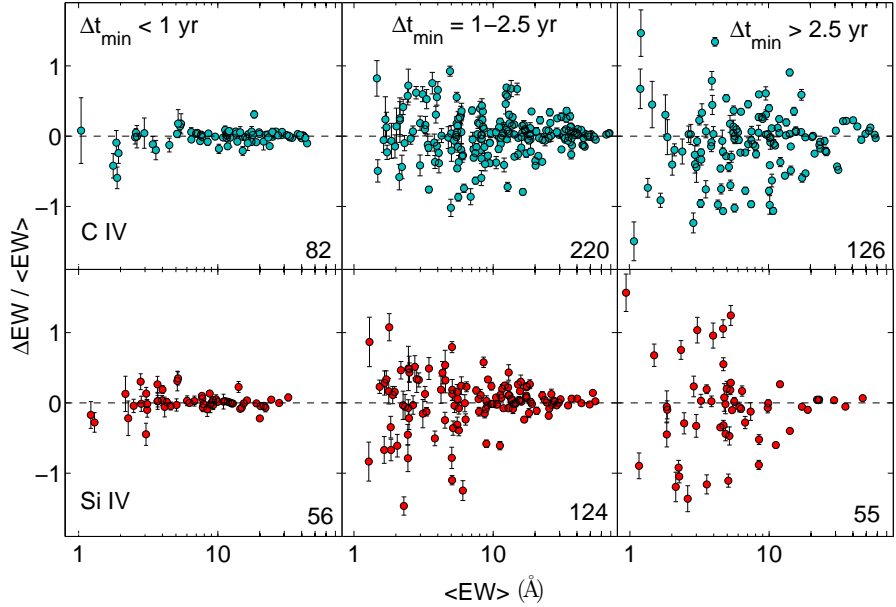}
\caption{Fractional EW variation, $\Delta$EW/$\langle$EW$\rangle$, 
vs. average EW over the two relevant epochs, $\langle$EW$\rangle$, 
for C\,{\sc iv} (blue) and Si\,{\sc iv} (red) BAL troughs for three different 
timescales as labeled. The $x$-axis is logarithmic. The number of data 
points are given in the lower right of each panel.}
\label{varf13}
\end{figure*}

\begin{figure*}[t!]
\epsscale{0.8}
\plotone{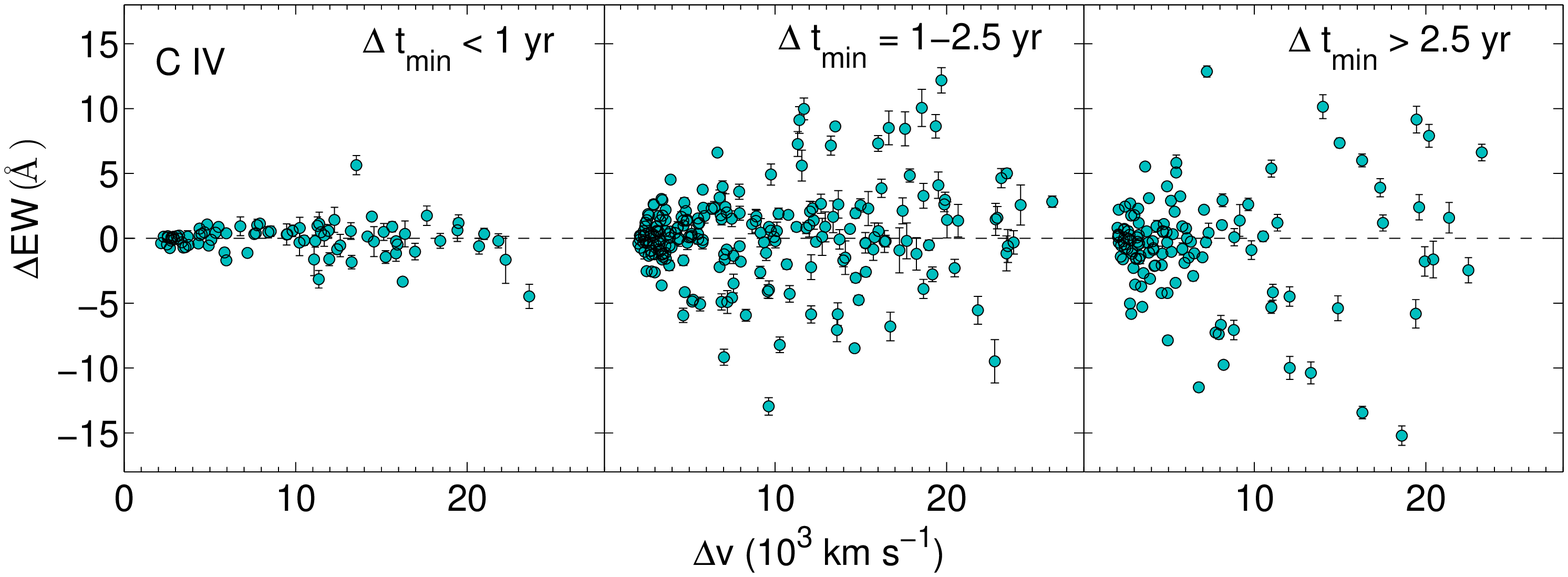}
\plotone{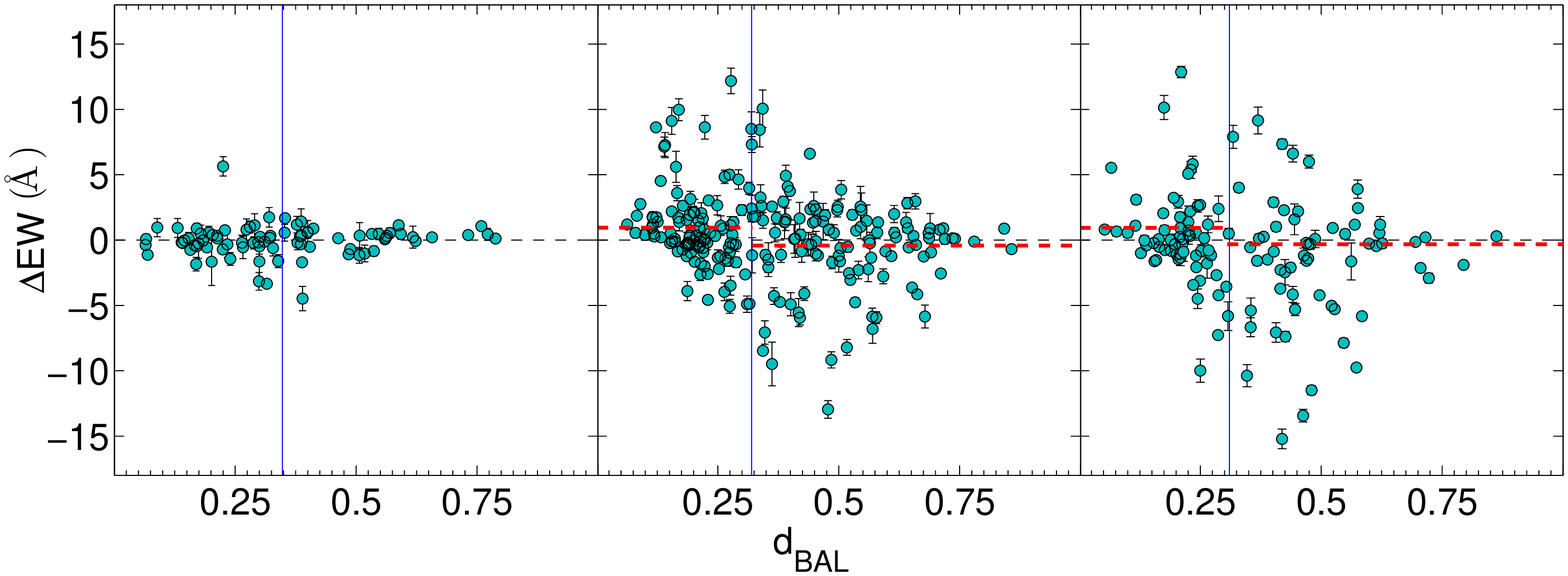}
\plotone{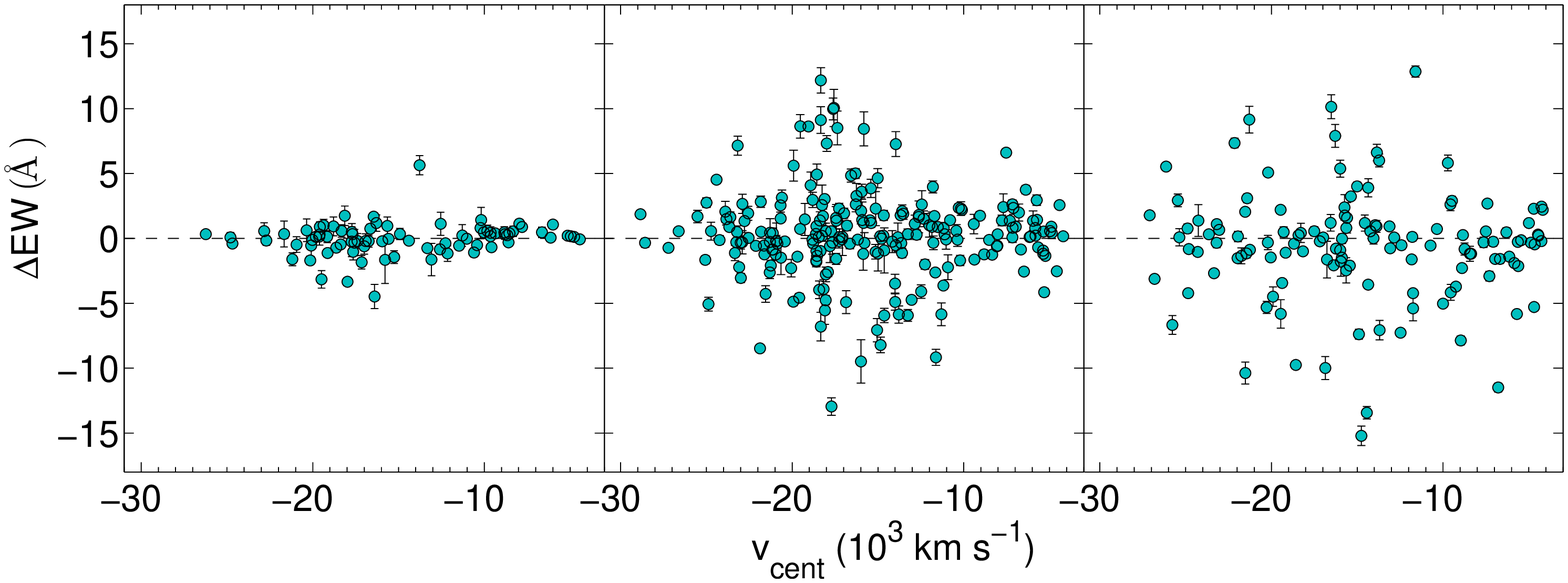}
\caption{C\,{\sc iv} BAL-trough EW variation, $\Delta$EW, as a function of 
BAL-trough width, $\Delta v$ (top panels), average depth of BAL 
troughs, $d_{\rm{BAL}}$ (middle panels), and centroid velocity, $v_{\rm{cent}}$ 
(bottom panels) for variations on  three different timescales as labeled. Vertical 
blue lines in the middle panels indicate the median $d_{\rm{BAL}}$ values and 
horizontal dashed red lines show the average  $\Delta$EW values for BAL 
troughs. }
\label{varf14}
\end{figure*}

Studies by \citet{lundgren07} and \citet{gibson08} suggested a significant 
correlation between the fractional variation of EW, 
$\Delta$EW/$\langle$EW$\rangle$, and  $\langle$EW$\rangle$. To investigate 
this correlation in a large sample, in Figure~\ref{varf13} we show 
$\Delta$EW/$\langle$EW$\rangle$ as a function of $\langle$EW$\rangle$ for 
C\,{\sc iv} and Si\,{\sc iv} BAL troughs for three timescales. In agreement with 
these previous studies, we find that the correlations between  
$|\Delta$EW/$\langle$EW$\rangle|$ and $\langle$EW$\rangle$ for C\,{\sc iv} 
BAL troughs are highly significant ($>$99.9\%) for all three  timescale ranges. 
The test results demonstrate  that the correlations between 
$|\Delta$EW/$\langle$EW$\rangle|$ and $\langle$EW$\rangle$ for  Si\,{\sc iv} 
BAL troughs are also highly significant ($>$ 99.9\%) for variations on moderate 
and long timescales. The significance level of the correlation is 98.4\% for 
variations of  Si\,{\sc iv} troughs on short timescales. Figure~\ref{varf13} shows 
that the fractional variations of small-EW BAL troughs are larger compared to 
large-EW BAL troughs. Consistent with our results, the studies by 
\citet{lundgren07} and \citet{gibson08} found the 
$|\Delta$EW/$\langle$EW$\rangle|$ vs. $\langle$EW$\rangle$ correlation. 
However, they likely did not see significant correlation for $|\Delta$EW$|$ vs. 
$\langle$EW$\rangle$ due to the small sample size. 

Given that the width and depth of a BAL trough determine its EW, we assess 
the contributions of these two components to the significant correlation found 
between $|\Delta$EW$|$ and $\langle$EW$\rangle$ for C\,{\sc iv} BAL 
troughs. We search for correlations between BAL-trough width, $\Delta v$, and 
BAL-trough depth, $d_{\rm{BAL}}$ (defined in Section~\ref{bal2}), vs. $|\Delta$EW$|$ 
for variations on short, moderate, and long  timescales (see Figure~\ref{varf14}). 

The Spearman test results show significant ($>$99.9\%) correlations between 
$|\Delta$EW$|$  and $\Delta v$ for all three timescale ranges indicating that 
wider BAL troughs tend to vary more than narrower ones; this is perhaps as 
expected since wider BAL troughs might have a better chance of containing 
variable regions. Unlike BAL-trough width, we find no significant correlation 
between $|\Delta$EW$|$  and $d_{\rm{BAL}}$. However, we do note from 
Figure~\ref{varf14} that the EWs of the deepest BAL troughs (those with  
$d_{\rm{BAL}} > 0.6$) appear to vary less than shallower ones. An even 
larger sample will be required to investigate this behavior reliably since we 
have limited trough statistics at large values of $d_{\rm{BAL}}$. This apparent 
behavior could be due to more saturated absorption at large $d_{\rm{BAL}}$ 
values; BAL absorption is often saturated  and, due to partial covering, saturated 
BALs are usually found as non-black absorption \citep[e.g.,][]{hamann98,arav99}.  

The  $\Delta$EW distribution as a function of $d_{\rm{BAL}}$  suggests that 
shallow BAL troughs on average increase their EWs and deep BAL troughs on 
average decrease their EW on moderate and long timescales (see 
Figure~\ref{varf14}). We search for a correlation between $\Delta$EW and 
$d_{\rm{BAL}}$ using the Spearman test and find likely correlations for both 
moderate and long timescales ($P$ = 2.6\% and $P = 5.6$\%, respectively). 
To show this trend, we also compare the mean $\Delta$EW values for BAL 
troughs that are deeper and shallower than the median BAL trough 
depth. The median BAL-trough depth is $0.32$ for moderate timescales 
and $0.31$ for long timescales.  We found the mean $\Delta$EW values to be 
$0.95\pm0.30$~\AA\, for BAL troughs with $d_{\rm{BAL}} < 0.32$ and 
$-0.42\pm0.34$~\AA\, for BAL troughs with $d_{\rm{BAL}} > 0.32$ on moderate 
timescales. Similarly, for variations on long timescales the mean $\Delta$EW 
values are $0.94 \pm 0.29$~\AA\, and $-0.32 \pm 0.36$~\AA\, for BAL troughs 
shallower and deeper than $d_{\rm{BAL}} = 0.31$, respectively. These results 
indicate that the statistical variation of a BAL trough depends not just on its EW 
and on the timescale between spectra, but also on its depth.

It has been found in previous studies that weak BAL troughs generally can 
achieve higher velocities than strong BAL troughs 
\citep[e.g.,][]{gibson09,cap11,ak12}; our data also show this trend. 
Considering our results indicate that weak BAL troughs tend to have smaller 
EW variations, we search for correlations between $|\Delta$EW$|$ and central 
velocity of a BAL trough, $v_{\rm{cent}}$ (see Section~\ref{bal2} for definition). 
Figure~\ref{varf14} displays  $\Delta$EW for C\,{\sc iv} BAL troughs as a 
function of $v_{\rm{cent}}$ for variations on three different timescales. Spearman  
test results show no significant correlations for variations on short, moderate, 
and long timescales.

\begin{figure}[h!]
\epsscale{1.2}
\plotone{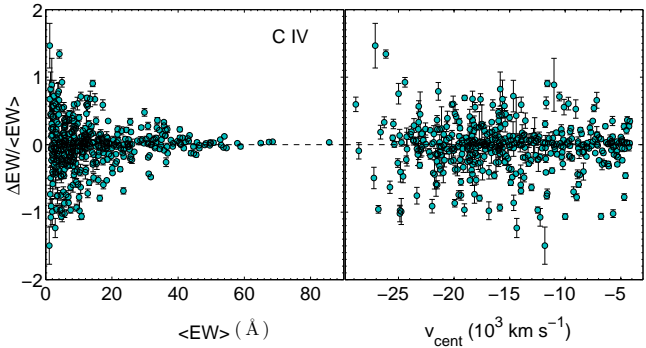}
\caption{C\,{\sc iv} BAL-trough $\Delta \rm {EW}/\langle \rm {EW} \rangle$ as 
a function of $\langle \rm {EW} \rangle$ (left) and $v_{\rm{cent}}$ (right) for 
variations on timescales of more than 1~yr.  }
\label{vare4}
\end{figure}

BAL variability has been noted to be fractionally stronger among weaker troughs 
\citep[e.g.,][]{lundgren07,gibson08,cap11}. However, since the average outflow
velocity is higher for the population of weak BAL troughs than for strong troughs, 
this finding could, in principle, ultimately be a velocity effect. Owing to limited 
sample sizes, it has been difficult to assess if trough weakness or trough 
velocity is the primary driver of increased fractional BAL variability
\citep[e.g.,][]{cap11,ak12}. To assess this with our larger sample, in 
Figure~\ref{vare4} we plot $\Delta \rm {EW}/\langle \rm {EW} \rangle$ 
vs. both $\langle \rm {EW} \rangle$ and $v_{\rm{cent}}$. 
Spearman test results show that there is a clear correlation between 
 $|\Delta \rm {EW}/\langle \rm {EW} \rangle|$  and $\langle \rm {EW} \rangle$
(significance level of $>99.9$\%), but no significant correlation 
between $|\Delta \rm {EW}/\langle \rm {EW} \rangle|$ and $v_{\rm{cent}}$. 
Therefore, our results establish that trough weakness is indeed the primary 
driver of increased fractional BAL variability.

We also search for correlations between $|\Delta$EW$|$  and $\Delta v$, 
$d_{\rm{BAL}}$, and $v_{\rm{cent}}$ for Si\,{\sc iv} BAL troughs on three different 
timescales using Spearman  tests. The test results show highly significant 
($>$99.9\%) correlations between $|\Delta$EW$|$ and $\Delta v$ for Si\,{\sc iv} 
BAL troughs on all three timescales. Similar to the results for C\,{\sc iv} BAL 
troughs, we found no significant correlation between $|\Delta$EW$|$ and 
$d_{\rm{BAL}}$   or $|\Delta$EW$|$ and $v_{\rm{cent}}$ for Si\,{\sc iv} BAL 
troughs.

\subsection{Comparison of C\,{\sc iv}  vs. Si\,{\sc iv} EW Variations}
\label{csibal}

Owing to the differences in  ionization potentials (45.1~eV for Si\,{\sc iv} and 
64.5~eV for C\,{\sc iv}), abundances, and consequently optical depths, 
comparisons of BAL-trough variations between C\,{\sc iv} and 
Si\,{\sc iv} BAL troughs have an important role in assessing possible causes 
of BAL variations. To investigate variations of C\,{\sc iv} and 
Si\,{\sc iv} BAL troughs from the same absorbing material, we compare EWs 
and EW variations of BAL troughs of these two ions at corresponding velocities.

As is well known from previous studies \citep[e.g.,][]
{wey91,barlow93,gibson09,gibson10,cap11,cap12}, C\,{\sc iv} BAL troughs are 
not always accompanied by  Si\,{\sc iv} troughs. Therefore, we search for a 
C\,{\sc iv} BAL trough in overlapping velocity ranges for each Si\,{\sc iv} BAL 
trough in two-epoch spectra of each quasar that sample minimum timescales 
of more than 1~yr, $\Delta t_{\rm{min,1}}$. Given that the Si\,{\sc iv} BAL region 
can be  contaminated by other lines between $-$13000 and $-$30000 km~s$^{-1}$ 
(see Section~\ref{vfrac}), in this comparison we search for accompanying C\,{\sc iv} 
BAL troughs for a total of 136 Si\,{\sc iv} BAL troughs with 
$-3000 > v_{\rm{cent}}> -13000$~km~s$^{-1}$.

\begin{figure}[t!]
\epsscale{1.05}
\plotone{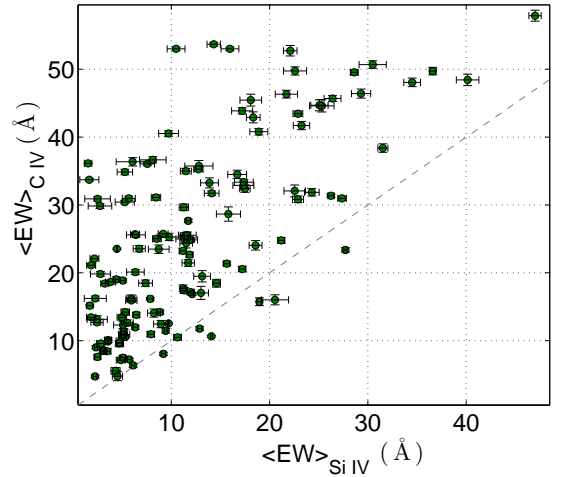}
\caption{Comparison of C\,{\sc iv} and Si\,{\sc iv} BAL-trough EWs for troughs 
having corresponding velocities.  The dashed line indicates equal strengths of 
these two absorption troughs. C\,{\sc iv} BAL troughs tend to be stronger than 
Si\,{\sc iv} BAL troughs.}
\label{varf15}
\end{figure}

Figure~\ref{varf15} compares the  average EW values,  $\langle$EW$\rangle$, 
for  C\,{\sc iv} and Si\,{\sc iv} BAL troughs in overlapping velocity ranges. 
Consistent with previous studies \citep[e.g.,][]{gibson09,gibson10,cap12}, we 
find that  Si\,{\sc iv} absorption tends to be weaker than C\,{\sc iv} absorption 
present at the same velocities. The vast majority  of Si\,{\sc iv} BAL troughs 
($\approx96$\%, 130 of 136) have EWs smaller than the corresponding 
C\,{\sc iv} troughs. 
 
%%%

A number of previous studies have searched for correlations between the 
variations of BAL troughs of different ions. \citet{gibson10} studied variations 
of nine C\,{\sc iv} and Si\,{\sc iv} BAL troughs. Although they were not able to 
find any correlation between EW variations of C\,{\sc iv} and Si\,{\sc iv} BAL 
troughs, they showed a strong correlation for fractional variations (see their 
Figures~11 and 12).  \citet{gibson10} showed that fractional variations of Si\,{\sc iv} 
absorption tend to be larger than those of C\,{\sc iv}. \citet{cap12} monitored the 
C\,{\sc iv} and Si\,{\sc iv} BAL variability relationship using multi-epoch observations 
of a sample of 24 quasars. Although they were unable to demonstrate any 
significant correlation between absolute variations of C\,{\sc iv} and Si\,{\sc iv} 
troughs (see their Figures~6 and 7), they noted a weak trend toward greater 
fractional change for  Si\,{\sc iv}.

\begin{figure*}[t!]
\epsscale{0.85}
\plotone{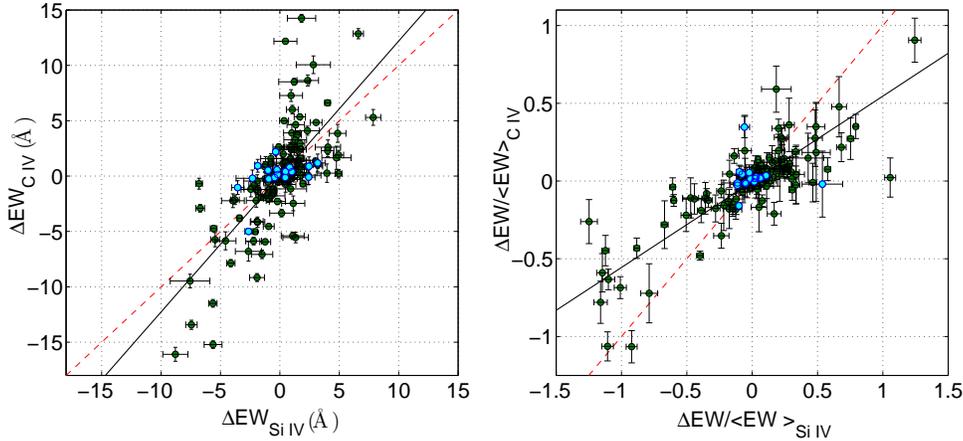}
\caption{Comparison of the variability of C\,{\sc iv} vs.\ Si\,{\sc iv} BAL troughs. 
The left panel compares EW variations for BAL troughs having corresponding 
velocities, and  the right panel  compares fractional EW variations. The light blue 
circles indicate Si\,{\sc iv} and C\,{\sc iv} BAL troughs with similar EWs as defined 
in the text. The dashed line in each panel has a slope of unity and zero offset.  
C\,{\sc iv} and Si\,{\sc iv} trough variations are clearly correlated; the solid black 
lines in each panel show the best-fit relations.  }
\label{varf16}
\end{figure*}

Figure~\ref{varf16} presents comparisons of $\Delta$EW and fractional  
$\Delta$EW measurements for C\,{\sc iv} and Si\,{\sc iv} BAL troughs in 
overlapping velocity ranges. Consistent with previous studies, the EW variations 
of C\,{\sc iv} and Si\,{\sc iv} BAL troughs almost always occur in the same direction, 
and both the $\Delta$EW and $\Delta$EW/$\langle$EW$\rangle$ of C\,{\sc iv} 
and Si\,{\sc iv} BAL troughs are clearly correlated. The Spearman  test results 
demonstrate that the correlations are highly significant ($>99.9\%$) for both 
measurements. We also found that   both the $|\Delta$EW$|$ and 
$|\Delta$EW/$\langle$EW$\rangle |$ of C\,{\sc iv} and Si\,{\sc iv} BAL troughs 
are strongly correlated (at $>99.9\%$ significance), indicating that while the BAL 
troughs almost always vary in the same direction, the magnitudes of the variations 
are also correlated.

As is apparent from Figure~\ref{varf16}, both the $\Delta$EW and 
$\Delta$EW/$\langle$EW$\rangle$ correlations have significant intrinsic scatter. 
To determine the relationship between EW variations of C\,{\sc iv} and Si\,{\sc iv} 
BAL troughs, we use a Bayesian linear-regression model  that considers  
the intrinsic scatter of the sample \citep{kelly07}. The algorithm estimates 
regression parameters using random draws from the posterior distribution.  We 
calculate the mean and standard deviation of model parameters that are found 
using 10000 random draws from our sample. Using the linear-regression 
parameters from this algorithm, we found the following relations:
\begin{eqnarray}
\Delta\rm{EW}_{\rm{C\,IV}} = (1.223 \pm 0.106) \times \Delta\rm{EW}_{\rm{Si\,IV}} 
\nonumber \\ - (0.038 \pm 0.271) 
\label{eqv6}
\end{eqnarray}
\begin{eqnarray}
\frac{\Delta\rm{EW}}{\langle\rm{EW}\rangle}_{\rm{C\,IV}} = (0.551 \pm 0.032) \times 
\frac{\Delta\rm{EW}}{\langle\rm{EW}\rangle}_{\rm{Si\,IV}}  \nonumber \\ - (0.005 \pm 0.012)
\label{eqv7}
\end{eqnarray}
The standard deviation of the intrinsic scatter is 3.034~\AA\, 
for the $\Delta$EW and  0.123 for the $\Delta$EW/$\langle$EW$\rangle$ distributions 
between C\,{\sc iv} and Si\,{\sc iv} BAL troughs.

Equation~\ref{eqv6} shows that the slope of the relationship between $\Delta$EW 
for  C\,{\sc iv} and Si\,{\sc iv} BAL troughs is close to unity. The dispersion of the 
data points increases for large variations of C\,{\sc iv} BAL troughs, indicating a 
possible non-linear component of correlation, especially for 
$|\Delta\rm{EW}_{\rm{C\,IV}}|~\gtrsim~10$~\AA. 
The $\Delta$EW/$\langle$EW$\rangle$ relation between C\,{\sc iv} and Si\,{\sc iv}
BAL troughs  shows that the fractional EW variation of C\,{\sc iv} BAL troughs is  
about half of the fractional EW variation of Si\,{\sc iv} BAL troughs at a 
corresponding velocity. This result is consistent with the findings of  \citet{gibson10} 
and \citet{cap12}. 

In Section~\ref{ewew}, we showed that weak BAL troughs tend to have 
larger fractional EW variability compared to strong BALs. In order to assess if 
the fractional EW of Si\,{\sc iv} BAL troughs is more variable because, in general, 
Si\,{\sc iv} BAL troughs are weaker than C\,{\sc iv} BAL troughs (see Figure~\ref{varf15}),
we examine Si\,{\sc iv} and C\,{\sc iv} BAL troughs with similar EWs.  
We select a subset of BAL troughs where 
\hbox{0.8 $<$ EW$_{\rm{Si\,IV}}$ / EW$_{\rm{C\,IV}} < 1.2$} (a total of 17 cases)  
and repeat the linear-regression modeling. 
Analogous to Equation \ref{eqv6}, we found the relation 
between $\Delta$EW for  these similar-strength C\,{\sc iv} and Si\,{\sc iv} BAL troughs to be 
\hbox{$\Delta\rm{EW}_{\rm{C\,IV}} = (0.486 \pm 0.267) \times \Delta\rm{EW}_{\rm{Si\,IV}} 
+ (0.159 \pm 0.413) $}, indicating that Si\,{\sc iv} troughs vary more in EW compared to 
C\,{\sc iv} troughs of matched EW. Unfortunately,  however, we could not derive a useful analog to 
Equation~\ref{eqv7} since the fitted slope was poorly constrained by the data, and thus
further work is needed to address this matter.

Physically, we do expect the fractional variations of Si\,{\sc iv} to be usually equal 
to or larger than those of C\,{\sc iv}. BAL troughs have lower column densities and 
covering factors in Si\,{\sc iv} than in C\,{\sc iv} \citep[e.g.,][]{bls13}. This situation arises 
primarily because the peak ionization fraction of Si\,{\sc iv} is a factor of 2.5 lower 
than that of C\,{\sc iv} and occurs at a slightly lower ionization parameter 
\citep[e.g., Figure~4 of][]{hamann00}. Thus, C\,{\sc iv} and Si\,{\sc iv} absorption do not 
arise from exactly the same parts of an absorbing structure. When an absorbing 
cloud has a column high enough for the ionization inside it to decrease to the level 
where C\,{\sc iv} is found, the cloud will produce C\,{\sc iv} absorption; when the 
cloud has an even higher column, then both C\,{\sc iv} and Si\,{\sc iv} absorption 
will be produced. As a consequence, C\,{\sc iv} and Si\,{\sc iv} absorption are likely 
to be on parts of their curves of growth where we see either EW changes in 
C\,{\sc iv} with no detectable Si\,{\sc iv} absorption, small EW changes in saturated 
C\,{\sc iv} and large EW changes in unsaturated Si\,{\sc iv} (i.e., larger fractional 
change in Si\,{\sc iv}),  or small EW changes in both C\,{\sc iv} and Si\,{\sc iv} due 
to saturation (equal fractional changes). The foregoing reasoning applies 
regardless of the origin of variations in C\,{\sc iv} and Si\,{\sc iv} absorption along 
the line-of-sight. Our data do rule out a scenario where all trough variations are 
due to the motion across the line-of-sight of clouds of uniform optical depth.

\subsection{Coordination of EW Variations in BAL Quasars with 
Multiple Troughs}\label{multiple}

Quasars with multiple BAL troughs of the same ion provide  an opportunity to 
investigate connections between distinct BAL troughs at different velocities. 
Using observations of BAL quasars with multiple troughs,  \citet{ak12} 
demonstrated that variations in the EWs of multiple C\,{\sc iv} BAL troughs are 
strongly correlated; when one BAL trough in a quasar spectrum disappears the 
other  troughs present usually (11 out of 12 BAL troughs) weaken even for 
velocity offsets as large as 10000--15000~km~s$^{-1}$. Consistent with this, 
\citet{cap12} show that BAL variations at different velocities in the same ion 
almost always show coordinated variations. 

\begin{figure}[t!]
\epsscale{1.1}
\plotone{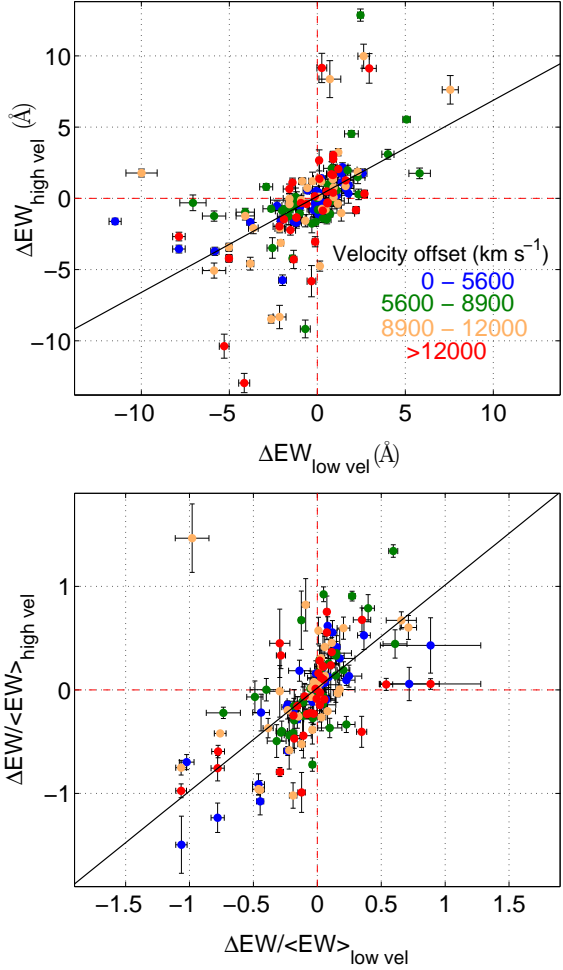}
\caption{Comparison of EW variations (upper panel) and fractional EW variations 
(lower panel) of distinct troughs for BAL quasars  with multiple C\,{\sc iv} troughs 
for variations on timescales of more than 1 yr. Colors indicate the velocity 
separation between troughs; the relevant velocity ranges have been chosen to 
provide an equal number of data points in each range. The solid black lines in 
each panel show the best-fit relations. Variations of distinct troughs are correlated 
for C\,{\sc iv}, although there is substantial scatter in the correlations. }
\label{varf17}
\end{figure}

For the purpose of investigating coordinated variations of multiple BAL troughs 
of the same ion, we select 107 quasars with multiple C\,{\sc iv} BAL troughs 
from our main sample. We compare EW variations of the lowest-velocity BAL 
trough with those of the  other BAL troughs (i.e., BAL troughs at higher velocities) 
present in the same pair of spectra. We detect a total of 137 higher-velocity BAL 
troughs in the spectra of these 107 quasars. Figure~\ref{varf17} shows the EW 
variation of the lowest-velocity BAL trough, $\Delta$EW$_{\rm low~vel}$, as a 
function of EW variations for higher-velocity distinct BAL troughs, 
$\Delta$EW$_{\rm high~vel}$, for variations on timescales of more than 1~yr. 
Similarly, we compare fractional EW variations for the lowest and higher-velocity 
BAL troughs in Figure~\ref{varf17}. Variations of distinct troughs are correlated, 
although there is substantial scatter in the correlations. The Spearman  test  shows 
that both of the apparent correlations for C\,{\sc iv} BAL troughs are highly ($>$ 99.9\%) 
significant. 

In order to assess correlations for magnitudes of $\Delta$EW and 
$\Delta$EW/$\langle$EW$\rangle$ between the lowest-velocity and higher-velocity 
BAL troughs, we run the Spearman test again using the absolute values of the 
measurements and found highly significant ($>$ 99.9\%) correlations.  

To quantify the relationship between EW variations of the lowest-velocity  and 
higher-velocity BAL troughs, we use the Bayesian linear regression model  of  
\citet{kelly07}, considering the intrinsic scatter of the sample.  Using  this algorithm, 
we found the following relations:
\begin{eqnarray}
\Delta\rm{EW}_{\rm{high\,vel}} = (0.674 \pm 0.094) \times 
\Delta\rm{EW}_{\rm{low\,vel}} \nonumber \\ + (0.141 \pm 0.246) 
\label{eq8}
\end{eqnarray}
\begin{eqnarray}
\frac{\Delta\rm{EW}}{\langle\rm{EW}\rangle}_{\rm{high\,vel}} = (0.996 \pm 0.103) 
\times \frac{\Delta\rm{EW}}{\langle\rm{EW}\rangle}_{\rm{low\,vel}} 
\nonumber \\ + (0.018 \pm 0.031)
\label{eq9}
\end{eqnarray}
The  standard deviation of the intrinsic scatter is 2.769~\AA\, for the 
$\Delta$EW and  0.324 for the $\Delta$EW/$\langle$EW$\rangle$ distributions 
for multiple  C\,{\sc iv} BAL troughs.

Equation~\ref{eq8} indicates that BAL troughs at higher velocities tend to have 
smaller EW variations compared to the lowest-velocity BAL troughs. This result is 
expected considering that small-EW BAL troughs can achieve higher velocities, 
and they also tend to have small $\Delta$EW values (see Section~\ref{ewew} and 
Figure~\ref{varf12}). The slope of Equation~\ref{eq9} is close to unity, 
indicating that multiple BAL troughs present in the same two-epoch spectra show 
similar fractional EW variations. This result is even more interesting given that 
weak BAL troughs tend to have larger $\Delta$EW/$\langle$EW$\rangle$ 
compared to  strong ones (see Section~\ref{ewew} and Figure~\ref{varf13}).

Owing to their large fractional variations, disappearing or emerging 
BAL troughs (see Section~\ref{disem2}) stand out in the lower panel of Figure~\ref{varf17}. 
The higher-velocity BAL trough of the quasar SDSS~J143948.06+042112.8, 
 an emergence candidate  with 
\hbox{${\Delta\rm{EW}}/{\langle\rm{EW}\rangle}_{\rm{high\,vel} }= 1.46$}, is found 
as an outlier at the top-left of the plot. The other emergence candidate, the higher-velocity 
BAL trough of the quasar SDSS~J151312.41+451033.9, is found at the extreme of the 
correlation with \hbox{${\Delta\rm{EW}}/{\langle\rm{EW}\rangle}_{\rm{high\,vel} } = 1.34$}. 
Similarly, the higher-velocity BAL trough of   SDSS~J095901.24+550408.2 that 
satisfies the disappearance criteria used in \citet{ak12} is shown at
the lower-left of the plot
(\hbox{${\Delta\rm{EW}}/{\langle\rm{EW}\rangle}_{\rm{high\,vel} } = -1.5$}). 

Table~\ref{vart1} presents the number of additional BAL troughs that vary in the 
same or opposite direction as the lowest-velocity BAL trough. We found that 
$78.1 \pm 8.3\%$ of the additional C\,{\sc iv} BAL troughs show correlated 
variations with the lowest-velocity BAL trough. A combinatorial probability 
calculation shows that  the probability of obtaining such a result for 
independently varying BAL troughs is $\approx10^{-12}$. 
Assuming that C\,{\sc iv} BAL trough pairs show a mixture of perfectly
correlated and perfectly uncorrelated variations,
then some apparently correlated variations will in fact be
uncorrelated variations that appear correlated just by chance.
Accounting for those cases using Table 6, the fraction
of C IV BAL troughs with variations arising from
a mechanism correlated between troughs is 77/137, or 
$56.2 \pm 7.2\%$.

\begin{figure*}[t!]
\epsscale{0.65}
\plotone{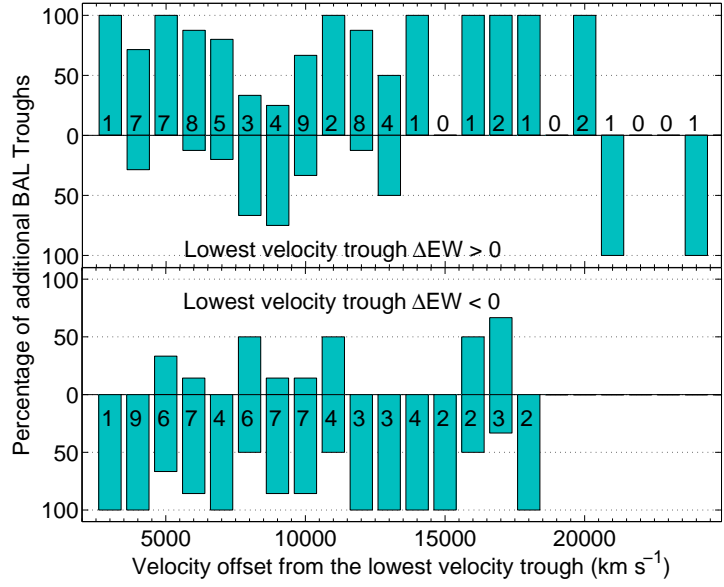}
\caption{Percentage of additional C\,{\sc iv} BAL troughs at a given velocity offset 
with increasing strength (above zero level in each panel; $\Delta\,{\rm EW} > 0$) 
and decreasing strength (below zero level; $\Delta\,{\rm EW} < 0$). The upper
panel is for cases where the lowest-velocity BAL trough strengthens 
($\Delta$EW$_{\rm low~vel}>0$), and the lower panel is for cases where it
weakens ($\Delta$EW$_{\rm low~vel}<0$). The numbers for each bar 
show the total number of additional BAL troughs found at the relevant velocity 
offset. Even BAL troughs separated from the lowest velocity
trough by \hbox{15000--20000~km~s$^{-1}$} show generally correlated variations.}
\label{varf18}
\end{figure*}

To determine the velocity range over which coordinated variations of multiple 
troughs occur, we calculate the velocity offset  of those BAL troughs at higher 
velocities from the lowest-velocity BAL trough present in the same pair of spectra. 
Figure~\ref{varf18} shows the percentage of additional C\,{\sc iv} BAL troughs at 
a given velocity offset with increasing strength ($\Delta\,{\rm EW} > 0$) and 
decreasing strength ($\Delta\,{\rm EW} < 0$) in comparison  with the 
$\Delta$EW variation of the lowest-velocity BAL trough. We find that even BAL 
troughs separated from the lowest-velocity trough by 
\hbox{15000--20000~km~s$^{-1}$} show generally correlated variations. 
Figure~\ref{varf17} indicates that BAL troughs with small velocity offsets plausibly 
have similar scatter about the correlations as troughs with large velocity 
offsets.
To assess if troughs with small velocity offsets show better coordination than 
those with large velocity offsets, we investigate the fraction of BAL troughs 
showing coordinated variations as a function of velocity offset using the 
velocity ranges given in Figure~\ref{varf17}. We found that the fraction
of higher-velocity BAL troughs showing coordinated variations is constant 
at $\approx 78\%$ within the error bars, indicating no strong dependence 
upon velocity offset.

We also examine the coordination of multiple troughs for the Si\,{\sc iv}  transition. 
Similar to C\,{\sc iv}  BAL troughs, the $\Delta$EW and  
$\Delta$EW/$\langle$EW$\rangle$ variations for Si\,{\sc iv} are correlated; the 
Spearman  test indicates that the significance level is $>99.9\%$ for both 
correlations. As for C\,{\sc iv} BAL troughs, multiple Si\,{\sc iv} BAL troughs also 
tend to vary in the same direction. The Si\,{\sc iv} BAL-trough region, however, is 
subject to contamination by some other lines in the velocity range of 
\hbox{13000--30000~km~s$^{-1}$} (see Section~\ref{vfrac}). Considering such 
contamination, we do not perform further examinations for multiple Si\,{\sc iv} 
troughs.

\subsection{EW Variations as a Function of Quasar Properties}\label{qparam}

In this section, we study BAL-trough EW variations  as a function of quasar 
properties, including luminosity, Eddington luminosity ratio, redshift, and radio 
loudness. 

\subsubsection{Luminosity}\label{lbol}

Previous studies \citep[e.g.,][]{kaspi05, bentz09} have  investigated the 
relationship between luminosity and broad-line region (BLR) size in AGN. In 
these studies the derived  power-law relation indicates that the size of the BLR  
increases with increasing quasar luminosity. Given that the BAL region often lies 
outside the  BLR (e.g., BALs often absorb BLR emission), this relation may also 
be indicative of an increasing size and decreasing orbital (transverse) velocity of 
the BAL region with increasing luminosity. A dependence of the transverse BAL 
velocity, and/or continuum source region size, on luminosity can potentially 
change the variability of BAL outflows as a function of luminosity. Furthermore, 
\citet{lb02} found a correlation between outflow velocity and quasar luminosity, 
indicating that BAL troughs in luminous quasar spectra can achieve higher 
velocities. Later, a study by \citet{ganguly07} investigated outflows and the physical 
properties of quasars using a sample of 5088 quasars. In addition to finding 
consistent results with \citet{lb02} on the luminosity-outflow relation, 
\citet{ganguly07} showed that the fraction of BAL quasars (relative to all quasars) 
increases with increasing luminosity. Therefore, we investigate BAL-trough 
EW variations as a function of  the quasar bolometric luminosity,  $L_{\rm {Bol}}$. 
As mentioned in  Section~\ref{ss2} $L_{\rm {Bol}}$ values are taken from \citet{shen11}.

\begin{figure*}[t!]
\epsscale{0.8}
\plotone{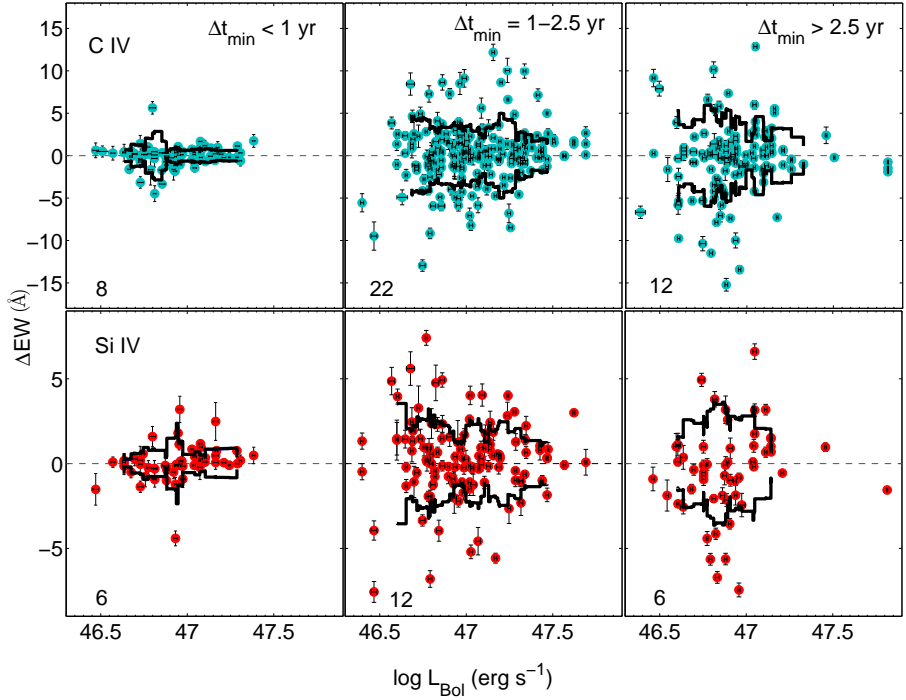}
\caption{EW variability of C\,{\sc iv} (upper panels) and Si\,{\sc iv} (lower panels) 
BAL troughs as a function of quasar bolometric luminosity, $L_{\rm Bol}$, 
for three different rest-frame timescales as labeled. The black curves in each 
panel show the running rms computed with a sliding window; the window width, 
in terms of number of troughs, is given in the lower left of each panel. If BAL 
variability decreased (increased) strongly with luminosity, one would expect to 
see a larger (smaller) scatter of the data points at low luminosities. We do not 
find significant evidence for such behavior on any of the sampled timescales, 
at least over the order-of-magnitude in luminosity with significant coverage.}
\label{varf19}
\end{figure*}

In Figure~\ref{varf19}, we show the dependence of $\Delta$EW for C\,{\sc iv} and  
Si\,{\sc iv}  BAL troughs upon $L_{\rm {Bol}}$ for variations on  three different 
timescales.  In each panel of Figure~\ref{varf19}, we show  standard-deviation 
curves calculated using a sliding window for luminosity-ordered data points 
where the mean EW error in each window is statistically removed from the curve.  
The number of data points contained in each window is given in the lower left 
of each panel of Figure~\ref{varf19}. If BAL variability decreased with luminosity, 
one would expect to see a larger scatter of the data points at low luminosities as 
a result of EW variations in both directions (and vice versa for increasing variability 
with luminosity). In Figure~\ref{varf19}, the standard deviation curves for  C\,{\sc iv}  
and Si\,{\sc iv} on moderate and long timescales may be indicative of BAL variability 
decreasing with luminosity.

To assess formally any correlation between $L_{\rm {Bol}}$ and $|\Delta$EW$|$, 
we use the Spearman  test. Table~\ref{vart2} lists the test probability results 
along with the number of data points in each timescale range. Generally, we do 
not find significant  evidence for correlations between $L_{\rm {Bol}}$ and 
$|\Delta$EW$|$ or $L_{\rm {Bol}}$ and $|\Delta$EW/$\langle$EW$\rangle|$. 
The one exception is for C\,{\sc iv} troughs on moderate timescales  
(99.8\% significance); this is also the case where we have the best trough 
statistics. However, given the number of trials in our correlation tests, we do not 
regard this one case as strong evidence for luminosity dependence.

\subsubsection{Eddington Luminosity Ratio and SMBH Mass} \label{ledd}

\citet{pk04} suggested that the presence of radiatively driven disk winds is highly 
sensitive to the ratio of the quasar bolometric luminosity to the Eddington 
luminosity, $L_{\rm Bol}/L_{\rm Edd}$. Therefore, quasar BAL properties are 
plausibly expected to depend upon $L_{\rm Bol}/L_{\rm Edd}$. Indeed, studies 
by \citet{lb02} and \citet{ganguly07} found a correlation between the outflow 
velocity of quasar winds and $L_{\rm Bol}/L_{\rm Edd}$. To assess the effect of 
$L_{\rm Bol}/L_{\rm Edd}$ on the variability of the BAL region, we investigate 
correlations between BAL-trough EW variations and $L_{\rm Bol}/L_{\rm Edd}$. 
\citet{shen11} calculated $L_{\rm Bol}/L_{\rm Edd}$ values using the virial 
black-hole mass which is estimated using the C\,{\sc iv} emission line for 
$z \geq 1.9$ quasars. As is well known 
\citep[e.g.,][and references therein]{shen08, shen11,shen13}, 
$L_{\rm Bol}/L_{\rm Edd}$ calculations using the C\,{\sc iv} emission line show 
a larger scatter compared to other emission lines such as H$\beta$ and 
Mg\,{\sc ii}. Therefore, uncertainties in the $L_{\rm Bol}/L_{\rm Edd}$ 
measurement may hide any underlying correlations.

\begin{figure*}[t!]
\epsscale{0.8}
\plotone{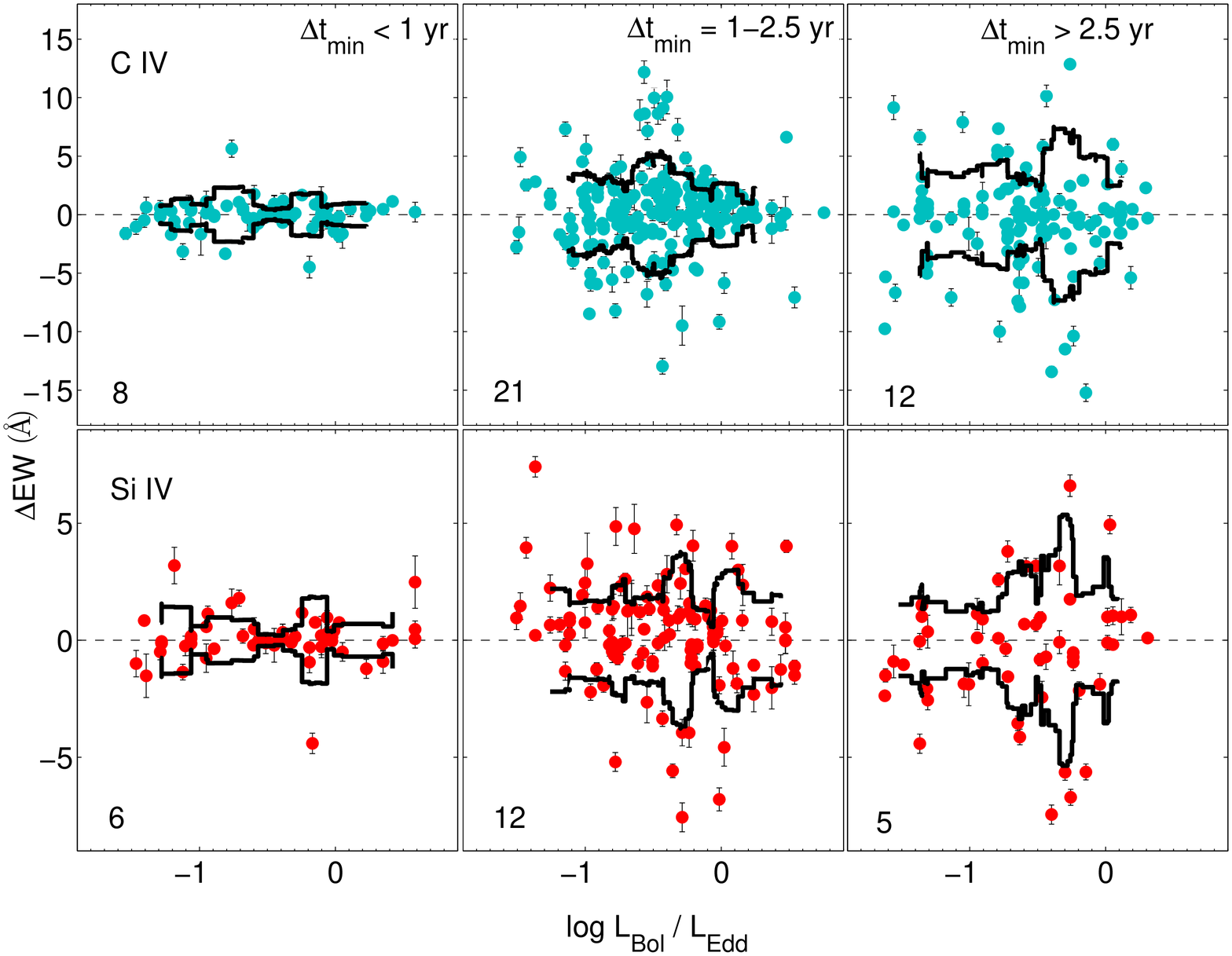}
\caption{Same as Figure~13 but for Eddington-normalized luminosity, 
$L_{\rm Bol}/L_{\rm Edd}$. We do not find any significant evidence 
for changes in BAL variability with $L_{\rm Bol}/L_{\rm Edd}$.}
\label{varf20}
\end{figure*}

Figure~\ref{varf20} shows  $\Delta$EW for C\,{\sc iv}  and Si\,{\sc iv} BAL 
troughs  as a function of $L_{\rm Bol}/L_{\rm Edd}$ on short, moderate and 
long timescales. Similar to Figure~\ref{varf19}, we show the standard-deviation 
curves computed with a sliding window in each panel of  Figure~\ref{varf20}.  As 
for luminosity, if BAL variability decreased (increased) with Eddington luminosity 
ratio, one would expect to see a larger (smaller) scatter of the data points at low 
Eddington-normalized luminosities. However, the standard-deviation curves do 
not indicate such a trend for C\,{\sc iv} and Si\,{\sc iv} BAL troughs on the sampled 
timescales.

We run  Spearman tests to assess connections between BAL EW variations 
and $L_{\rm Bol}/L_{\rm Edd}$. The test results generally show no evidence of 
significant correlations between  $L_{\rm Bol}/L_{\rm Edd}$ and $|\Delta$EW$|$ 
or $L_{\rm Bol}/L_{\rm Edd}$ and $|\Delta$EW/$\langle$EW$\rangle|$ for 
C\,{\sc iv} and Si\,{\sc iv} BAL troughs on the sampled timescales (see 
Table~\ref{vart2}). However, as for luminosity, the small probability for  
C\,{\sc iv} troughs on moderate timescales might be indicative of  a correlation 
between $L_{\rm Bol}/L_{\rm Edd}$ and $|\Delta$EW$|$. This result is consistent 
with the test result between $L_{\rm Bol}$ and $|\Delta$EW$|$ considering the 
strong relation between $L_{\rm Bol}$ and $L_{\rm Bol}/L_{\rm Edd}$.   

In addition, we assess any correlation between EW variations of BAL troughs 
and SMBH mass estimates from \citet{shen11}. The Spearman test results show 
no evidence of significant correlations between $M_{\rm BH}$ and $|\Delta$EW$|$ 
or  $M_{\rm BH}$ and $|\Delta$EW/$\langle$EW$\rangle|$ for C\,{\sc iv} and 
Si\,{\sc iv} BAL troughs on the sampled timescales (see Table~\ref{vart2}).

\subsubsection{Redshift} \label{redshift}

It is also of interest to assess any dependence of BAL variability upon redshift; 
e.g., since \citet{allen11} reported a relation between the fraction of BAL quasars 
and redshift. Therefore, we investigate  correlations between quasar redshifts 
and EW variations on short, moderate, and long timescales. Plots similar to 
Figures~\ref{varf19} and \ref{varf20} generally do not show significant relations 
with redshift, consistent with the results in Table~\ref{vart2}.

\subsubsection{Radio Loudness} \label{radio}

Do BAL troughs in radio-quiet and radio-loud quasars vary differently?  Given 
that the presence of  powerful radio jets in a quasar appears to affect its wind 
properties \citep[e.g.,][]{becker00,shankar08,miller09, miller12,welling13}, different BAL 
variability might be expected for radio-loud quasars. 

\begin{figure*}[t!]
\epsscale{0.8}
\plotone{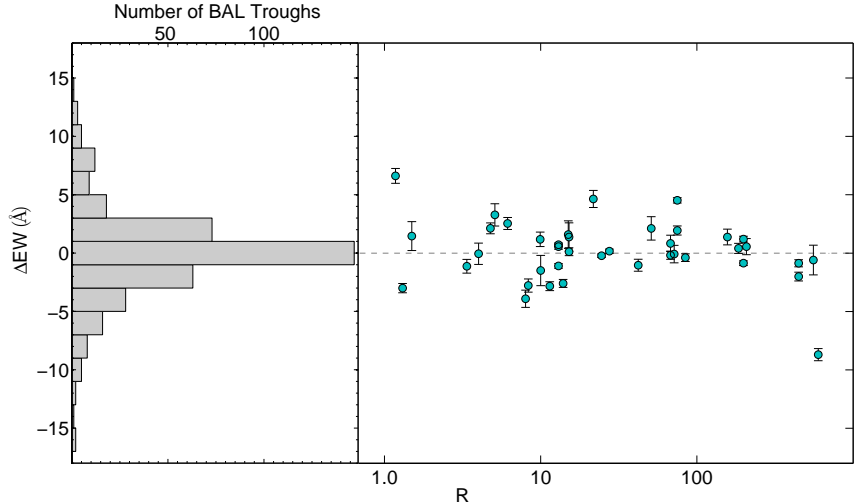}
\caption{EW variation for C\,{\sc iv} BAL troughs as a function of radio-loudness 
parameter, $R$, for variations on timescales of more than 1~yr. Individual data 
points are plotted for quasars with radio detections in the FIRST catalog, while a 
histogram is shown along the left-hand side for quasars lacking radio detections. 
We do not find strong evidence for changes in BAL variability with $R$. }
\label{varf21}
\end{figure*}

To address this issue, we first examine $\Delta$EW and 
$\Delta$EW/$\langle$EW$\rangle$ as a function of $R$ for BAL variations on 
timescales of more than 1~yr; here $R$ is the radio-loudness parameter calculated by 
\citet{shen11} using VLA FIRST observations (see Section~\ref{ss2}). 
Radio emission is detected from 33 of our 
main-sample quasars, nine of which have $R>100$ and thus are radio-loud. 
We identified  40 C\,{\sc iv} and 28 Si\,{\sc iv} BAL troughs in multi-epoch spectra 
of these 33 quasars. In Figure~\ref{varf21}, we show the distribution of 
$\Delta$EW for the 40 C\,{\sc iv} BAL troughs as a function of $R$.  A 
Spearman  test shows no indication of a significant correlation between 
$|\Delta$EW$|$ and $R$. Similarly, we find no significant correlation between 
$|\Delta$EW/$\langle$EW$\rangle|$ and $R$. We repeat the correlation 
tests for those nine radio-loud quasars with $R > 100$ and found no significant 
correlations.

Second, we compare  BAL-trough EW variations on matched timescales for 
radio-detected ($R \gtrsim 1$) and radio-non-detected  ($R \lesssim 1$)  quasars. 
We randomly select  a  sample 
of C\,{\sc iv} BAL troughs of radio-non-detected quasars that have matching timescales 
with the 40 C\,{\sc iv} BAL troughs, and then run a two-sample KS test comparing 
the $\Delta$EW distributions of these two samples. The KS test results show no 
inconsistency between the two $\Delta$EW distributions. Similarly, we do not find 
significant inconsistency from the comparison of the 
$\Delta$EW/$\langle$EW$\rangle$ distributions. 

We also repeat the matched-sample selection and KS tests for those nine 
radio-loud quasars with $R > 100$. We find no inconsistency between 
BAL-trough variability of radio-loud and radio-non-detected quasars from the 
comparison of the $\Delta$EW and $\Delta$EW/$\langle$EW$\rangle$ 
distributions for C\,{\sc iv} BAL troughs.

\section{Discussion} \label{disc}

In Section 4, we presented our main observational results with generally limited 
physical discussion. In this section, we therefore examine the physical implications 
of  our most notable findings for quasar winds. 

 \subsection{The Frequency of BAL Variations}\label{dvar}

Our large sample demonstrates that C\,{\sc iv} and Si\,{\sc iv} BAL variations 
are common on multi-year rest-frame timescales. About \hbox{50--60}\% of 
both C\,{\sc iv} and Si\,{\sc iv} BALs vary detectably over \hbox{1--3.7}~yr, 
with the exact derived fraction depending upon the approach taken in variable 
BAL-trough identification. The fraction of quasars showing BAL-trough 
variability is even higher, since some BAL quasars possess more than one 
trough. Using the two-epoch observations of our sample, we find that 
$62.2^{+4.9}_{-4.6}$\% (181/291) and $59.1^{+6.3}_{-5.7}$\% (107/181) of 
BAL quasars show C\,{\sc iv} and Si\,{\sc iv} trough variability, respectively. 
Our results are derived from a much larger sample of quasars than past work, 
and they also do not suffer from systematic uncertainty owing to multi-counting 
biases (see Footnote~\ref{fot19}) or preferential observation of BAL quasars 
known to show trough variability 
\citep[e.g., see Section~3.3 of][]{cap13}. However, our results are generally consistent 
with past studies. For example, \citet{cap13} found that $\approx 55$\% of their 
sample of BAL quasars showed C\,{\sc iv} trough variability on a timescale of 
2.5~yr (see their Figure~12). \citet{gibson08} found that 12 of their 13 
($\approx 92$\%) BAL quasars showed C\,{\sc iv} trough variability, although 
they sampled longer timescales of \hbox{3--6}~yr where a higher percentage 
of variable BAL quasars is expected.

\begin{figure*}[t!]
\epsscale{1}
\plotone{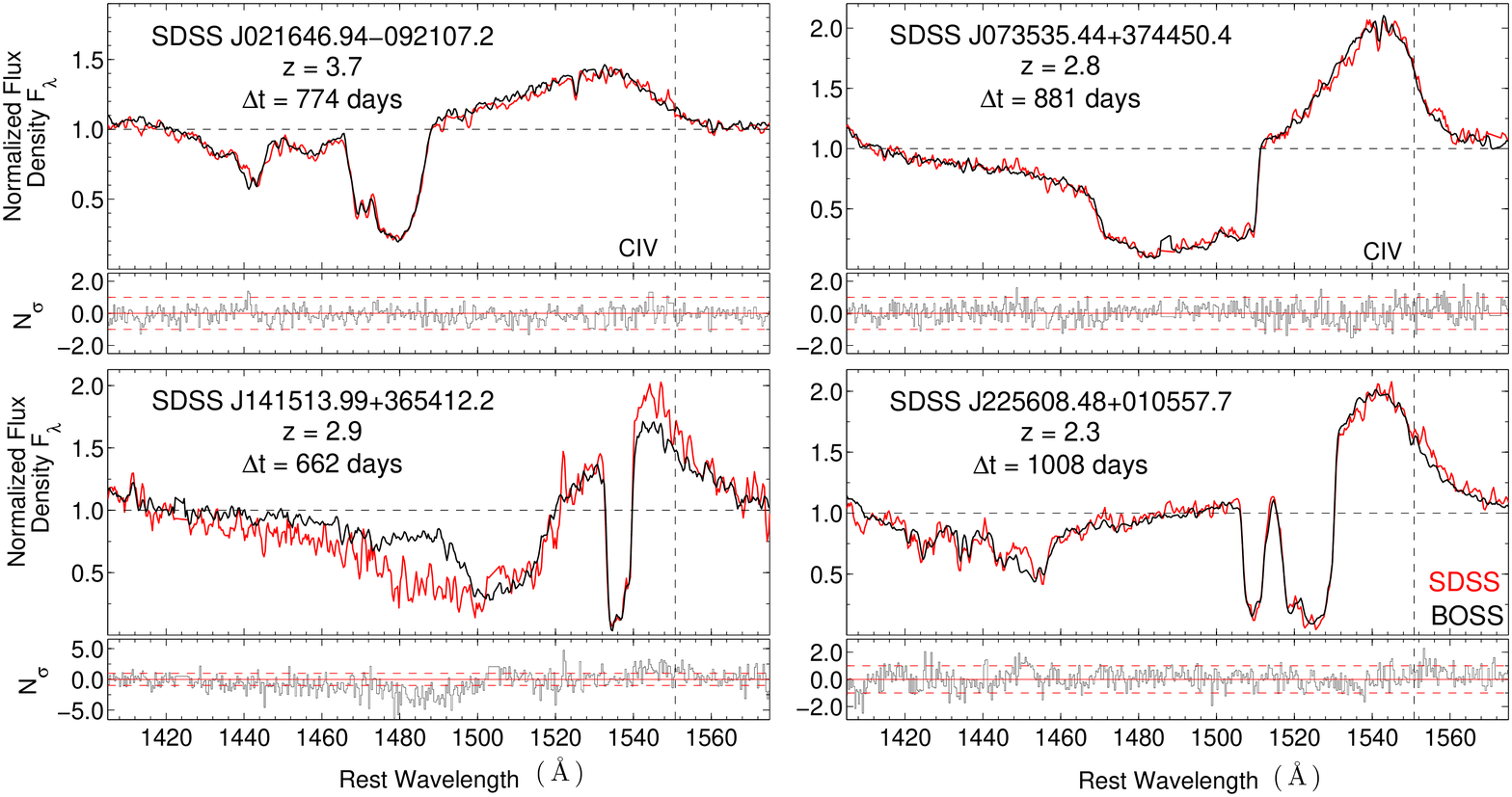} %10092 10158 11446 11973
\caption{Two-epoch spectra from SDSS (red) and BOSS (black) of quasars with 
BAL troughs that remain remarkably stable over multiple years in the rest
frame; in some cases additional distinct BAL troughs that vary are present as 
well. These are examples of quasars that may have some BAL absorption arising 
on large physical scales. 
The dashed vertical lines show the C\,{\sc iv} emission-line  rest wavelength. The 
lower section of each panel shows $N_{\sigma}$ values for SDSS vs. BOSS 
observations where the dashed red lines show the $\pm 1\sigma$ level.}
\label{vare9}
\end{figure*}

The high incidence of BAL variability that we find on multi-year timescales is 
generally supportive of models where most BAL absorption arises within about 
an order-of-magnitude of the radial distance of the launching region; i.e., 
\hbox{10--1000} light days (see Section~\ref{vintro}). However,  our results are still 
consistent with a significant fraction of BAL absorption arising on larger scales. 
Indeed, some of the troughs in our sample remain remarkably stable over 
multiple years in high-quality spectra (e.g., see Figure~\ref{vare9}). This result  
is consistent with independent findings that some BAL absorption arises on 
\hbox{$\approx100$--3000~pc} scales \citep[e.g.,][]{faucher12,arav13,borquet13},
although lack of variability does not strictly require that the absorption arises on 
these size scales. In some cases, we find both remarkably stable as well as variable 
troughs in the same object (e.g., SDSS~J141513.99+365412.2 and 
SDSS~J225608.48+01557.7 in Figure~\ref{vare9}), suggesting that BAL 
absorption may arise over a wide range of size scales. Those BAL troughs 
believed to be formed on \hbox{$\approx100$--3000~pc} scales should be 
intensively monitored for variability as an independent check of their distance 
from the SMBH. 

\subsection{Constraints Upon BAL Lifetimes}\label{dcon}
               
In Section~\ref{EWt}, we examined the behavior of $\Delta \rm {EW}$  and 
$\Delta \rm {EW}/\langle \rm {EW} \rangle$ for  BAL troughs as a function 
of rest-frame timescale. One implication of these results is that we can make 
basic order-of-magnitude estimates of the average ``lifetime'' of a BAL trough; 
here the lifetime represents the time over which a BAL trough is seen along 
our line-of-sight and not necessarily the physical lifetime of the absorbing 
structure (e.g., a structure could remain intact but move out of our line-of-sight). 
We use Equation~\ref{eqv4} considering that a BAL trough will disappear when 
the magnitude of its EW variation is equal to its  EW, $|\Delta \rm{EW}| = \rm{EW}$. 
In our current sample the measured EWs are in the range \hbox{0.3--87.5~\AA\,} 
with a median of $10.9$~\AA.  Given that the strengthening and weakening of BAL 
troughs occur at similar rates (see Section~\ref{vdist}), solution of Equation~\ref{eqv4} 
for $|\Delta \rm{EW}| = 10.9$~\AA\,  gives half of an average BAL lifetime (the 
other half corresponds to the BAL strengthening from 0~\AA\, to 10.9~\AA). 
Therefore, this approach reveals that  
\hbox{$\langle t_{\rm{BAL}}\rangle = 1600^{+2100}_{-900}$~yr}. 
The large error bars of our lifetime estimate  largely arise because of the significant 
difference between the sampled timescales and the derived lifetime. Although the 
measurements underlying Equation~\ref{eqv4} span a factor of $\approx10^3$ in 
timescale, the lifetime appears a factor of $\sim 10^3$ longer still (and thus, of 
course, the extrapolation used to estimate a lifetime needs to be treated with caution).
Our estimate of the average BAL lifetime is  consistent with the lower limits for BAL 
lifetime in \citet{gibson08,gibson10}  and \citet{hall11}. 
We note that this average lifetime is long compared to the orbital time of the 
accretion disk at the expected wind launching radius of 10--100 light days (for a 
$10^9~\rm{M}_{\odot}$ SMBH, $t_{\rm{orb}}$ is $\sim 50$~yr). It is also long, or 
at best comparable to, the orbital time at the radius of the BLR of $\approx 1$ light 
year (e.g., \citeauthor{kaspi07} 2007; $t_{\rm orb} \sim 500$~yr).

One should keep in mind that BAL lifetimes derived with the above approach 
depend significantly upon EW, and thus the lifetimes derived here are in 
agreement with those estimated by \citet{ak12} using a qualitatively different 
approach. Using a sample of 21 examples of disappearing BAL troughs, 
\citet{ak12} found the average BAL lifetime  to be about a century; the 
measured  EWs of these 21 BAL troughs in the first-epoch spectra  were 
\hbox{2.2--10.6~\AA\,}, with a median of $4.7$~\AA. 
Applying  the approach based upon Equation~\ref{eqv4}, we find 
\hbox{$\langle t_{\rm{BAL}}\rangle = 60^{+38}_{-22}$~yr} 
for these 21 disappearing BAL troughs, consistent with \citet{ak12}. 

\subsection{Relation of BAL Disappearance and Emergence to General 
BAL Variability}\label{disem2}

In this section, we  assess where BAL disappearance and emergence 
events lie within the distribution of BAL EW variability.  Since our sample-selection 
criteria (see Section~\ref{ss2}) limit the redshift,  SN$_{1700}$, and  $\rm{BI'}$ more 
strictly than in \citet{ak12}, our main sample includes only three of the disappearing 
C\,{\sc iv} BAL troughs described in \citet{ak12}. None of these three  
has a Si\,{\sc iv} BAL trough at corresponding velocities at 
any epoch. Figure~\ref{varf8} displays the three examples of disappearing 
C\,{\sc iv} BAL troughs. \citet{ak12} found that disappearing BAL troughs tend to 
have relatively small EWs in their first-epoch spectra; the three examples of 
disappearing C\,{\sc iv} BAL troughs lie between 0 and $-$5\,\AA\, in the  
$\Delta$EW distribution. In the distribution of fractional EW variations, all three 
of the disappearing C\,{\sc iv} BAL troughs lie at  the negative extreme of the 
distribution, although they are not a distinct population. These examples of BAL 
disappearance indicate that disappearance is an extreme example of general BAL 
variability, rather than a qualitatively  distinct phenomenon.

We have visually inspected the BAL troughs whose EWs decreased by at least 
a factor of five between the two epochs (i.e., BAL troughs with 
\hbox{$\Delta$EW/$\langle$EW$\rangle < -1.33$}). Our main sample includes 
one additional case of  C\,{\sc iv} BAL disappearance that satisfies the 
disappearance criteria used in \citet{ak12}; this trough, in the quasar 
SDSS~J095901.24+550408.2, is plotted in the upper-right panel of 
Figure~\ref{varf8} at  \hbox{$\Delta$EW/$\langle$EW$\rangle=-1.5$}. 
 %10592
Two of the main-sample quasars (SDSS~J092522.72+370544.1 and  
SDSS~J112055.79+431412.5), %10475, 10861
whose  Si\,{\sc iv} BAL trough EWs decreased by at least a factor of five 
between the two epochs, can be classified as candidates for quasars with 
disappearing Si\,{\sc iv} BAL troughs. The corresponding C\,{\sc iv} BAL troughs 
of these two quasars do not disappear. 

Visual inspection of the BAL troughs whose EWs increased by at least a factor 
of five between the two epochs (i.e.,  BAL troughs with 
\hbox{$\Delta$EW/$\langle$EW$\rangle > 1.33$}) shows that two of the 
C\,{\sc iv} BAL troughs (for the quasars  SDSS~J143948.06+042112.8 and 
SDSS~J151312.41+451033.9) %11535,11632
and two of the Si\,{\sc iv} BAL troughs 
(for the quasars SDSS~J145045.42$-$004400.2 and 
SDSS~J160202.40+401301.4) %11567, 11748 
can be classified as candidates for BAL-trough emergence. These emergence 
events lie at the positive extreme of the distribution of fractional EW variations, 
but again do not appear to be a distinct population. The detailed analysis of 
emergence is beyond the scope of this study, but we plan to address this topic 
in  future work. 

\subsection{A Random-Walk Model for The Evolution of BAL Troughs
}\label{ranw}

\begin{figure}[t!]
\epsscale{1}
\plotone{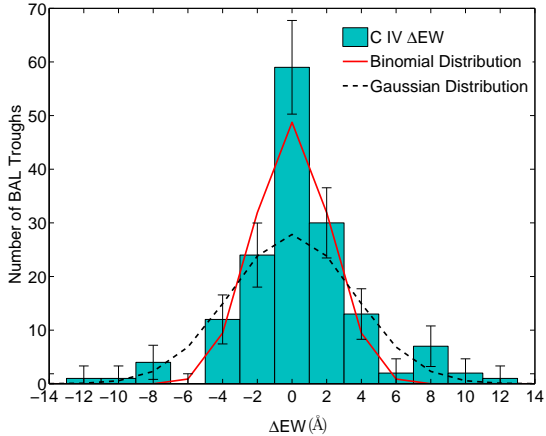}
\caption{Distribution of C\,{\sc iv} BAL-trough EW variations, $\Delta$EW, 
for variations on timescales of 2.0--2.5~yr and a binomial distribution calculated 
for a  random-walk model with six steps and a step size of 1.65~\AA\,.
We bin our  $\Delta$EW data at a bin width of 2~\AA, and  calculate the counting 
errors following \citet{gehrel}.  The black dashed line shows the best-matching 
Gaussian distribution. }
\label{varf9}
\end{figure}

Modelers of quasar winds have not yet been able to make quantitative predictions 
of how BAL-trough EWs should evolve over time, and thus it is not possible to use 
our observational data on this topic to test wind models directly. Therefore, as an 
alternative, we test and constrain a ``toy'' model where long-term EW variations 
occur as a series of discrete events. Assuming that a BAL-trough EW varies by a 
fixed amount $\delta$EW after a fixed time step $\delta t$, we use a simple 
one-dimensional (unbiased) random-walk model to characterize the EW evolution. 
Therefore, we assume that over a period of time, $T$, a BAL trough undergoes 
$n = T/ \delta t$ changes to its EW. For the purpose of defining $T$, we select our 
data to examine  BAL-trough EW variations on timescales of  2.0--2.5~yr (see 
Section~\ref{vdist} for the motivation for this interval choice). Therefore, we assume each 
BAL trough evolved with the same number of steps over a time $T = 2.25$~yr (i.e., 
the mean of the sampled timescales in this interval). The $\Delta$EW distribution 
for C\,{\sc iv} BAL-trough variations on 2.0--2.5~yr timescales is shown in 
Figure~\ref{varf9}.

\begin{figure}[t!]
\epsscale{1}
\plotone{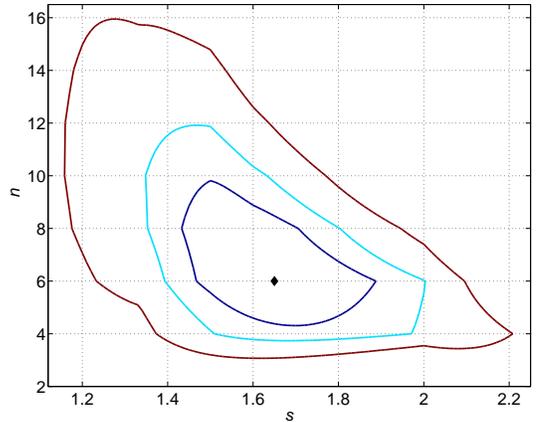}
\caption{68.3\% (dark blue), 90\% (light blue), and 99\% (dark red) confidence 
intervals (corresponding to $\Delta \chi^2 =$ 2.30, 4.61, and 9.21) for the two 
parameters of the random-walk model describing  C\,{\sc iv} BAL-trough EW 
variations, $\Delta$EW,  on timescales of  2.0--2.5~yr.  The black diamond  
indicates the best-fitting model with the number of steps, $n=6$, and the size of 
each step, $s=1.65$~\AA. }
\label{varf10}
\end{figure}

A random-walk model produces a binomial distribution for $\Delta$EW 
determined by two parameters: the number of steps, $n$, and the size of each 
step, $s$. To determine the random-walk model that best matches our data, we 
produced a set of binomial distributions using  different numbers of steps and step 
sizes, and  performed a $\chi^2$ test between our data and each distribution. A 
random-walk model with $n = 6$ and $s = 1.65$~\AA\, yields the minimum 
$\chi^2$ for C\,{\sc iv} BAL troughs on timescales  of 2.0--2.5~yr. The best-fitting 
parameters suggest that C\,{\sc iv} BAL troughs  in this model tend to evolve in a 
small number of steps on 2.0--2.5~yr timescales. 
Figure~\ref{varf9} shows the binomial distribution that best 
matches the $\Delta$EW distribution. The \hbox{$P(\chi^2,\nu) = 64.3\%$} for the 
best-fitting model indicates that the data are consistent with the random-walk 
model.   We calculate  confidence intervals for the model parameters using 
numerical $\Delta \chi^2$ confidence-region estimation for two parameters 
of interest \citep[see Section~ 15.6.5 of ][]{press}. In Figure~\ref{varf10}, we  show 
confidence intervals of 68.3\%, 90\%, and 99\% for two parameters of interest. 
The best-fitting fixed time step is 
$\delta t =137^{+46}_{-53}$~days;  we calculate uncertainties on this parameter 
at  68.3\% confidence.

A binomial distribution produced by a random-walk model closely approximates 
a Gaussian distribution for a large number of steps ($n \gtrsim 20$); for a smaller 
number of steps the binomial distribution remains non-Gaussian.  Consistent with 
the fact that the $\Delta$EW distribution for C\,{\sc iv} BAL troughs on 2.0--2.5~yr 
timescales is  non-Gaussian (see Section~\ref{vdist}),  the results in Figure~\ref{varf10} 
show that our data on the $\Delta$EW distribution for C\,{\sc iv} BAL troughs reject 
any Gaussian approximation (i.e., large values of $n$ are disfavored). As for the 
$\Delta$EW distributions in  Section~\ref{vdist}, the binomial model distribution also has 
a stronger central peak compared to the Gaussian approximation. 

Similarly, we use a random-walk model to describe the evolution of Si\,{\sc iv} 
BAL-trough EWs. The best-fitting model parameters  for Si\,{\sc iv} BAL-trough 
EW variations on timescales of 2.0--2.5~yr are $n = 6$ and $s = 1.56$~\AA\, 
with \hbox{$P(\chi^2,\nu) = 25.2\%$}. The best random-walk model parameters 
for both C\,{\sc iv} and Si\,{\sc iv} BAL troughs are almost identical; this result is 
broadly consistent with the correlated C\,{\sc iv} and Si\,{\sc iv} BAL-trough 
variations discussed in Section~\ref{csibal}.

\begin{figure*}[t!]
\epsscale{0.9}
\plotone{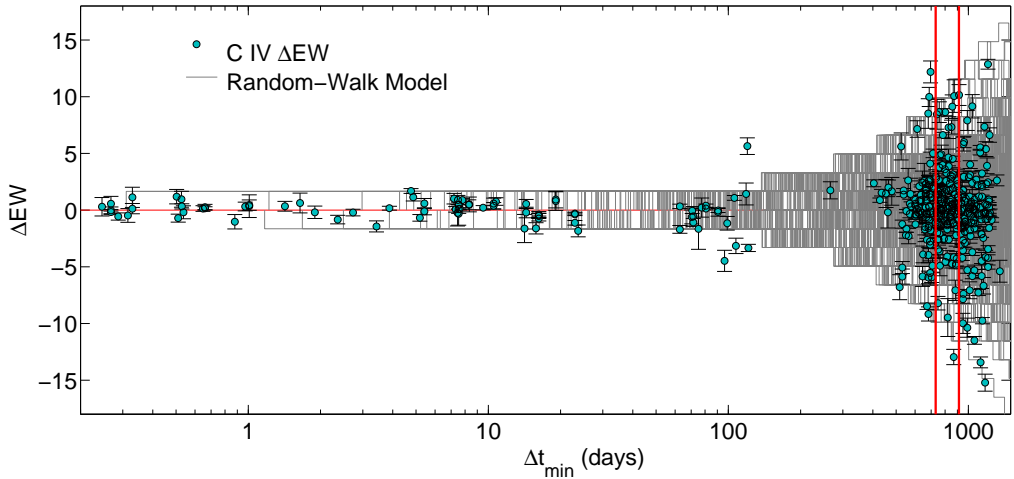}
\caption{Comparison between EW variations for 428 C\,{\sc iv} BAL troughs 
in our main sample  and the best-fitting random-walk model. Vertical red bars 
show a timescale range  of 2.0--2.5~yr. }
\label{varf11}
\end{figure*}

Using the best-fitting random-walk model ($n=6$, $s=1.65$~\AA) calculated for  
C\,{\sc iv} BAL-trough EW variations on timescales  of 2.0--2.5 yr, we simulated 
EW variations for a sample of C\,{\sc iv} troughs; we selected the simulated 
sample to have the same size as our main sample (428). Figure~\ref{varf11} 
shows a comparison between the simulations and observed EW variations on 
timescales of 5.9 hr to 3.7 yr. Given that the fixed time step is larger than the 
shorter timescales sampled by our data, we randomly select the start time  
of each simulated trough and preserved $\delta t =137$~days for each step. 
As can be seen in Figure~\ref{varf11}, the best-fitting random-walk model on 
timescales of 2.0--2.5 yr
appears to represent acceptably the range of EW variations on a larger range of 
timescales. We also compared sliding-window standard deviation curves computed 
for the data and the model. These curves are consistent on timescales of 
100--1400 days, and on shorter timescales the data are too limited for proper 
comparison. 

The simple random-walk model examined above should only be considered to 
be an approximation of a more complex reality; e.g., BAL EWs surely do not 
evolve via discrete steps after remaining constant for a fixed interval. Nevertheless, 
the basic apparent success of this model suggests that some of its features are of 
importance. For example, the derived characteristic timescale of $\approx 137$~days 
may represent the length of time that BAL EWs tend to increase/decrease 
monotonically. This timescale also suggests that the distribution of BAL EW values 
will only approximate a Gaussian for monitoring periods of $\gtrsim7.5$~yr in the rest frame.

\subsection{Shielding Gas Changes as a Driver of Coordinated Multi-Trough 
Variability? }
\label{shield}

Our large sample has allowed us to establish, beyond question, the coordinated 
variability of multiple troughs in quasar spectra \citep{cap12,ak12}. This 
phenomenon is found for both C\,{\sc iv} and Si\,{\sc iv} troughs; e.g., for C\,{\sc iv} 
the derived probability of this phenomenon being due to chance is 
$\approx~10^{-12}$. Our investigations of coordinated multi-trough variability show 
that it is present even for velocity offsets as large as 
\hbox{15000--20000~km~s$^{-1}$}.

Coordinated multi-trough variability might appear somewhat surprising in the 
context of quasar-wind models (see Section~\ref{vintro}), and it can provide insights  
into the drivers of BAL variability more generally. Absorption troughs at widely 
separated velocities should be formed in outflow streamlines that are widely 
physically separated and thus largely unrelated. Some mechanism capable of 
acting over a wide range of radii must be present to enforce the coordinated 
multi-trough variability. Consideration of current quasar-wind models suggests 
a possible enforcement mechanism is substantial changes in the amount of 
``shielding gas'' along the line-of-sight \citep[e.g.,][]{misawa07,ak12}; the 
shielding gas prevents BAL outflows from being overionized by nuclear extreme 
UV (EUV) and X-ray photons \citep[e.g.,][]{murray95,proga00}. Shielding-gas 
variability is likely according to  quasar-wind simulations \citep[e.g.,][]
{proga00,sim10,sim12}, and it may have been detected in a couple of BAL 
quasars \citep[e.g.,][]{gal04,saez12}. It can arise due to internal gas motions 
and/or accretion-disk rotation at the relatively small radii where the shielding 
gas is located. Changes in the column density of shielding gas can increase 
or decrease the level of ionizing EUV/X-ray radiation reaching the BAL wind. 
In response to this variation, absorption components at different velocities can 
rise or fall in ionization level together, leading to coordinated multi-trough 
variations.\footnote{The absorption depth in less-saturated regions of absorption 
would naturally be more responsive to ionization changes than that in
highly saturated regions, and such velocity dependent saturation effects
might explain how C\,{\sc iv} and Si\,{\sc iv} variations usually occur in 
relatively narrow portions of BAL troughs (see Section~\ref{width}).} 
Of course, this same mechanism might be applicable to single-trough BAL quasars 
as well and thus much of BAL variability in general. 

One method for testing this scenario would be to examine the continuum variability
of quasars showing coordinated multi-trough variations. Ideally one would measure
the relevant EUV/X-ray ionizing continuum directly, but measurements of the more
straightforwardly accessible rest-frame UV continuum might suffice. Unfortunately, 
limitations in the spectrophotometric calibration of the BOSS spectra do not allow
this test to be implemented straightforwardly at present, but it should be possible
in the future as BOSS calibration efforts proceed \citep{margala11,suzuki13}. 

It is also perhaps possible that EUV/X-ray luminosity variations could primarily drive 
the BAL variability rather than shielding-gas variations. This would require that some 
EUV/X-ray radiation is not blocked from reaching the wind by the shielding gas.

\section{Summary and Future Work} \label{vsummary}

We have studied the variability of C\,{\sc iv} and Si\,{\sc iv} BAL troughs using a 
systematically observed sample of 291 BAL quasars. We have utilized 699 
high-quality spectra of these  quasars obtained by SDSS-I/II and BOSS. The 
main basic observational results of our study are the following: 

\begin{enumerate}
\item
Using a BAL-trough definition designed for the investigation of trough variability 
in multi-epoch observations, we have identified 428 distinct C\,{\sc iv} and 
235 distinct Si\,{\sc iv} BAL troughs in the multi-epoch spectra of our 291 
``main-sample'' quasars. The sampled rest-frame timescales range between 
5.9~hr and 3.7~yr with a median of 2.1~yr. See Section~\ref{bal2}.

\item
We identify variable BAL troughs using two different approaches and find, 
consistent with earlier work, that BAL variability is common on multi-year
timescales. About \hbox{50--60}\% of both C\,{\sc iv} and Si\,{\sc iv} BAL 
troughs detectably vary on rest-frame timescales of \hbox{1--3.7}~yr. 
Our large sample size also allows us to quantify the fraction of BALs 
that vary by a given amount in EW. The cumulative fraction of BAL troughs 
varying by a given $|\Delta$EW$|$ threshold drops with increasing 
$|\Delta$EW$|$, but it remains significant even for $|\Delta$EW$|$ values 
as large as 5~\AA. See Sections~\ref{vr2} and \ref{vfrac}.

\item
C\,{\sc iv} BAL variability is found across a wide range of outflow velocity. 
The percentage of C\,{\sc iv} BAL regions showing variability remains relatively 
constant at \hbox{30--40}\% at velocities from $-3000$~km~s$^{-1}$ to 
$-25000$~km~s$^{-1}$, and it then rises at higher velocities. The dependence of
Si\,{\sc iv} BAL variability upon outflow velocity is more difficult to 
constrain owing to contamination by unrelated emission lines. See Section~\ref{vfrac}.

\item
We show, consistent with earlier work, that C\,{\sc iv} and Si\,{\sc iv} BAL 
variations usually occur in relatively narrow portions of BAL troughs 
(i.e., troughs do not vary monolithically in depth). The variable portions 
typically span $<30$\% of the trough velocity width. Narrow BAL 
troughs tend to have variable portions that span a larger fraction of
their velocity width than broad BAL troughs. See Section~\ref{width}.

\item
The incidence of variability within a BAL trough does not depend significantly 
upon the relative velocity within the trough. See Section~\ref{width}.

\item
The EW and fractional EW variations of C\,{\sc iv} and Si\,{\sc iv} BAL troughs 
increase with sampled rest-frame timescale, and our large sample allows 
improved measurement of this effect over a range of $\approx 10^3$ in 
timescale. We find that the rate-of-change of BAL EW is larger on 
\hbox{0.01--1}~yr than on \hbox{1--3.7}~yr timescales. See Section~\ref{EWt}.

\item
Even in this large sample, the distributions of EW variations and fractional EW 
variations for C\,{\sc iv} and Si\,{\sc iv} BAL troughs show no detectable
deviation from symmetry about zero. This indicates that BALs do not strengthen
and weaken at significantly different rates. Both the $\Delta$EW and 
$\Delta$EW/$\langle$EW$\rangle$ distributions appear non-Gaussian. 
See Section~\ref{vdist}.

\item
We have assessed, in several timescale ranges, correlations between
BAL EW variations and BAL-profile properties, including average EW,
velocity width, depth, and outflow velocity. Weak BAL troughs tend
to have smaller absolute EW variations but larger fractional EW
variations than strong troughs. Decomposing EW into 
velocity-width-based and depth-based components, BAL EW variability 
depends more strongly upon velocity width.  We found no correlation 
between absolute EW variations and trough outflow velocity. See Section~\ref{ewew}.

\item
The $\Delta$EW and $\Delta$EW/$\langle$EW$\rangle$ values of C\,{\sc iv} BAL 
troughs correlate with those of Si\,{\sc iv} troughs corresponding in velocity. 
The $\Delta$EW correlation has significant scatter, perhaps suggesting 
a non-linear correlation. The $\Delta$EW/$\langle$EW$\rangle$ correlation shows 
that the fractional EW variation of C\,{\sc iv} BAL troughs is about half that 
of Si\,{\sc iv} troughs. See Section~\ref{csibal}.

\item
Our large sample clearly establishes that when a BAL quasar shows multiple 
troughs of the same ion, these troughs usually strengthen or weaken together. 
We explore this phenomenon for C\,{\sc iv}, finding that $\approx 78$\% of 
high-velocity troughs vary in a coordinated manner with the lowest-velocity trough. 
Coordinated trough variations are found even for velocity offsets as large as 
\hbox{15000--20000}~km~s$^{-1}$. See Section~\ref{multiple}.

\item
We have examined if BAL EW variations on several timescales depend upon 
quasar properties, including
$L_{\rm Bol}$, $L_{\rm Bol}/L_{\rm Edd}$, $M_{\rm BH}$, redshift, and 
radio loudness. Within the ranges of these properties spanned by our
sample, we do not find any strong dependences. There may be a hint 
that C\,{\sc iv} EW variations depend on 
$L_{\rm Bol}$ and/or $L_{\rm Bol}/L_{\rm Edd}$. 
See Section~\ref{qparam}.

\end{enumerate}

Considering the observational results above, we discuss implications for quasar 
outflows. The main implications are the following: 

\begin{enumerate}

\item
The high observed frequency of BAL variability on multi-year timescales is 
generally supportive of models where most BAL absorption arises at radii 
of \hbox{10--1000} light days. However, a significant minority of BAL troughs 
are remarkably stable and may be associated with absorption on 
larger scales. See Section~\ref{dvar}.

\item
Our measurements of EW variations as a function of timescale can be used
to infer an (EW-dependent) average lifetime for a BAL trough along our line-of-sight of 
$\langle t_{\rm{BAL}}\rangle = 1600^{+2100}_{-900}$~yr. This average 
lifetime is long compared to the orbital time of the accretion disk 
at the wind-launching or BLR radius. See Section~\ref{dcon}.

\item
Comparison of the $\Delta$EW and $\Delta$EW/$\langle$EW$\rangle$ distributions 
with previously identified examples of BAL disappearance indicates that 
BAL disappearance is an extreme type of general BAL variability, rather than 
a qualitatively distinct phenomenon. The same appears to apply for BAL
emergence. See Section~\ref{disem2}.

\item
We examine the extent to which a simple one-dimensional random-walk model can
explain the evolution of BAL-trough EWs. The binomial distribution resulting 
from this model can acceptably fit the $\Delta$EW distribution for C\,{\sc iv} 
BAL troughs on timescales of \hbox{2.0--2.5}~yr, where we have the best trough
statistics. The best-fitting model has a step size of 1.65~\AA\ and six 
random-walk steps, corresponding to a rest-frame step timescale of \hbox{$\approx137$~days}. 
The small derived number of steps is consistent with the non-Gaussian nature of 
the $\Delta$EW distribution, and we derive consistent random-walk parameters 
when considering the $\Delta$EW distribution for Si\,{\sc iv} troughs. 
A simulation of randomly walking trough EWs shows the best-fitting model 
derived for timescales of \hbox{2.0--2.5}~yr also acceptably represents 
EW variations over a wider range of timescales. See Section~\ref{ranw}.

\item
Coordinated trough variability for BAL quasars showing multiple troughs implies
the existence of a coordinating mechanism capable of affecting outflow
streamlines spread over a wide range of radii. The mechanism may be 
changes in the shielding gas that lead to changes in the level of EUV/X-ray 
radiation reaching the streamlines. Such changes in shielding gas may be 
a driver of BAL variability more generally. See Section~\ref{shield}. 

\end{enumerate}

The ongoing BOSS ancillary project on quasar-wind variability continues to
enlarge the sample of BAL quasars with high-quality spectra spanning multiple
years, and we hope to obtain an additional \hbox{1--2} epochs of observation
for the associated BAL quasars as part of the Time Domain Spectroscopic Survey
(TDSS) of the SDSS-IV.\footnote{The current planning for SDSS-IV is briefly
described at http://www.sdss3.org/future/} These data should enable multiple
further studies of BAL variability, and here we highlight a few selected ways
that the main results found above might be advanced. First, in this paper we
have focused upon variability of the relatively high-ionization transitions
of C\,{\sc iv}  and Si\,{\sc iv}, and it would be worthwhile to extend large-sample 
analyses to other transitions including Mg\,{\sc ii}, Al\,{\sc iii}, Fe\,{\sc ii}, and 
Fe\,{\sc iii} \citep[e.g.,][]{hall11,vivek12}. The outflow structure along
the line-of-sight may differ significantly when such lower ionization
transitions are present. Second, additional data can improve constraints
upon the fraction of BALs showing variability, BAL lifetimes, and our
simple random-walk model for BAL EW variability. Ideally, one would also like
to test our measurements of BAL variability against specific predictions made by
advancing models of quasar winds. Third, our results on coordinated
trough variability for BAL quasars showing multiple troughs indicate that
large-scale investigations of BAL vs. continuum variability are important
to pursue. Continuum variability investigations in the rest-frame UV should
be possible once the BOSS spectrophotometric calibration is improved, and
limited X-ray continuum monitoring should be possible with targeted 
observations and the \textit{eROSITA} all-sky survey \citep[e.g.,][]{merloni12}.
Fourth, a large-scale investigation of BAL vs. emission-line variability
should give unique insights into the quasar-wind contribution to
high-ionization line production. 
Finally, the full utilization (see Section~\ref{bal2}) and gathering of
additional observational epochs will allow assessment of how various
types of BAL troughs evolve, and perhaps coevolve, over timescales of
days-to-years. Additional epochs will also allow a large-scale search
for BAL acceleration events which have generally proved elusive to date
 \citep[e.g.,][]{gibson08,gibson10,cap12}.

\acknowledgments

We gratefully acknowledge financial support from National Science Foundation 
grant AST-1108604 (NFA, WNB, DPS) and from NSERC (PBH). 
We thank K. Dawson, M. Eracleous, and D. Schlegel for helpful discussions. 
We also thank the anonymous referee for constructive feedback. 

Funding for SDSS-III has been provided by the Alfred P. Sloan Foundation, 
the Participating Institutions, the National Science Foundation, and the U.S. 
Department of Energy Office of Science. The SDSS-III web site is 
http://www.sdss3.org/.

SDSS-III is managed by the Astrophysical Research Consortium for the 
Participating Institutions of the SDSS-III Collaboration including the University 
of Arizona, the Brazilian Participation Group, Brookhaven National Laboratory, 
University of Cambridge, Carnegie Mellon University, University of Florida, the 
French Participation Group, the German Participation Group, Harvard University, 
the Instituto de Astrofisica de Canarias, the Michigan State/Notre Dame/JINA 
Participation Group, Johns Hopkins University, Lawrence Berkeley National 
Laboratory, Max Planck Institute for Astrophysics, Max Planck Institute for 
Extraterrestrial Physics, New Mexico State University, New York University, 
Ohio State University, Pennsylvania State University, University of Portsmouth, 
Princeton University, the Spanish Participation Group, University of Tokyo, 
University of Utah, Vanderbilt University, University of Virginia, University of 
Washington, and Yale University.

\clearpage

\begin{deluxetable*}{lrccc} 
\tabletypesize{\footnotesize}
\tablecolumns{5} 
\tablewidth{0pc} 
\tablecaption{Sample-Based Studies of BAL Quasar Variability} 
\tablehead{ 
\colhead{ Reference}  & \colhead{\# of Quasars} & \colhead{$\Delta t$ Range }  
 & \colhead{\# of Epochs }  \\
\colhead{ }  & \colhead{} & \colhead{(yr) }  
 & \colhead{ }  }
\startdata 
\citet{barlow93} & 23 &0.2--1.2 & 2--6 \\
\citet{lundgren07} & 29 & 0.05--0.3 & 2  \\
\citet{gibson08} & 13 & 3.0--6.1 & 2  \\
\citet{gibson10} & 14& 0.04--6.8& 2--4\\
\citet{cap11,cap12,cap13} & 24&0.02--8.7& 2--13\\
\citet{vivek12}\tablenotemark{a} & 5 & 0.01--5& 4--14 \\
\citet{haggard12} & 17 & 0.001--0.9 & 6 \\
\citet{ak12}\tablenotemark{b}  & 19 & 1.1--3.9 & 2--4  \\
\citet{welling13}\tablenotemark{c} & 46& 0.2--16.4 & 2--6& \\
This study & 291 & 0.0006--3.7& 2--12\\
Full BOSS Ancillary & 2105 & 0.0006--6 & 2--12
\enddata
\label{varet1}
\tablenotetext{a}{Fe low-ionization BAL quasars} 
\tablenotetext{b}{Quasars with disappearing BAL troughs}
\tablenotetext{c}{Radio-loud BAL quasars}
\end{deluxetable*}

{
\begin{table*}
\caption{C\,{\sc iv} BAL Troughs}  \label{balc}
\begin{center}
{\scriptsize 
\begin{tabular}{ccrrrrccrrrrrrrrrrrrrrrrrrrrrrrrrrrrrrrrr}
\tableline\tableline \\[-0.3em]
 C\,{\sc iv}& {Quasar Name} & \multicolumn{1}{c}{RA} & \multicolumn{1}{c}{Dec} &\multicolumn{1}{c} {$z$} 
&\multicolumn{1}{c}{$\sigma_z$} & {LoBAL Flag}\tablenotemark{a} & {Timescale Flag}\tablenotemark{b} 
& {Plate[1]} & {MJD[1]} &  \nodata\\
 {Trough ID} &  & J2000 & J2000 & & &  &  &  & &  &  \\
 \\[-0.3em]
\tableline\\[-0.6em]
C1 & J001502.26+001212.4 & 3.75943 & 0.20346 & 2.8525 & 0.00055 & 0 & 1 &  389 & 51795 &  \\
C2 & J001502.26+001212.4 & 3.75943 & 0.20346 & 2.8525 & 0.00055 & 0 & 1 &  389 & 51795 &  \\
C3 & J003135.57+003421.2 & 7.89823 & 0.57257 & 2.2364 & 0.00026 & 0 & 1 &  689 & 52262 &  \\
C4 & J003135.57+003421.2 & 7.89823 & 0.57257 & 2.2364 & 0.00026 & 0 & 1 &  689 & 52262 &  \\
C5 & J003517.95+004333.7 & 8.82481 & 0.72604 & 2.9169 & 0.00055 & 0 & 1 & 1086 & 52525 &  \\
\\[-0.6em]
\hline \\[0.6em]
\end{tabular}

\begin{tabular}{rrrrrrrrrrrrrrrrrrrrrrrrrrrrrrrrrrrrrrrrr}
\tableline\tableline \\[-0.3em]

{Fiber[1]} & {SN$_{1700}$[1]} & {Plate[2]} & {MJD[2]} & {Fiber[2]}& {SN$_{1700}$[2]} 
&\multicolumn{1}{c}{$\Delta t$} &\multicolumn{1}{c}{$v_{\rm max}$} &
\multicolumn{1}{c}{$v_{\rm min}$} & \multicolumn{1}{c}{$v_{\rm cent}$[1]} & 
\multicolumn{1}{c}{$v_{\rm cent}$[2] }&  \nodata \\

  & & &  &  &  & \multicolumn{1}{c}{(days)} &  \multicolumn{1}{c}{(km\,s$^{-1}$)} &  
 \multicolumn{1}{c}{(km\,s$^{-1}$)} & \multicolumn{1}{c}{(km\,s$^{-1}$)} &  
 \multicolumn{1}{c}{(km\,s$^{-1}$)}&  & \\

\\[-0.3em]
\tableline\\[-0.6em]

465 & 10.62 & 4218 & 55479 & 818 & 21.43 & 956.2525 & $-$23238.4 & $-$19878.0 &   $-$21752.0 & $-$21650.5 &  \\ 
465 & 10.62 & 4218 & 55479 & 818 & 21.43 & 956.2525 & $-$10478.3 &   $-$5841.4 &   $-$8427.4 & $-$8285.5  &   \\ 
502 & 15.06 & 3587 & 55182 & 570 & 24.52 & 902.2298 & $-$21045.9 & $-$18737.9 & $-$19810.5 & $-$19820.2   &  \\ 
502 & 15.06 & 3587 & 55182 & 570 & 24.52 & 902.2298 & $-$12539.9 &   $-$4858.5 &    $-$9649.1 & $-$9361.5 &   \\ 
481 & 12.27 & 3587 & 55182 & 722 & 14.60 & 678.3438 & $-$20339.7 & $-$16881.9 & $-$18683.1 & $-$18626.6   &  \\ 
 \\[-0.6em]
\hline \\[0.6em]
\end{tabular}

\begin{tabular}{rrrrrrrrrrrrrrrrrrrrrrrrrrrrrrrrrrrrrrrrr}
\tableline\tableline \\[-0.3em]
$d_{\rm BAL}$[1] & $\sigma_{d_{\rm BAL}}$[1] & $d_{\rm BAL}$[2] & $\sigma_{d_{\rm BAL}}$[2] & EW[1] & 
$\sigma_{\rm EW}$[1]   & EW[2] & $\sigma_{\rm EW}$[2] & $\Delta \rm{EW}$ & $\sigma_{\Delta EW}$ & 
$\frac {\Delta\rm{EW}} {\langle \rm{EW} \rangle}$ & $\sigma_{\frac {\Delta\rm{EW}}  {\langle \rm{EW} \rangle}}$  
 & $N_{\rm VR}$\tablenotemark{c}& 
 \nodata \\

& & & &(\AA\,) & (\AA\,)& (\AA\,) & (\AA\,)&  (\AA\,) & (\AA\,)&  & \\
\\[-0.3em]
\tableline 
0.209 & 0.0227 & 0.134 & 0.0061 & 3.71 & 0.59  & 2.36 & 0.16 & $-$1.34 & 0.61 &$-$0.4424 & 0.2147     & 1 \\ 
0.464 & 0.0337 & 0.436 & 0.0298 & 11.51 & 0.56 & 10.33 & 0.13 & $-$1.19 & 0.58 & $-$0.1086 & 0.0611   & 0 \\ 
0.136 & 0.0242 & 0.213 & 0.0220 & 1.65 & 0.13  & 2.52 & 0.07 & 0.87 & 0.15 & 0.4150 & 0.1025          & 0 \\ 
0.690 & 0.0253 & 0.732 & 0.0246 & 26.86 & 0.20 & 28.47 & 0.10 & 1.61 & 0.22 & 0.0581 & 0.0109         & 2 \\ 
0.225 & 0.0222 & 0.224 & 0.0124 & 3.99 & 0.37  & 3.94 & 0.25 & $-$0.05 & 0.44 & $-$0.0120 & 0.1549    & 0 \\ 
\\[-0.6em]
\hline \\[0.6em]
\end{tabular}

\begin{tabular}{rrcrrrrrrrrrrrrrrrrrrrrrrrrrrrrrrrrrrrrrrr}
\tableline\tableline \\[-0.3em]
$\sum \Delta v_{\rm VR}$ & $f_{\Delta v}$ & Corresp. Si\,{\sc iv}\tablenotemark{d} &$\log L_{\rm {Bol}}$ & $\sigma_{\log L_{\rm {Bol}}}$ & $\log M_{\rm BH}$ &
$\sigma_{\log M_{\rm BH}}$ & $\log \frac{L_{\rm Bol}}{L_{\rm Edd}}$ &  \multicolumn{1}{c}{$M_i$} &  \multicolumn{1}{c} {$R$} \\

\multicolumn{1}{c}{(km\,s$^{-1}$)} & &Trough ID & \multicolumn{1}{c}{(erg\,s$^{-1}$)} & \multicolumn{1}{c}{(erg\,s$^{-1}$)} 
& \multicolumn{1}{c}{(M$_\odot$)}  & \multicolumn{1}{c}{(M$_\odot$)} \\
\\[-0.3em]
\tableline 

 274.7 & 0.082 & \nodata&46.97 & 0.020 & 9.34 & 0.121 & $-$0.468 & $-$27.682 & $-1$ \\
0 & 0 &S1& 46.97 & 0.020& 9.34 & 0.121 & $-$0.468 & $-$27.682 & $-1$ \\
0 & 0 &\nodata& 47.04 & 0.007 & 9.38 & 0.117 & $-$0.441 & $-$27.654 & $-1$ \\
1655.7 & 0.216 &S2& 47.04 & 0.007 & 9.38 & 0.117 & $-$0.441 & $-$27.654 & $-1$ \\
0 & 0 &\nodata &46.88 & 0.014 & 9.86 & 0.182 & $-$1.076 & $-$27.610 & $-1$ \\  
\\[-0.6em]
\hline 
\end{tabular}}
\footnotetext[1]{1 for low-ionization BAL quasars, 0 otherwise}
\footnotetext[2]{1 for timescales of more than 1~yr ($\Delta t_{\rm min,1}$), 0 otherwise ($\Delta t_{\rm min}$)}
\footnotetext[3]{Number of variable regions found in the BAL trough}
\footnotetext[4]{Corresponding Si\,{\sc iv} BAL trough ID as given in Table~\ref{balsi}}
\end{center}
\tablecomments{(This table is available in its entirety in a machine-readable form 
in the online journal. A portion is shown here for guidance regarding its form and content.)}
\tablecomments{Throughout this table [1] indicates the first-epoch spectra and [2] indicates the second-epoch spectra.}
\label{balc}
\end{table*}

}

{
\begin{table*}
\caption{Si\,{\sc iv} BAL Troughs}
\begin{center}
{\scriptsize 
\begin{tabular}{ccrrrrccrrrrrrrrrrrrrrrrrrrrrrrrrrrrrrrrr}
\tableline\tableline \\[-0.3em]
 Si\,{\sc iv} & {Quasar Name} & \multicolumn{1}{c}{RA} & \multicolumn{1}{c}{Dec} &\multicolumn{1}{c} {$z$} 
&\multicolumn{1}{c}{$\sigma_z$} & {LoBAL Flag}\tablenotemark{a} & {Timescale Flag}\tablenotemark{b} 
& {Plate[1]} & {MJD[1]} &  \nodata\\
{Trough ID}  &  & J2000 & J2000 & & &  &  &  & &  &  \\
  \\[-0.3em]
\tableline\\[-0.6em]

S1 & J001502.26+001212.4 & 3.75943 & 0.20346 & 2.8525 & 0.00055 & 0 & 1 & 389 & 51795 & \\
S2 & J003135.57+003421.2 & 7.89821 & 0.57260 & 2.2364 & 0.00026 & 0 & 1 & 689 & 52262 & \\
S3 & J003517.95+004333.7 & 8.82481 & 0.72604 & 2.9169 & 0.00055 & 0 & 1 & 1086 & 52525 & \\
S4 & J004613.54+010425.7 & 11.55643 & 1.07381 & 2.1492 & 0.00028 & 0 & 1 & 691 & 52199 & \\
S5 & J005419.99+002727.9 & 13.58329 & 0.45776 & 2.5143 & 0.00025 & 0 & 1 & 394 & 51913 & \\
\\[-0.6em]
\hline \\[0.6em]
\end{tabular}

\begin{tabular}{rrrrrrrrrrrrrrrrrrrrrrrrrrrrrrrrrrrrrrrrr}
\tableline\tableline \\[-0.3em]

{Fiber[1]} & {SN$_{1700}$[1]} & {Plate[2]} & {MJD[2]} & {Fiber[2]}& {SN$_{1700}$[2]} 
&\multicolumn{1}{c}{$\Delta t$} &\multicolumn{1}{c}{$v_{\rm max}$} &
\multicolumn{1}{c}{$v_{\rm min}$} & \multicolumn{1}{c}{$v_{\rm cent}$[1]} & 
\multicolumn{1}{c}{$v_{\rm cent}$[2] }&  \nodata \\

  & & &  &  &  & \multicolumn{1}{c}{(days)} &  \multicolumn{1}{c}{(km\,s$^{-1}$)} &  
 \multicolumn{1}{c}{(km\,s$^{-1}$)} & \multicolumn{1}{c}{(km\,s$^{-1}$)} &  
 \multicolumn{1}{c}{(km\,s$^{-1}$)}&  & \\

\\[-0.3em]
\tableline\\[-0.6em]

465 & 10.62 &4218 & 55479 & 818 & 21.43 & 956.2525  & $-$12074.1  & $-$5872.0  & $-$8972.6  & $-$8908.4 &  \\
502 & 15.06 &3587 & 55182 & 570 & 24.52 & 902.2298  & $-$12521.9  & $-$4933.6  & $-$8905.7  & $-$8831.1 &  \\ 
 481 & 12.27&3587 & 55182 & 722 & 14.60 & 678.3438  & $-$5224.5  & $-$3000.0  & $-$4286.6  & $-$4296.7 &     \\ 
 460 & 23.70&3589 & 55186 & 866 & 28.94 & 948.4810  & $-$20051.0  & $-$9612.2  & $-$14741.9  & $-$14841.3 & \\ 
 511 & 31.12&4224 & 55481 & 726 & 25.32 & 1015.2949  & $-$11686.2  & $-$5045.8  & $-$8284.2  & $-$8293.7 &    \\ 
 \\[-0.6em]
\hline \\[0.6em]
\end{tabular}

\begin{tabular}{rrrrrrrrrrrrrrrrrrrrrrrrrrrrrrrrrrrrrrrrr}
\tableline\tableline \\[-0.3em]
$d_{\rm BAL}$[1] & $\sigma_{d_{\rm BAL}}$[1] & $d_{\rm BAL}$[2] & $\sigma_{d_{\rm BAL}}$[2] & EW[1] & 
$\sigma_{\rm EW}$[1]   & EW[2] & $\sigma_{\rm EW}$[2] & $\Delta \rm{EW}$ & $\sigma_{\Delta EW}$ & 
$\frac {\Delta\rm{EW}} {\langle \rm{EW} \rangle}$ & $\sigma_{\frac {\Delta\rm{EW}}  {\langle \rm{EW} \rangle}}$  
 & $N_{\rm VR}$\tablenotemark{c}& 
 \nodata \\

& & & &(\AA\,) & (\AA\,)& (\AA\,) & (\AA\,)&  (\AA\,) & (\AA\,)&  & \\

\\[-0.3em]
\tableline 
0.221 & 0.0159 & 0.139 & 0.0059 &  6.41 & 0.68    &3.97 & 0.12  & $-$2.44 & 0.69  & $-$0.4707 & 0.1292 &   1 \\ 
 0.311 & 0.0158 & 0.361 & 0.0175 &  10.84 & 0.24  &12.60 & 0.12 & 1.76 & 0.27 & 0.1499 & 0.0315 &             1 \\ 
0.439 & 0.0456 & 0.432 & 0.0437 &  4.76 & 0.21    &4.69 & 0.13  & $-$0.07 & 0.25  & $-$0.0144 & 0.0731 &   0 \\ 
0.223 & 0.0058 & 0.171 & 0.0082 &  10.46 & 0.29&8.16 & 0.17  & $-$2.30 & 0.34  & $-$0.2467 & 0.0484 &     3 \\ 
0.200 & 0.0087 & 0.174 & 0.0107 &  6.11 & 0.16   &5.33 & 0.12  & $-$0.78 & 0.20  & $-$0.1371 & 0.0475 &    3 \\  
\\[-0.6em]
\hline \\[0.6em]
\end{tabular}

\begin{tabular}{rrcrrrrrrrrrrrrrrrrrrrrrrrrrrrrrrrrrrrrrr}
\tableline\tableline \\[-0.3em]
$\sum \Delta v_{\rm VR}$ & $f_{\Delta v}$ & Corresp. C\,{\sc iv}\tablenotemark{d}& $\log L_{\rm {Bol}}$ & $\sigma_{\log L_{\rm {Bol}}}$ & $\log M_{\rm BH}$ &
$\sigma_{\log M_{\rm BH}}$ & $\log \frac{L_{\rm Bol}}{L_{\rm Edd}}$ &  \multicolumn{1}{c}{$M_i$} &  \multicolumn{1}{c} {$R$} \\

\multicolumn{1}{c}{(km\,s$^{-1}$)} & & Trough ID &\multicolumn{1}{c}{(erg\,s$^{-1}$)} & \multicolumn{1}{c}{(erg\,s$^{-1}$)} 
& \multicolumn{1}{c}{(M$_\odot$)}  & \multicolumn{1}{c}{(M$_\odot$)} \\
\\[-0.3em]
\tableline 

276.1 & 0.045 & C2&46.97 & 0.020 & 9.342 & 0.121  & $-$0.468  & $-$27.682  & $-$1 \\
345.1 & 0.046 & C4&47.04 & 0.007 & 9.380 & 0.117  & $-$0.441  & $-$27.654  & $-$1 \\
         0 & 0 & \nodata&46.88 & 0.014 & 9.855 & 0.182  & $-$1.076  & $-$27.610  & $-$1 \\
1858.4 & 0.178 &\nodata &47.08 & 0.009 & 9.220 & 0.220  & $-$0.245  & $-$28.404 & 11.4 \\
1311.2 & 0.198 & C10&47.07 & 0.005 & 9.652 & 0.043  & $-$0.681  & $-$28.209  & $-$1 \\
\\[-0.6em]
\hline 
\end{tabular}
\footnotetext[1]{1 for low-ionization BAL quasars, 0 otherwise}
\footnotetext[2]{1 for timescales of more than 1~yr ($\Delta t_{\rm min,1}$), 0 otherwise ($\Delta t_{\rm min}$)}
\footnotetext[3]{Number of variable regions found in the BAL trough}
\footnotetext[4]{Corresponding C\,{\sc iv} BAL trough ID as given in Table~\ref{balc}. As explained in Section~\ref{csibal}, 
we search for accompanying C\,{\sc iv} BAL troughs for Si\,{\sc iv} troughs with 
$-3000 > v_{\rm{cent}}> -13000$~km~s$^{-1}$. We found that five Si\,{\sc iv} troughs, S3, S13, S201, S202, and S237, have 
an accompanying C\,{\sc iv} trough that does not satisfy our BAL-trough identification criteria. }}
\end{center}
\tablecomments{(This table is available in its entirety in a machine-readable form in 
the online journal. A portion is shown here for guidance regarding its form and content.)}
\tablecomments{Throughout this table [1] indicates the first-epoch spectra and [2] indicates the second-epoch spectra.}
\label{balsi}
\end{table*}
}

{
\begin{table*}
\caption{C\,{\sc iv} BAL-Trough Variable Regions}
\begin{center}
{\scriptsize 
\begin{tabular}{ccrrcrrrrrr}
\tableline\tableline \\[-0.3em]
C\,{\sc iv}& {Quasar Name} & {MJD[1]} & {MJD[2]} & {C\,{\sc iv}}\tablenotemark{a} &
\multicolumn{1}{c}{$v_{\rm max}$} &
\multicolumn{1}{c}{$v_{\rm min}$} &
\multicolumn{1}{c}{$v_{\rm max,VR}$} &
\multicolumn{1}{c}{$v_{\rm min,VR}$} &
\multicolumn{1}{c}{$\Delta v_{\rm VR}$} &
\multicolumn{1}{c}{$v_{\rm nrt}$} \\
 
 { VR ID}  &  &  &  & {Trough ID}&\multicolumn{1}{c}{(km\,s$^{-1}$)} & \multicolumn{1}{c}{(km\,s$^{-1}$)} &
 \multicolumn{1}{c}{(km\,s$^{-1}$)} & \multicolumn{1}{c}{(km\,s$^{-1}$)}& 
 \multicolumn{1}{c}{(km\,s$^{-1}$)} &  \\

 \\[-0.3em]
\tableline\\[-0.6em]

CV1 & J001502.26+001212.4    & 51795 & 55479  &C1 & $-$23238.4  & $-$19878.0  & $-$23160.0  & $-$22885.4 & 274.7  & $-$0.85 \\
CV2 & J003135.57+003421.2    & 52262 & 55182  &C4 & $-$12539.9  & $-$4858.5  & $-$11608.3  & $-$10642.5 & 965.8  & $-$0.51 \\
CV3 & J003135.57+003421.2    & 52262 & 55182  &C4 & $-$12539.9  & $-$4858.5  & $-$10504.5  & $-$9814.5 & 690.0  & $-$0.18 \\
CV4 & J004323.43$-$001552.4 & 52261 & 55184  &C6 & $-$21405.5  & $-$10137.6  & $-$16564.3  & $-$15186.6 & 1377.7 & 0 \\
CV5 & J004323.43$-$001552.4 & 52261 & 55184  &C6 & $-$21405.5  & $-$10137.6  & $-$15048.8  & $-$12153.5 & 2895.3 & 0.40 \\

\\[-0.6em]
\hline \end{tabular}
}
\footnotetext[1]{BAL trough IDs as given in  Table~\ref{balc}}
\end{center}
\tablecomments{(This table is available in its entirety in a machine-readable form in 
the online journal. A portion is shown here for guidance regarding its form and content.)}
\tablecomments{Throughout this table [1] indicates the first-epoch spectra and [2] indicates the second-epoch spectra.}
\label{varrc}
\end{table*}
}

{
\begin{table*}
\caption{Si\,{\sc iv} BAL-Trough Variable Regions}
\begin{center}
{\scriptsize 
\begin{tabular}{ccrrcrrrrrr}
\tableline\tableline \\[-0.3em]
 Si\,{\sc iv}& {Quasar Name} & {MJD[1]} & {MJD[2]} & {Si\,{\sc iv}}\tablenotemark{a} &
\multicolumn{1}{c}{$v_{\rm max}$} &
\multicolumn{1}{c}{$v_{\rm min}$} &
\multicolumn{1}{c}{$v_{\rm max,VR}$} &
\multicolumn{1}{c}{$v_{\rm min,VR}$} &
\multicolumn{1}{c}{$\Delta v_{\rm VR}$} &
\multicolumn{1}{c}{$v_{\rm nrt}$} \\
 
 { VR ID} &  &  &  &{Trough ID} &\multicolumn{1}{c}{(km\,s$^{-1}$)} & \multicolumn{1}{c}{(km\,s$^{-1}$)} &
 \multicolumn{1}{c}{(km\,s$^{-1}$)} & \multicolumn{1}{c}{(km\,s$^{-1}$)}& 
 \multicolumn{1}{c}{(km\,s$^{-1}$)} &  \\
 \\[-0.3em]
\tableline\\[-0.6em]

SV1 & J001502.26+001212.4 & 51795 & 55479  &S1 & $-$12074.1  & $-$5872.0  & $-$8365.8  & $-$8089.7 & 276.1  & $-$0.24 \\
SV2 & J003135.57+003421.2 & 52262 & 55182  &S2 & $-$12521.9  & $-$4933.6  & $-$9137.9  & $-$8792.8 & 345.1 & 0.02 \\
SV3 & J004613.54+010425.7 & 52199 & 55186  &S4 & $-$20051.0  & $-$9612.2  & $-$19915.7  & $-$18952.7 & 963.0 & 0.91 \\
SV4 & J004613.54+010425.7 & 52199 & 55186  &S4 & $-$20051.0  & $-$9612.2  & $-$18126.9  & $-$17507.4 & 619.5 & 0.60 \\
SV5 & J004613.54+010425.7 & 52199 & 55186  &S4 & $-$20051.0  & $-$9612.2  & $-$11649.9  & $-$11374.0 & 275.9  & $-$0.61 \\

\\[-0.6em]
\hline 
\end{tabular}
}
\footnotetext[1]{BAL trough IDs as given in Table~\ref{balsi}}
\end{center}
\tablecomments{(This table is available in its entirety in a machine-readable form in 
the online journal. A portion is shown here for guidance regarding its form and content.)}
\tablecomments{Throughout this table [1] indicates the first-epoch spectra and [2] indicates the second-epoch spectra.}
\label{varrsi}
\end{table*}
}

\begin{deluxetable*}{ccccc} 
\tabletypesize{\footnotesize}
\tablecolumns{3} 
\tablewidth{0pc} 
\tablecaption{Variability Comparison Between the Lowest Velocity BAL 
Trough and Additional BAL Troughs } 
\tablehead{ 
&\\[1em]
\colhead{}    &  \multicolumn{2}{c}{Number of  Additional C\,{\sc iv} Troughs }  \\ [0.5em]
\cline{2-3} \\ [0.5em]
\colhead{} & \colhead{$ \Delta\,EW < 0$}   & \colhead{$ \Delta\,EW > 0$} \\ }
\startdata \\[1em]
$\Delta$EW$_{\rm low~vel}<0$ & 57 & 13 \\[0.6em]
$\Delta$EW$_{\rm low~vel}>0$   & 17 & 50 \\ \\[0.9em]
\enddata
\label{vart1}
\end{deluxetable*}

\begin{deluxetable*}{cccccccc} 
\tabletypesize{\footnotesize}
\tablecolumns{8} 
\tablewidth{0pc} 
\tablecaption{Spearman Rank Correlation Test Probabilities for Quasar Properties vs. BAL-Trough EW Variations} 

\tablehead{ \\[1em]
\colhead{}    &  \multicolumn{3}{c}{C\,{\sc iv}} &   \colhead{}   & 
\multicolumn{3}{c}{Si\,{\sc iv}} \\[1em]
\cline{2-4} \cline{6-8} \\[1em]
\colhead{} & \colhead{$ < 1$ yr}   & \colhead{1--2.5 yr}    & \colhead{ $> 2.5$ yr} & 
\colhead{}    & \colhead{$ < 1$ yr}   & \colhead{1--2.5 yr}    & \colhead{$> 2.5$ yr}
\\[1em]}
\startdata 
\\[1em]
 \multicolumn{8}{c}{$L_{\rm {Bol}}$} \\[1em] \hline \\[1em]
$|\Delta\,EW|$   & 78.4 &  0.2 & 5.2 & & 65.3 &  5.5 &  79.8\\[1.3em]
 
$|\frac{\Delta\,EW}{\langle EW \rangle}|$  & 31.5 & 83.2 & 67.1 & & 44.3 & 90.6 & 11.1\\[1.2em]
         \hline\\[1em]

\multicolumn{8}{c}{$L_{\rm Bol}/L_{\rm Edd}$} \\[1em] \hline \\[1em]
$|\Delta\,EW|$   & 54.6 & 0.1 & 35.1 & & 65.6 &  81.5 &  96.2\\[1.3em]
 
$|\frac{\Delta\,EW}{\langle EW \rangle}|$  & 26.3 & 2.4 & 25.1 & & 39.5 & 92.5 & 25.4\\[1.2em]
         \hline\\[1em]
         \multicolumn{8}{c}{$M_{\rm BH}$} \\[1em] \hline \\[1em]
$|\Delta\,EW|$   & 32.9 & 11.4 & 66.4 & & 56.1 &  85.1 &  66.6\\[1.3em]
 
$|\frac{\Delta\,EW}{\langle EW \rangle}|$  & 45.3 & 1.2 & 32.7 & & 38.5 & 85.1 & 7.8\\[1.2em]
         \hline\\[1em]
\multicolumn{8}{c}{$z$} \\[1em] \hline \\[1em]
$|\Delta\,EW|$   & 70.4 &  1.2 & 12.5 & & 49.9 &  31.1 & 6.8\\[1.3em]
 
$|\frac{\Delta\,EW}{\langle EW \rangle}|$  & 8.4 & 3.4 & 48.3 & & 32.9 & 16.9 & 30.9\\[1.2em]
         \hline\\[1em]
 Number of data points      & 82 & 220 & 126 & & 56  & 124  & 55 \\[1em]
\enddata 
\tablecomments{Numbers in this table show the Spearman rank correlation test probabilities as percentages.}
\label{vart2}
\end{deluxetable*}

\end{document}